\documentclass[twocolumn,amssymb,superscriptaddress,floatfix,showpacs,nofootinbib]{revtex4-1}

\usepackage{siunitx}
\usepackage{array}
\usepackage{amsmath,amssymb,color}
\usepackage{slashed}
\usepackage{setspace}
\usepackage{graphicx}
\usepackage{amssymb,amsfonts,multirow}
\usepackage{placeins}
\usepackage{pifont}
\usepackage{epsfig}
\usepackage{amssymb,url,bm}
\usepackage{graphicx}
\usepackage{xcolor,colortbl}
\usepackage{hyperref}
\usepackage{color}
\usepackage{mathtools,booktabs}

\usepackage{tikz-feynman}

\newcolumntype{G}{>{\columncolor{LightGray}}c}
\newcolumntype{R}{>{\columncolor{LightRed}}c}
\newcolumntype{C}{>{$}c<{$}}
\newcolumntype{?}{!{\vrule width 1pt}}
\newcolumntype{`}{!{\vrule width 1.5pt}}

\newcommand\al[1]{\alpha_{{}_{#1}}}

\definecolor{Gray}{gray}{0.85}
\definecolor{LightGray}{gray}{0.93}
\definecolor{LightGreen}{rgb}{0.88, 1, 0.88}
\definecolor{LightCyan}{rgb}{0.88,1,1}
\definecolor{LightRed}{rgb}{1, 0.85, 0.85}
\definecolor{LightYellow}{rgb}{1, 1, 0.85}
\definecolor{Yellow}{rgb}{1,1,0.05}
\definecolor{LightBlue}{rgb}{0.87, 0.94, 1}
\definecolor{white}{gray}{1}
\definecolor{black}{gray}{0}

\usepackage{cancel}

\long\def\del #1 \enddel { }

\definecolor{Black}{gray}{0}
\definecolor{Gray}{gray}{0.85}
\definecolor{LightGray}{gray}{0.93}
\definecolor{LightGreen}{rgb}{0.88, 1, 0.88}
\definecolor{LightCyan}{rgb}{0.88,1,1}
\definecolor{LightRed}{rgb}{1, 0.85, 0.85}
\definecolor{LightYellow}{rgb}{1, 1, 0.85}
\definecolor{LightBlue}{rgb}{0.87, 0.94, 1}
\definecolor{white}{gray}{1}

\AtBeginDocument{
\heavyrulewidth=.16em
\lightrulewidth=.1em
\cmidrulewidth=.03em
\belowrulesep=.4ex
\belowbottomsep=0pt
\aboverulesep=.4ex
\abovetopsep=0pt
\cmidrulesep=\doublerulesep
\cmidrulekern=.5em
\defaultaddspace=.5em
}

\def\beq{\begin{equation}}
\def\eeq{\end{equation}}

\def\bea{\arraycolsep .1em \begin{eqnarray}}
\def\eea{\end{eqnarray}}

\def\eq#1{(\ref{#1})}

\def\s0#1#2{\mbox{\small{$ \frac{#1}{#2} $}}}
\def\0#1#2{\frac{#1}{#2}}

\def\grgl{\:\hbox to -0.2pt{\lower2.5pt\hbox{$\sim$}\hss}{\raise3pt\hbox{$>$}}\:}
\def\klgl{\:\hbox to -0.2pt{\lower2.5pt\hbox{$\sim$}\hss}{\raise3pt\hbox{$<$}}\:}

\def\lsim{\mathrel{\rlap{\lower4pt\hbox{\hskip1pt$\sim$}}
    \raise1pt\hbox{$<$}}}                
\def\gsim{\mathrel{\rlap{\lower4pt\hbox{\hskip1pt$\sim$}}
    \raise1pt\hbox{$>$}}}                

\makeatletter
    \def\CT@@do@color{%
      \global\let\CT@do@color\relax
            \@tempdima\wd\z@
            \advance\@tempdima\@tempdimb
            \advance\@tempdima\@tempdimc
    \advance\@tempdimb\tabcolsep
    \advance\@tempdimc\tabcolsep
    \advance\@tempdima2\tabcolsep
            \kern-\@tempdimb
            \leaders\vrule
                    \hskip\@tempdima\@plus  1fill
            \kern-\@tempdimc
            \hskip-\wd\z@ \@plus -1fill }

\usepackage{tikz}

\begin{document}
\;\newpage
\hfill DO-TH 22/03
\aboverulesep = 0mm
\belowrulesep = 0mm

${}$\vskip0.80cm

\title{Fixed Points in Supersymmetric Extensions of the Standard Model}
\author{Gudrun Hiller}
\email{ghiller@physik.tu-dortmund.de}
\affiliation{TU Dortmund University, Department of Physics, Otto-Hahn-Str.4, D-44221 Dortmund, Germany}
\author{Daniel F.~Litim}
\email{d.litim@sussex.ac.uk}
\affiliation{\mbox{Department of Physics and Astronomy, U Sussex, Brighton, BN1 9QH, U.K.}}
\author{Kevin Moch}
\email{kevin.moch@udo.edu}
\affiliation{TU Dortmund University, Department of Physics, Otto-Hahn-Str.4, D-44221 Dortmund, Germany}

\begin{abstract}
We search  for weakly interacting  fixed points in  extensions of the minimally supersymmetric standard model (MSSM).
Necessary conditions  lead to three distinct classes of  anomaly-free extensions  involving either new quark singlets, new quark doublets, or a fourth generation.
While interacting fixed points arise prolifically in asymptotically free theories, their existence is significantly constrained as soon as some of the non-abelian gauge sectors are infrared free.
Performing a scan over  $\sim 200$k  different MSSM extensions using matter field multiplicities and the number of superpotential couplings as free parameters, we  find  mostly infrared conformal fixed points, and a  small subset with  ultraviolet ones. 
All settings predict  low-scale supersymmetry-breaking and a violation of $R$-parity.
Despite of residual interactions, the running of couplings out of asymptotically safe fixed points  is  logarithmic as in asymptotic freedom.
Some fixed points can be matched to the Standard Model  though the  matching scale  comes out  too low.
Prospects for higher matching scales and  asymptotic safety  beyond the MSSM are indicated.

\end{abstract}

\maketitle

\tableofcontents

\section{\bf Introduction}
\label{sec:introduction}

Supersymmetry (SUSY)  continues to be an important driver for particle physics and model building.
 Over the past decades, a plethora of supersymmetric extensions  have been constructed and scrutinised both  in theory and experiment as  appealing templates for  the next Standard Model (SM).
Thus far, however, the LHC has returned null results  \cite{ParticleDataGroup:2020ssz}, thereby strengthening earlier understandings from LEP  \cite{Barbieri:2000gf}. 
Clearly, this state of affairs requires to rethink model building paradigms and incentives, as well as vanilla parameter spaces  for  masses  and couplings in the ongoing quest for SUSY at colliders and beyond,~e.g.~\cite{Arkani-Hamed:2004ymt,Baer:2020kwz}.

New directions for model building have arisen recently from the theory frontier 
thanks to the  discovery of particle theories with interacting ultraviolet (UV)   fixed points~\cite{Bond:2016dvk,Bond:2018oco,Litim:2014uca,Bond:2017tbw,Bond:2019npq,Bond:2021tgu,Bond:2017lnq,Bond:2017suy}.
UV fixed points are key for a fundamental definition of quantum field theory,
in particular when asymptotic freedom is absent \cite{Bond:2016dvk,Bond:2018oco}.
 Without supersymmetry, they  have by now been observed abundantly in settings with simple~\cite{Litim:2014uca,Bond:2017tbw,Bond:2019npq,Bond:2021tgu} or semi-simple~\cite{Bond:2017lnq}
 gauge groups. 
 Yukawa interactions are key
  for theories  to become ``asymptotically safe" (a term  originally coined for 
 the field-theoretic UV completion of  gravity~\cite{Weinberg:1980gg}) and lead to salient  features
 such as   the taming of Landau poles, vacuum stability, 
  power-law running, and  full conformal symmetry in the high-energy limit.  

 In a recent stream of works \cite{Bond:2017wut,Kowalska:2017fzw,Bissmann:2020lge,Hiller:2019mou,Hiller:2020fbu,Bause:2021prv} these new model building ideas
  have been used to construct concrete extensions of the SM, with further  benefits: 
UV-safe SM extensions  can broadly be probed at colliders~\cite{Bond:2017wut,Kowalska:2017fzw}, introduce a characteristic novel type of flavor phenomenology \cite{Bissmann:2020lge},  and explain naturally  the discrepancies with the SM predictions in today's data on the  electron and muon anomalous magnetic moments~\cite{Hiller:2019mou,Hiller:2020fbu}, or the intriguing  flavor anomalies evidenced in rare $B$-meson decays~\cite{Bause:2021prv}, besides stabilizing the Higgs.

\begin{figure}
\centering
\includegraphics[width = 0.45\textwidth]{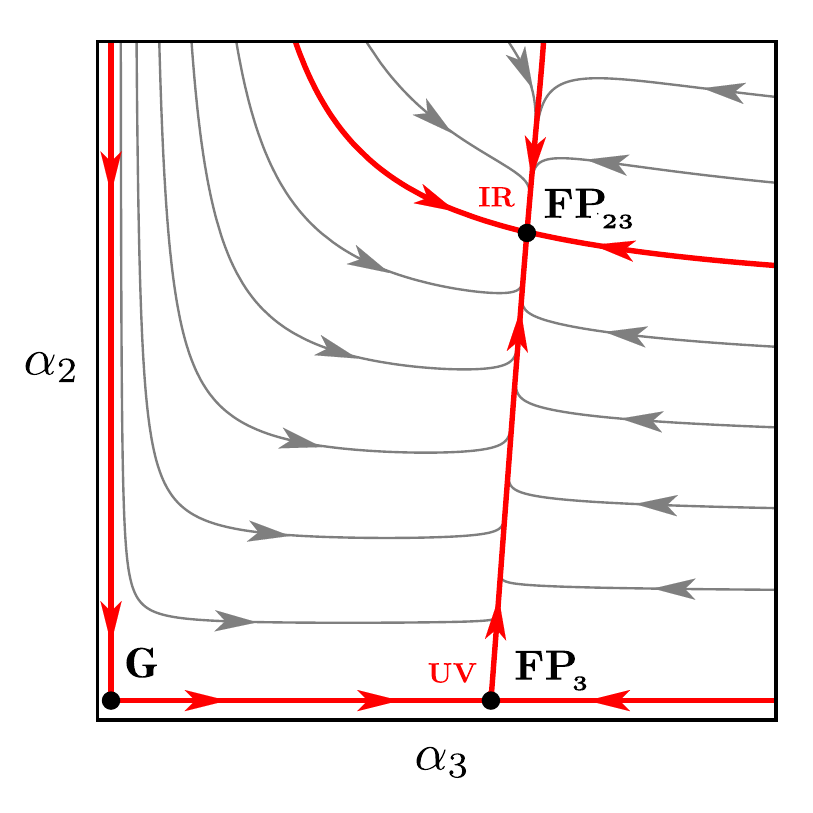}
\caption{Template phase diagram for an MSSM-like  quantum field theory with an interacting UV fixed point  in the plane of the weak $(\alpha_2)$ and strong $(\alpha_3)$ gauge couplings, also showing an interacting IR fixed point, the free gaussian fixed point (G), and sample renormalisation group trajectories with arrows pointing towards the IR.  Note that the gaussian must be a ``saddle" with both relevant and irrelevant perturbations. Plot adopted from \cite{Bond:2017suy}.}
\label{fig:templateRG}
\end{figure}

With supersymmetry, it was long believed 
that UV completions beyond asymptotic freedom may not exist~\cite{Martin:2000cr,Intriligator:2015xxa}. 
However,  a recent discovery~\cite{Bond:2017suy} has shown otherwise:
Yukawa  interactions  (tri-linear superpotential terms) continue to be  key \cite{Bond:2016dvk,Bond:2018oco}, except that gauge groups can no longer be simple and the gaussian must be a ``saddle" (see Fig.~\ref{fig:templateRG}).  Accordingly, one is led to stable, unitary, and asymptotically safe SUSY theories with superconformal symmetry in the high-energy limit~\cite{Bond:2017suy}.

In this paper,
we investigate whether  concrete and weakly coupled superconformal theories in the UV  can be found which connect with the known TeV-scale particle phenomenology at low energies.
For this, the minimally supersymmetric SM (MSSM) provides an ideal starting point: 
 its weak gauge sector is unstable and the scenario of  Fig.~\ref{fig:templateRG}  is naturally in reach,
 it offers  basic ingredients  for asymptotically safe SUSY theories such as several gauge groups and trilinear superpotential couplings \cite{Bond:2017suy},  
 it is phenomenologically acceptable and consistent with SM observations at low energies, 
 and
  it provides ample room for extensions which can be dialed-through in search for fixed points.

While we are particularly interested in UV fixed points, we will also search for IR (infrared) fixed points, which may coexist or arise independently.
Finding weakly interacting UV and IR fixed points in supersymmetric theories  is also of  interest because they corresponds to non-trivial superconformal field theories  \cite{Luty:2012ww}. 
 Moreover, IR fixed points and quasi IR fixed points have been known to exist in the MSSM for a long time, and have been explored in model building, including for third generation fermion masses
\cite{Allanach:1996nj,Lanzagorta:1995ai,Kobayashi:1996zu,Codoban:1999fp,Aulakh:2008sn,Abel:1998yi,Huang:2000rn,Nevzorov:2013ixa,Casas:1998vh,Barger:1993vu,Bardeen:1993rv}.

 The outline of this paper is as follows. In Sec.~\ref{sec:RG} we review the  renormalisation group equations for supersymmetric gauge theories with matter, and discuss necessary conditions and general features of perturbative fixed points. In Sec.~\ref{sec:MSSM}, we investigate UV and IR fixed points of the MSSM with conserved or broken $R$-parity. We further explain the rationale for  several new types of MSSM extensions and determine their respective fixed points. 
 In Sec.~\ref{sec:ASanalysis}, we  focus on the prospects of matching MSSM extensions with interacting UV fixed points to the SM at low energies.
  In Sec.~\ref{sec:conclusion} we discuss our results and conclude.
Auxiliary information is provided in several appendices.

\section{\bf Renormalisation Group for Supersymmetry \label{sec:RG}}

We consider $\mathcal{N}=1$ supersymmetric gauge theories with product gauge group
\begin{equation}
G=\prod_a G_a
\label{eq:gaugefactors}
\end{equation}
and gauge couplings $g_a$, where the index $a$ runs over simple and abelian group factors. 
Throughout we scale loop factors into the definition of couplings  and introduce 
\beq\label{alphaa}
\alpha_a=g_a^2/(4\pi)^2\,.
\eeq
We also consider chiral superfields $\Phi_i$, which may or may not carry local gauge charges, 
and which may further interact through a superpotential. Mass terms are of no relevance for this section 
and are neglected. 
Instead, we   consider the most general superpotential with canonically marginal couplings but omit 
canonically irrelevant interactions, hence
\beq\label{W}
W(\Phi)=\frac16\, {Y}^{ijk}\,\Phi_i \,\Phi_j\,\Phi_k
\eeq
with Yukawa couplings ${Y}^{ijk}$, ${Y}_{ijk}=({Y}^{ijk})^*$.
Yukawa couplings are a necessity for asymptotically safe UV fixed points to occur at weak coupling~\cite{Bond:2016dvk}.
We are particularly interested in theories which display interacting fixed points in the IR or UV. Conformal critical points correspond to free or interacting fixed points, which can be found using the renormalisation group equations.

\subsection{Renormalisation Group}

In perturbation theory  the renormalisation of the gauge couplings up to two-loop level
 in the $\overline{\text{DR}}$ scheme is given by~\cite{Machacek:1983tz,Martin:1993zk}\footnote{At the loop-levels considered in this work there is no difference between the schemes $\overline{\text{DR}}$ and $\overline{\text{MS}}$~\cite{Martin:1993yx}.}
 \bea\label{gaugeYukawa2}
\mu\frac{ d\alpha_a}{d\mu}\equiv  \beta_a&=& \alpha_a^2\left(-B_a\, +C_{ab}\,\alpha_b 
- 2\,Y_{4,a}
\right) \,,
\eea
with indices $a,b$ always referring to gauge couplings. The one-loop coefficients $B_a$ and the two-loop gauge coefficients $C_{ab}$ are given by
\begin{align}
	B_a &= 6\, C_2^{G_a} - 2\, S^{R_a}_2\,,\label{OneLoopa}\\
	C_{ab} &= 4\,C_2^{G_a}\left(S^{R_a}_2- 3\, C_2^{G_a}\right)\delta_{ab}
	+8\,S^{R_a}_2\,C^{R_b}_2\,.\label{TwoLoopa}
\end{align}
The subscripts $a,b$ on
the quadratic Casimir $(C^G_2)$ and on the Dynkin index ($S^R_2$) of the matter fields indicate the subgroup of $G$. Using \eq{OneLoopa} we may rewrite the two-loop term as
\beq\label{Cab}
C_{ab} = 	8\,S^{R_a}_2\,C^{R_b}_2-2\,C_2^{G_a}\,B_a\,\delta_{ab}
\eeq
in terms of the one-loop terms. Evidently, the mixing terms are manifestly non-negative
($C_{ab}\geq 0$ for $a\neq b$) for any semi-simple supersymmetric gauge theory. 
Also, for $B_a<0$ we have $C_{aa}>0$ (no sum) in any quantum field theory~\cite{Bond:2016dvk}.

The Yukawa couplings \eq{W} contribute to the running of the gauge couplings \eq{gaugeYukawa2} starting at the two-loop level, with
\bea
\label{Y4a}
Y_{4,a}&=&Y_{ijk}\,Y^{ijk}\, {C}^{R_a}_2(k)/d(G_a)\,,
\eea
and $d(G_a)$ denotes the dimension of group $G_a$.
Non-renormalisation theorems for the superpotential guarantee that the exact flow for the Yukawa couplings  $\beta_Y=d Y/d\ln \mu$  is given by
\beq\label{Yijk}
\begin{aligned}
\mu\frac{ d}{d\mu} Y^{ijk}
= Y^{ij\ell}\,\gamma^k_\ell+(i\leftrightarrow k)+(j\leftrightarrow k)
\end{aligned}
\eeq
to any order in perturbation theory.
Here, $\gamma^k_\ell$ denote the anomalous dimension matrix of the chiral superfields. 
In perturbation theory, they read at one-loop
\beq\label{gammaa}
\begin{array}{rcl}
\displaystyle
\gamma{}^{(1)}{}^k_\ell&=& 
\displaystyle
\frac12 Y_{ij\ell}\,Y^{ijk}-2\,\alpha_a \,C^{R_a}_{2}(k)\,\delta ^k_\ell \,.
\end{array}
\eeq

\subsection{UV and IR Fixed Points}

An important consistency condition arises through the flow of the superpotential couplings \cite{Martin:2000cr}. 
Taking the  sum of absolute values squared of all superpotential couplings, $|Y|^2=Y_{ijk}\,Y^{ijk}$, and also using  \eqref{gaugeYukawa2}, \eqref{gammaa}
we find
\beq
\label{eq:semiYuk2}
\begin{aligned}
	\frac{1}{12}\partial_t 
	|Y|^2&=d(R)|\gamma(R)|^2
	\\ & +\alpha_a \,d(G_a) \left(Y_{4,a} 
- 4\,\alpha_b \, S^{R_a}_{2}\,C^{R_b}_{2}\right)\,,
\end{aligned}
\eeq
with $t= \ln \mu$, $d(R)$ the dimension of the   representation $R$, $\gamma(R)$ the chiral superfield anomalous dimension, and Yukawas  rescaled as $Y_{ijk}\to Y_{ijk}/4\pi$. 
A fixed point requires  the simultaneous vanishing of all beta functions.
For the gauge couplings, $\beta_a = 0$ implies  
\begin{align}\label{gaugefix}
	2 Y_{4,a}^* &= -B_a + C_{ab}\,\alpha_b^*\,,
\end{align}
see \eqref{gaugeYukawa2}. Expressions with an '$*$' -superscript are understood as being evaluated at a fixed point. Using   \eq{Cab}, \eqref{eq:semiYuk2}, and \eqref{gaugefix}, we then find from $\partial_t|Y|^2=0$ that
\begin{align}
d(R)|\gamma_*(R)|^2
& =\s012\, B_a\, \alpha^*_a \,d(G_a)
\left(1+2\,C_2^{G_a}\,\alpha^*_a\right)
\end{align}
must hold true for any fixed point.
Since the left-hand-side is by definition a positive number,
positivity of the weighted sum
\beq\label{necessary1}
d(G_a) \,B_a\, \alpha^*_a >0\quad\text{together with}\quad\alpha_a^*\ge0
\eeq
is a necessary condition for  interacting fixed points \cite{Bond:2017suy}. 
For theories with a single gauge group this implies that asymptotically non-free theories $(B_a<0)$ cannot develop interacting fixed points \cite{Martin:2000cr,Bond:2017suy}.
However, for theories with several gauge groups,  \eqref{necessary1} mandates that at least one of the gauge factors is asymptotically free \cite{Bond:2017suy}, illustrated in  Fig.~\ref{fig:templateRG}.

A useful simplification  arises through choices in the Yukawa sector  (see App.\ref{sec:appendixunnaturalYukawas} for more details),
in which case the set of Yukawa couplings $\{Y_{ij\ell}\}$ can be mapped onto a set $\{y_i\}$ 
such that the RG beta functions for the Yukawa couplings squared are proportional to themselves. 
We  may then introduce
the Yukawa couplings as 
\beq\label{alphai}
\alpha_i=y_i^2/(4\pi)^2\,,
\eeq
and the  beta functions \eq{Yijk} with \eq{gammaa} turn into
\begin{equation}
\mu\frac{ \partial \alpha_i}{\partial \mu}\equiv \beta_i=\alpha_{i}\left[\sum_j E_{ij}\alpha_j-\sum_a F_{ia}\alpha_a\right]\;,
\label{generalYukawabetas}
\end{equation}
characterised by the  one-loop matrix $E_{ij}$ from superpotential contributions and the one-loop matrix $F_{ia}$ of gauge field contributions.
Throughout, indices $i,j$  relate to Yukawa couplings while indices $a,b$ relate to gauge couplings (to avoid confusion, we also have written out the required summations explicitly). Moreover, the two-loop Yukawa term \eqref{Y4a} simplifies into a linear combination of the Yukawa couplings,
\beq\label{Y4D}
Y_{4,a}=\s012 \sum_i D_{ai}\,\alpha_i
\eeq
for some coefficients $D_{ai}>0$. In these conventions, the beta functions for the gauge couplings \eq{gaugeYukawa2} reads
\begin{equation}\label{gauge2}
\beta_a=\alpha_a^2\left[-B_a+\sum_b C_{ab}\,\alpha_{b}-\sum_i D_{ai}\,\alpha_i\right]\;,
\end{equation}
with the one-loop coefficients $B_a$, the two-loop matrix $C_{ab}$ of  gauge field contributions and the two-loop matrix $D_{ai}$ from  the superpotential. In general,  the elements of the matrices $D, E$ and $F$ are positive or zero.

At weak coupling, theories may display  Banks-Zaks (BZ)  fixed points 
and/or gauge-Yukawa (GY) fixed points~\cite{Bond:2016dvk}.
The former are always IR, while the latter can be IR or UV. BZ fixed points are the solutions to  $\beta_a/\alpha_a^2=0$ with vanishing superpotential couplings $\alpha_i=0$, leading to
\begin{equation}\label{FPBZ}
B_a=\sum_b C_{ab}\,\alpha_{b}^*
\end{equation}
for each of the non-vanishing gauge couplings. Further, gauge-Yukawa fixed points additionally have non-vanishing superpotential couplings.
In this case, assuming that the matrix $E$ can be inverted,
we can solve $\beta_i/\alpha_i=0$ using~\eqref{generalYukawabetas} to find
the nullcline relation 
\begin{equation}
\alpha_i^*=\sum_{j,a}(E^{-1})_{ij}\, F_{ja}\,\alpha_a^*\,,
\label{Yukawanullcline}
\end{equation}
relating the non-vanishing Yukawa couplings to the gauge couplings.
After inserting \eq{Yukawanullcline} into \eq{gauge2}, and demanding  that $\beta_a/\alpha_a^2=0$,
we find the fixed point condition
\begin{equation}\label{FPGY}
B_a=\sum_b C'_{ab}\, \alpha_{b}^*
\end{equation}
for each of the non-vanishing gauge couplings. The matrix $C'$ can be viewed as a Yukawa-shifted two-loop matrix,
\begin{equation}
C'_{ab}=C_{ab}-\sum_{i,j}D_{ai}\,(E^{-1})_{ij}\,F_{jb}\;,
\label{eq:modcoeff}
\end{equation}
which follows from \eq{Cab} and \eq{gauge2}  after inserting \eq{Yukawanullcline}. As such, the shift $C\to C'$ takes into account the fact that
the superpotential couplings achieve a simultaneous fixed point.

In the following it turns out to be  convenient to introduce a notation to differentiate between different types of fixed points. 
If gauge couplings $\alpha_a$, $\alpha_b,\cdots$ remain non-zero at a fixed point, we refer to it as 
${\rm FP}_{ab\cdots}$, where the indices
relate to the non-zero gauge couplings, see Tab.~\ref{tab:FPnames}. Additionally, Yukawa couplings may or may
not be non-zero.

\subsection{Two Gauge Sectors}

To be concrete, we discuss a model with two gauge couplings $\alpha_2$ and $\alpha_3$ and a superpotential, and with \eq{gaugeYukawa2}.
This serves as a template for the $SU(2)\times SU(3)$ sector of MSSM extensions which is the focus of the following sections.
We are interested in interacting UV or IR fixed point in settings where asymptotic freedom is absent. Hence, \eqref{necessary1} mandates 
\beq
\label{B3>0>B2}
B_3> 0 > B_2\,,
\eeq 
or the other way around, but not both $B_{2,3}<0$.  
With \eqref{B3>0>B2}, $\alpha_2$ is a marginally irrelevant coupling close to the Gaussian, but it may 
become (marginally) relevant close to an interacting fixed point $\alpha^*_3$.  
In this setting, the sole BZ fixed point \eq{FPBZ} is given by
\beq\label{alpha_3_3}
\alpha_3^*=\frac{B_3}{C_{33}}\,,\quad \alpha_2^*=0\,.
\eeq
The option $\alpha_2^*>0$ is not available because $B_2<0$ with  \eq{Cab} entails $C_{22}>0$. In turn, two options arise for GY fixed points \eq{FPGY}.
First, the GY fixed point
may be partially interacting $({\rm FP}_3)$, in which case
\beq\label{FP3}
\alpha_3^*=\frac{B_3}{C'_{33}}\,,\quad \alpha_2^*=0
\eeq
alongside a non-trivial superpotential coupling. It requires that the shifted two-loop coefficient $C'_{33}$ is positive.
For this partially interacting fixed point to become a UV fixed point, it is required that $\alpha_2$ becomes marginally relevant in its vicinity (see Fig.~\ref{fig:templateRG}). Expanding $\beta_2$ for small $\alpha_2$ and in the vicinity of the interacting fixed point, we find 
\beq\label{B2-1}
\beta_2\big|_{{\rm FP}_3}=B_{2,\rm eff}\,\alpha_2^2+{\cal O}(\alpha_2^3)\,,
\eeq
with the effective one-loop coefficient $B_{2,\rm eff}$  now given by
\beq\label{B2eff}
B_{2,\rm eff}=B_2 - C'_{23}\,\alpha_3^*\,.
\eeq
Hence,  the sufficient condition for the fixed point ${\rm FP}_3$ to be UV reads
\beq\label{B2eff>0>B2}
B_{2,\rm eff}>0>B_2\,.
\eeq
which requires $C'_{23}<0$. 

Second, the required fixed point
may be fully interacting $({\rm FP}_{23})$. Using  \eq{FPGY} one obtains
\begin{align}
\label{alpha3_23}
\alpha_3^*&=\frac{B_3\,C'_{22}-B_2\,C'_{32}}{C'_{33}\,C'_{22}-C'_{23}\,C'_{32}}\,,\\ 
\label{alpha2_23}
\alpha_2^*&=\frac{B_2\,C'_{33}-B_3\,C'_{23}}{C'_{33}\,C'_{22}-C'_{23}\,C'_{32}}
\end{align}
to leading order in perturbation theory. Whether fixed points of this type are UV or IR  depends on the eigenvalue spectrum of the stability matrix.
Fig.~\ref{fig:templateRG} illustrates the scenario in which the partially interacting GY fixed point ${\rm FP}_3$ is UV, and the fully interacting ${\rm FP}_{23}$ is IR. Most notably, $\alpha_2$ has become marginally relevant owing to interactions at ${\rm FP}_3$.

\begin{table}[t]
  \centering
 \scalebox{0.9}{
  \begin{tabular}{`ccccc`}
\toprule
\rowcolor{Yellow}
\bf Fixed Point   & \multicolumn{3}{c}{\ \ \bf Gauge Couplings\ \ }
&\bf Type\\ 
\midrule
$\text{FP}_0$&$\alpha_3^*=0$ &$\ \ \ \alpha_2^*=0\ \ \ $ &$\alpha_1^*=0$&  free\\
\rowcolor{LightGray}
$\text{FP}_1$&$\alpha_3^*=0$ &$\alpha_2^*=0$ &$\alpha_1^*>0$&  \  interacting\ \ \\
$\text{FP}_2$&$\alpha_3^*=0$ &$\alpha_2^*>0$ &$\alpha_1^*=0$&   interacting\\
\rowcolor{LightGray}
$\text{FP}_3$&$\alpha_3^*>0$ &$\alpha_2^*=0$ &$\alpha_1^*=0$&   interacting\\
$\text{FP}_{12}$&$\alpha_3^*=0$ &$\alpha_2^*>0$ &$\alpha_1^*>0$&   interacting\\
\rowcolor{LightGray}
$\text{FP}_{13}$&$\alpha_3^*>0$ &$\alpha_2^*=0$ &$\alpha_1^*>0$&   interacting\\
$\text{FP}_{23}$&$\alpha_3^*>0$ &$\alpha_2^*>0$ &$\alpha_1^*=0$&   interacting\\
\rowcolor{LightGray}
$\text{FP}_{123}$&$\alpha_3^*>0$ &$\alpha_2^*>0$ &$\alpha_1^*>0$&   interacting\\
\bottomrule
  \end{tabular}
}
  \caption{Classification of fixed points according to the values of gauge couplings.
 Fixed points may be UV provided one or more Yukawa couplings are non-vanishing at the fixed point.}
\label{tab:FPnames}
\end{table}

\section{\bf MSSM and Extensions}
\label{sec:MSSM}

In this section we investigate fixed points of the MSSM with conserved (Sec.~\ref{sec:RPC}) and broken $R$-parity (Sec.~\ref{sec:RPV}). 
We then put forward strategies for interacting fixed points in MSSM extensions based on additional matter fields and Yukawa interactions  (Sec.~\ref{sec:construction}), and analyse three characteristic types of extensions in full detail (Secs.~\ref{sec:model1} -- \ref{sec:model3}).

\subsection{MSSM with \texorpdfstring{$R$}{R}-Parity \label{sec:RPC}}

We consider the SM gauge group
\begin{equation}
G_\text{SM}=SU(3)_C\times SU(2)_L\times U(1)_Y
\label{eq:SMgauge}
\end{equation}
and denote the hypercharge,  the weak and strong gauge couplings as
$\alpha_1,\alpha_2,$ and $\alpha_3$,
respectively, with $\alpha_a=g_a^2/(4\pi)^2$ $(a=1,2,3)$ and $g_a$ the gauge couplings.
The (left-handed) chiral superfields of the MSSM are summarised in Tab.~\ref{tab:MSSM}.
Consequently, the MSSM one-loop and two-loop gauge beta coefficients in~\eqref{gauge2} are given by
\begin{align}
\begin{array}{llll}
B_1 =-22\;,& C_{11} =\frac{398}{9}\;,& C_{12} =18\;,& C_{13} =\frac{176}{3}\;,\\
B_2 =-2\;,& C_{21} =6\;,& C_{22} =50\;,& C_{23} =48\;,\\
B_3 =6\;,& C_{31} =\frac{22}{3}\;,& C_{32} =18\;,& C_{33} =28\;.
\end{array}
\label{eq:MSSMGaugeCoeffs}
\end{align}
Notice that $B_2<0$ and $B_1<0$ imply that the hypercharge and the weak gauge sector are not asymptotically free,
and imply that the running gauge couplings $\alpha_1$ and $\alpha_2$ may both terminate in UV Landau poles unless they 
run into a fixed point in the UV.

In principle, there may arise up to seven distinct classes of interacting  fixed points depending on whether one, two, or three of the gauge couplings remain non-zero at the fixed point. For want of language, we distinguish these using the terminology  of Tab.~\ref{tab:FPnames}. For example, the class of fixed points FP${}_3$ refers to all possible fixed points where the hypercharge and weak gauge couplings vanish, the strong gauge coupling remains non-zero, and none, some, or all of the Yukawa couplings are non-zero. Notice that for fixed points of any type to be UV, at least one of the Yukawa couplings must be non-zero.

\begin{table}[t]
 		\addtolength{\tabcolsep}{3pt}
		\setlength{\extrarowheight}{3pt}			
  \sisetup {
    per-mode = fraction
  }
 \scalebox{0.75}{
  \begin{tabular}{`ccccc`}
\toprule
		\rowcolor{Yellow}
	\bf Superfield&$\bm{SU(3)_C}$&$\bm{SU(2)_L}$&$\bm{U(1)_Y}$& \bf Multiplicity\\
\midrule 
quark doublets $Q$&\textbf{3}&\textbf{2}&$+\frac{1}{6}$&$3$\\
\rowcolor{LightGray}
up-quark singlets $\overline{u}$&$\overline{\textbf{3}}$&\textbf{1}&$-\frac{2}{3}$&$3$\\
down-quark singlets $\overline{d}$&$\overline{\textbf{3}}$&\textbf{1}&$+\frac{1}{3}$&$3$\\
\rowcolor{LightGray}
lepton doublets $L$&\textbf{1}&\textbf{2}&$-\frac{1}{2}$&$3$\\
lepton singlets $\overline{e}$&\textbf{1}&\textbf{1}&$+1$&$3$\\
\rowcolor{LightGray}
up-Higgs $H_u$&\textbf{1}&\textbf{2}&$+\frac{1}{2}$&$1$\\
down-Higgs $H_d$&\textbf{1}&\textbf{2}&$-\frac{1}{2}$&$1$\\
\toprule
  \end{tabular}
}
  \caption{Summary of left-handed superfields in the MSSM.}
\label{tab:MSSM}
\end{table}

Next, we turn to the superpotential  of the MSSM. Besides anomaly-cancellation, we also impose invariance under $R$-parity~\cite{Farrar:1978xj,Dreiner:1997uz}, characterised  by the global $\text{U}(1)$ symmetry
\begin{equation}
P_R=(-1)^{3(B-L)+2s}\,.
\end{equation} 
Here $B$ , $L$ and $s$ are baryon number, lepton number and spin, respectively. The $R$-parity conserving superpotential of the MSSM reads
\begin{align}
\begin{aligned}
W_\text{MSSM}=&\ Y_d^{ij}\overline{d}_iQ_{j}H_d+Y_u^{ij}\overline{u}_iQ_{j}H_u+Y_e^{ij}\overline{e}_iL_jH_d\\&+\mu H_u H_d\;,
\label{eq:superpotentialMSSM}
\end{aligned}
\end{align}
where $i,j=1,2,3$ correspond to  flavor degrees of freedom, while gauge indices have been suppressed. 
As such, the MSSM may have up to $N_Y=27$ in general complex-valued Yukawa couplings. 
In this work, we are mostly interested in the case where the Yukawa matrices $Y_e$, $Y_d$ and $Y_u$ in~\eqref{eq:superpotentialMSSM} are approximated by 
$Y_e\approx 0$, $Y_d\approx\text{diag}(0,0,y_b)$, $Y_u\approx\text{diag}(0,0,y_t)$ with $y_b$ and $y_t$ denoting the bottom and top Yukawa couplings, respectively. The $\mu$-term is a mass term and does not play any role in the high energy limit of the theory and can be ignored in our study. The MSSM superpotential therefore reads
  \begin{equation}
W_\text{MSSM}\approx y_b\overline{d}_3Q_{3}H_d+y_t\overline{u}_3Q_{3}H_u\;.
\label{eq:superpotentialMSSM2}
\end{equation}
It  constitutes the backbone for the MSSM and MSSM extensions studied in the following.

We now turn to the fixed points of the MSSM with  the $R$-parity conserving  superpotential~\eqref{eq:superpotentialMSSM2}. 
In addition to the  gauge beta functions we  have the Yukawa beta functions for the bottom and top couplings $\alpha_{b,t}=|y_{b,t}|^2/(4\pi)^2$, thus a total of three gauge and two Yukawa couplings,
\begin{equation}\label{basic}
\{\alpha_1,\alpha_2,\alpha_3,
\alpha_b,\alpha_t\}\;.
\end{equation}
The beta functions \eq{generalYukawabetas} for the bottom and top Yukawas are given by
\begin{align}
\begin{aligned}
\beta_b&=\alpha_b\bigg[12\alpha_b+2\alpha_t-\frac{14}{9}\alpha_1-6\alpha_2-\frac{32}{3}\alpha_3\bigg]\;,\\
\beta_t&=\alpha_t\bigg[12\alpha_t+2\alpha_b-\frac{26}{9}\alpha_1-6\alpha_2-\frac{32}{3}\alpha_3\bigg]\;.
\label{eq:MSSMtopbottombetas}
\end{aligned}
\end{align}
The bottom and top Yukawa nullclines \eq{Yukawanullcline}, defined as the relations of couplings along which the top and bottom Yukawa beta functions~\eqref{eq:MSSMtopbottombetas} vanish, are given by
\begin{align}
\begin{aligned}
&\alpha_b=\frac{29}{315}\alpha_1+\frac{3}{7}\alpha_2+\frac{16}{21}\alpha_3\,,\\&
\alpha_t=\frac{71}{315}\alpha_1+\frac{3}{7}\alpha_2+\frac{16}{21}\alpha_3\;.
\label{eq:MSSMnullcline}
\end{aligned}
\end{align}
Inserting these into the gauge beta functions~\eqref{gauge2} with the MSSM coefficients~\eqref{eq:MSSMGaugeCoeffs}, and also using the two-loop Yukawa corrections to the running of the gauge couplings in \eq{gauge2}
\begin{align}
\begin{aligned}
D_{1b}&=\frac{28}{3}\;,\;D_{1t}=\frac{52}{3}\;,\; \\ D_{2b}&=12\;,\;\; D_{2t}=12\;,\; \\ D_{3b}&=8\;,\;\;\;\; D_{3t}=8\ \;,
\end{aligned}
\end{align}
we are  able to identify fixed point candidates. Since the hypercharge and  $SU(2)$ beta functions are both asymptotically non-free ($B_1,B_2<0$), the only possibility for an interacting fixed point in perturbation theory requires $B_3\,\alpha_3^*>0$, see \eqref{necessary1}.  We find that all interacting fixed point candidates of the type $\text{FP}_{13}$, $\text{FP}_{23}$, or $\text{FP}_{123}$ invariably imply either $\alpha_1^*<0$ or $\alpha_2^*<0$, which is unphysical.
The only viable setting is a fixed point of the type  $\text{FP}_3$, with trivial $(\alpha_1,\alpha_2)|_*=(0,0)$ and non-trivial coordinates
\begin{align}
\begin{aligned}
\left.(\alpha_3,\alpha_b,\alpha_t)\right|_*&=(\s0{63}{166},\s0{24}{83},\s0{24}{83})\\&\approx(0.38,0.29,0.29)\;.
\end{aligned}
\end{align}
Notice that both Yukawa couplings come out non-zero and unified.
The effective one-loop coefficients $B_{1,\text{eff}}$ and $B_{2,\text{eff}}$~\eqref{B2eff>0>B2} are negative
\begin{equation}
\begin{array}{rcl}
B_{1,\text{eff}}&=&-\s0{1102}{83}\approx -13.3\; ,\\
B_{2,\text{eff}}&=&-\s0{4242}{83}\approx-51.1\;,
\end{array}
\end{equation}
implying that the gauge-Yukawa  fixed point of the MSSM is IR.
We have also explicitly checked that this conclusion is robust against including the tau Yukawa coupling, and against  further finite entries in $Y_d$, $Y_u$ and $Y_e$ of the MSSM superpotential~\eqref{eq:superpotentialMSSM}.\\
\hphantom\quad We conclude that the MSSM does not offer 
interacting UV fixed points to the leading orders in perturbation theory.
For  investigations of IR fixed points or quasi IR fixed points in the MSSM or MSSM GUTs, we refer to the studies in
\cite{Allanach:1996nj,Lanzagorta:1995ai,Kobayashi:1996zu,Codoban:1999fp,Aulakh:2008sn,Abel:1998yi,Huang:2000rn,Nevzorov:2013ixa,Casas:1998vh,Barger:1993vu,Bardeen:1993rv}.

\subsection{MSSM without \texorpdfstring{$R$}{R}-Parity \label{sec:RPV}}

We now turn to  the $R$-parity violating (RPV) MSSM with superpotential
\begin{align}
\begin{aligned}
W_\text{RPV}&=W_\text{MSSM}+\lambda'^{ijk}\overline{d}_iQ_jL_k+\frac{1}{2}\lambda^{ijk}L_iL_j\overline{e}_k\\&+\frac{1}{2}\lambda''^{ijk}\overline{u}_i\overline{d}_j\overline{d}_k+\mu'^{i}L_iH_u \;.
\label{eq:superpotentialRPV}
\end{aligned}
\end{align}
The first term in~\eqref{eq:superpotentialRPV} is the MSSM superpotential provided earlier. The second and third terms change lepton number by one unit, $\Delta L=1$, and the fourth term changes baryon number by one unit, $\Delta B=1$. The $\mu'$ term is a mass term irrelevant in the high energy limit. We therefore observe that the violation of $R$-parity results in lepton and baryon number violating processes, which may be relevant in processes like proton decay. Due to the non-observation of such processes, either the $\lambda$, $\lambda'$, $\mu'$ and $\lambda''$ couplings in~\eqref{eq:superpotentialRPV} have to be small or superpartner masses are large~\cite{Dawson:1985vr,Barbieri:1985ty,Barger:1989rk,Godbole:1992fb,Bhattacharyya:1995pq,Dreiner:1997uz,Domingo:2018qfg}.
The RPV MSSM may have up to $N_Y=108$ independent Yukawa couplings, four times as many as the $R$-parity preserving MSSM. Moreover, for each of the interacting fixed point classes of Tab.~\ref{tab:FPnames} may have up to $2^{N_Y}$ different fixed points, depending on which of the Yukawa couplings are vanishing or non-vanishing.

We now search for fixed points in the RPV MSSM. To avoid constraints due to proton decay,
we concentrate on the $\lambda'$ terms, with flavor indices $i,j,k=1,2,3$,
\begin{equation} \label{eq:RPV-super}
\lambda'^{ijk}\overline{d}_iQ_jL_k\;.
\end{equation}
Hence, in addition to the top and bottom Yukawa couplings of the MSSM, we retain the RPV Yukawa couplings  $\lambda'_{ijk}$. For the sequel, we define
\begin{equation}\label{ijk}
\alpha_{\lambda'_{ijk}}=\frac{|\lambda'_{ijk}|^2}{(4\pi)^2}\,.
\end{equation}
To avoid models with
non-linear  nullcline conditions (see App.\ref{sec:appendixunnaturalYukawas}) we limit ourselves here in the RPV MSSM and the MSSM extensions studied in Secs~\ref{sec:model1} -- \ref{sec:model3}  to superpotentials with 
permutation flavor  symmetries  or with dangerous terms switched off by selection rules.
Here, we introduce  for each lepton species $k=1,2,3$ a universal matrix $(\lambda'^k)_{ij}=\lambda'^{ijk}$, with
\begin{align}
\mathbf{\lambda}'^k=\lambda^\prime
\begin{pmatrix}
\mathbf{M}&&0\\
0&&0\\
\end{pmatrix} \, .
\label{eq:universalM}
\end{align}
Here,  $\mathbf{M}$  is a $2 \times2$ matrix with entries either one or zero. We denote by $I_1$ the number of times an entry '1'
appears in  $\mathbf{M}$, $1\leq I_1 \leq 4$, excluding the MSSM-limit ($I_1=0$) and the symmetry-breaking case with $I_1=3$. The number of remaining  different matrices is 11.
To avoid non-linear Yukawa nullclines 
 we do not allow  for third generation quark couplings in $\lambda'^k$ as top and bottom already appear in $W_\text{MSSM}$.
This leads to 
a set of $n=3I_1$ additional Yukawa couplings \eq{ijk} which we denote as \footnote{We label the $\alpha_i$ starting from $i= 4$ because the symbols $\alpha_{1,2,3}$ are already taken for the gauge couplings. \label{foot:4}}
\begin{equation*}
\text{RPV Yukawa couplings:   }\{\alpha_4,\cdots,\alpha_{n+4}\}\;,
\end{equation*}
thus leading to a total of $3$ gauge and $2+n$ Yukawa couplings.
In our analysis ($I_1\le4$), we have retained up to $n \leq 12$ RPV Yukawa couplings. The evolution of the Yukawa couplings are controlled by \eq{eq:MSSMtopbottombetas} together with
\begin{align}\label{YukawaRPV}
\begin{aligned}
\beta_{\lambda'}&=\alpha_{\lambda'}\bigg[36I_1\alpha_{\lambda'}-\frac{14}{9}\alpha_1-6\alpha_2-\frac{32}{3}\alpha_3\bigg]\;.
\end{aligned}
\end{align}
Here, due to the symmetries of~\eqref{eq:universalM},  the one-loop beta functions for $\alpha_{\lambda'_{ijk}}\rightarrow \alpha_{\lambda'}$ are identical, although their  values do not need to be identical due to possibly different initial conditions.\footnote{This is similar to the running of the top and bottom Yukawas \eq{eq:MSSMtopbottombetas} which becomes identical for $\alpha_1=0$.}

Fixed points can be found from inserting the nullcline of \eq{YukawaRPV}  into the gauge beta functions~\eqref{gauge2} with the MSSM coefficients~\eqref{eq:MSSMGaugeCoeffs}, and also noting that the two-loop Yukawa contributions to the running of the gauge couplings take the form
\begin{align}
\begin{aligned}
&D_{1b}=\frac{28}{3}\;,\;&D_{1t}&=\frac{52}{3}\;,\; &D_{1\lambda^\prime}&=28 I_1\;,\\
&D_{2b}=12\;,\; &D_{2t}&=12\;,\; &D_{2\lambda^\prime}&=36I_1\;,\\[.5ex]
&D_{3b}=8\;,\;&D_{3t}&=8\;,\; &D_{3 \lambda^\prime}&=24I_1\;.
\end{aligned}
\end{align}
We find  that the only viable interacting fixed point is of the FP${}_3$ type, with trivial $(\alpha_1,\alpha_2)|_*=(0,0)$ and
\begin{align}
\begin{aligned}
(\alpha_3,\alpha_b,\alpha_t,\alpha_{\lambda'})|_*&=(\s0{189}{274},\s0{72}{137},\s0{72}{137},\s0{28}{137 I_1})
\\&\approx(0.69,0.53,0.53,\s0{0.20}{I_1})\;.
\end{aligned}
\end{align}
Notice that $\alpha_{\lambda'}^*$ now stands for  any of the different RPV Yukawas, all of which take the same finite fixed point value. 
The fixed point is  IR attractive and some of its couplings are large. We  observe that the RPV MSSM is not offering interacting UV fixed points.
\begin{table*}
  		\addtolength{\tabcolsep}{3pt}
		\setlength{\extrarowheight}{3pt}			
 \centering
  \sisetup {
    per-mode = fraction
  }
  \scalebox{0.85}{
   \begin{tabular}{`c?ccc?c?ccc`}
\toprule
\rowcolor{Yellow}
\bf Superfield&$\bm{SU(3)_C}$&$\bm{SU(2)_L}$&$\bm{U(1)_Y}$&\bf MSSM& \bf Extension~I &\bf Extension~II &\bf Extension~III \\
\midrule
quark doublet $Q$&\textbf{3}&\textbf{2}&$+\s016$&$3$&$3$&$4$&$4$\\
\rowcolor{Gray}
anti-quark doublet $\overline{Q}$&$\overline{\textbf{3}}$&$\overline{\textbf{2}}$&$-\s016$&$0$&0&$1$&0\\
up-quark $\overline{u}$&$\overline{\textbf{3}}$&\textbf{1}&$-\s023$&$3$&$3+n_u$&$3$&$4$\\
\rowcolor{Gray}
down-quark $\overline{d}$&$\overline{\textbf{3}}$&\textbf{1}&$+\frac13$ &$3$&$3+n_d$&$3$&$4$\\
anti-up-quark $u$&$\textbf{3}$&\textbf{1}&$+\s023$&$0$&$n_u$&$0$&$0$\\
\rowcolor{Gray}
anti-down-quark $d$&$\textbf{3}$&\textbf{1}&$-\s013$&$0$&$n_d$&$0$&$0$\\
lepton doublet $L$&\textbf{1}&\textbf{2}&$-\s012$&$3$&$3+n_L$&$3+n_L$&$4+n_L$\\
\rowcolor{Gray}
anti-lepton doublet $\overline{L}$&\textbf{1}&$\overline{\textbf{2}}$&$+\s012$&$0$&$n_L$&$n_L$&$n_L$\\
lepton singlet $\overline{e}$&\textbf{1}&\textbf{1}&$+1$&$3$&3&3&$4$\\
\rowcolor{Gray}
up-Higgs $H_u$&\textbf{1}&\textbf{2}&$+\s012$&$1$&1&1&$1$\\
down-Higgs $H_d$&\textbf{1}&\textbf{2}&$-\s012$&1&1&$1$&$1$\\
\rowcolor{Gray}
gauge singlets $S$&\textbf{1}&\textbf{1}&0&$0$&$0$&$n_S$&$0$\\
\bottomrule
  \end{tabular}
}
  \caption{Field content of different types of MSSM extensions in comparison with the MSSM, also showing gauge charges of superfields under the SM gauge groups.}
\label{tab:model123particles}
\end{table*}

\subsection{Constructing  MSSM Extensions \label{sec:construction}}

Next, we turn to extensions of the MSSM and the prospect  for  interacting  UV or IR fixed points. 
One may hope to find an  interacting UV  fixed point provided that the Gaussian  corresponds to a saddle point \cite{Bond:2017suy}. 
Hence, at least one gauge sector should remain asymptotically free while another one should be infrared free. In our setting, the hypercharge one-loop gauge coefficient $(B_1)$ is always negative, and remains negative in any extension. Further, for the MSSM, we are in the scenario where  the non-Abelian gauge factors obey \eq{B3>0>B2}. Hence, the one-loop gauge coefficient of the weak coupling  $(B_2)$ is negative, and will remain negative in any extension. On the other hand,  the one-loop coefficient of the strong coupling in the MSSM reads  $B_3=6$, thus leaving room  for a finite number of additional colored superfields.

Specifically, each additional superfield in the representation $R$ of $SU(3)_C$ lowers $B_3$ by $2S_3(R)$. For the fundamental or anti-fundamental representation holds $S_3=\frac12$, of which one needs one each  to avoid gauge anomalies.
The sextet and anti-sextet representations have $S_3=\frac52$, yet gauge anomaly cancellation dictates to include at least two of these, yielding a wrong-sign $B_3$ of $-4$. The representation with the next higher Dynkin index is the adjoint, which is real with $S_3=3$, however, a single one of them leads already to $B_3=0$. All other, higher  representations have $S_3>3$ and  are  therefore not viable. We conclude that
 there are only two possibilities  to add  colored BSM superfields  which  keep $B_3$ positive,  either one pair,   or two 
pairs of (fundamental, anti-fundamental) $SU(3)_C$ chiral superfields.
These arguments do not limit the number of colorless fields, such as leptons.

We are therefore led to three  types of MSSM extensions:
\begin{itemize}
\setlength{\itemindent}{.38cm}
\item[ {Type I:}] {\bf New quark singlets and new leptons.} On top of the  MSSM fields, type I models display $n_u$ additional pairs of up-quark singlets $(u,\overline{u})$,  $n_d$ new down-quark singlets $(d,\overline{d})$, and $n_L$ additional lepton doublet pairs ($L$, $\overline{L}$). 
\setlength{\itemindent}{.52cm}
\item[ {Type II:}] {\bf New quark doublets and new leptons.} These models display two additional quark doublets ($Q_4,\overline{Q}_1$),  $n_L$ additional lepton and anti-lepton doublet pairs ($L$, $\overline{L}$). 
\setlength{\itemindent}{.6cm}
\item[ {Type III:}] {\bf A fourth generation and new leptons.} These extensions display a fourth generation with new superfields ($Q_4$, $\overline{u}_4$, $\overline{d}_4$, $L_4$), and $n_L$ new lepton and anti-lepton doublet pairs ($L$, $\overline{L}$). 
\end{itemize}
In addition, we also have the liberty to  add  $n_S$ gauge singlet fields $S_i$  and suitable Yukawa couplings involving MSSM and BSM matter fields. We find that the impact of singlets for fixed points is  subleading except in  type II models, which is why we include them there and only there.
The field content of  the MSSM extensions is summarised in Tab.~\ref{tab:model123particles}, also showing the SM gauge charges of matter fields.
Note that  we are not concerned with the $U(1)_Y$ sector, which remains infrared free. This is viable  phenomenologically as long as  the $U(1)_Y$ Landau pole arises beyond the Planck scale. Extensions which also aim at stabilising $U(1)_Y$ will be considered elsewhere.
In the following, we  investigate the availability of interacting  fixed points for  each of these settings in detail.

\subsection{New Quark Singlets and Leptons}
\label{sec:model1}

We begin with the  first type of MSSM extension by  adding BSM quark singlets as well as lepton doublets to the MSSM. 
The BSM particle content (see Tab~\ref{tab:model123particles}) is characterised, respectively,  by the number of beyond-MSSM  $(u, \bar u)$, $(d, \bar d)$ and $(L ,\bar L)$ pairs, 
\begin{equation}
n_u\;,\; n_d\;,\; n_L\;,
\label{eq:model1flavorparams}
\end{equation}
Asymptotic freedom of the strong force is lost for $n_u+n_d\ge 3$. A priori, no upper limits apply on  $n_L$.
The  number of matter fields beyond the MSSM is 
\begin{equation}
N_\text{BSM}=N_{q,\text{BSM}}+N_{L,\text{BSM}} \,.
\label{eq:modelclass1NBSM}
\end{equation}
where  $N_{q,\text{BSM}}=2(n_u+n_d)$ and $N_{L,\text{BSM}}=2 n_L$ are the new quark singlets and  lepton doublets, respectively, and where we count fermions and anti-fermions separately.
The most general  gauge invariant and perturbatively renormalisable superpotential then reads
\begin{align}
\begin{aligned}
W_1=&\ Y^{ijk}\overline{d}_iQ_jL_k+\overline{Y}^{ijk}\overline{u}_iQ_j\overline{L}_k\\&+x_b \, y_b\overline{d}_3Q_3H_d+x_t \,  y_t\overline{u}_3Q_3H_u\;,
\label{eq:model1superpotential}
\end{aligned}
\end{align}
with top and bottom Yukawas denoted by $y_t$ and $y_b$, and BSM Yukawas $Y^{ijk}$ and $\overline{Y}^{ijk}$. Here $i,j,k$ denote flavor indices, while gauge indices are suppressed. The parameters $x_{b,t}\in\{0,1\}$ allow us to switch the bottom and top Yukawas on and off. In terms 
of the BSM matter field multiplicities~\eqref{eq:model1flavorparams}, the superpotential~\eqref{eq:model1superpotential} has up to
 \begin{equation}
 N_{Y}^\text{general}=3(3+n_d)(n_L+3)+3(3+n_u)n_L+2
 \label{eq:model1AnzYukawasgeneral}
 \end{equation}
   independent Yukawa couplings.
In the fixed point search, we focus on a subset of all possible non-zero Yukawa couplings, parameterized by the following set of integers
\begin{equation}
x_b\,,\; x_t\,,\;I_{12}\,,\;I_{13}\,,\;I_{1d}\,,\;I_{2d}\,,\;I_{3d}\,,\;I_{1u}\,,\;I_{2u}\,,\;I_{3u}\,.
\label{eq:model1integers}
\end{equation}
These  integers, if positive,  indicate which type of  Yukawa couplings in~\eqref{eq:model1superpotential} are taken to be non-zero, 
and how many of them are retained. Specifically,
we are interested in superpotentials~\eqref{eq:model1superpotential}  which retain
\begin{align}
\begin{aligned}
\text{\underline{type of monomial}}& : \text{\underline{how many of them}}\\
y_b\,\overline{d}_3\,  Q_3\, H_d &: x_b\;,\\
y_t\, \overline{u}_3\, Q_3\, H_u &: x_t\;,\\
y_4\, \overline{d}_i\, Q_1\, L_k  + y_6\,\overline{d}_i\, Q_2\,  L_{k'} &:I_{12}\;,\\
y_5\,  \overline{d}_i\, Q_1\, L_k &: I_{1d}\;,\\
y_7 \, \overline{d}_i\, Q_2\, L_k &: I_{2d}\;,\\
y_8 \,\overline{d}_i\, Q_1\, L_k+y_9\, \overline{d}_i\,Q_3\,L_{k'} &: I_{13}\;,\\
y_{10} \, \overline{d}_i\,Q_3\, L_k &: I_{3d}\;,\\
y_{11} \, \overline{u}_i\,Q_1 \overline{L}_k &: I_{1u}\;,\\
y_{12}\, \overline{u}_i\,Q_2 \overline{L}_k &: I_{2u}\;,\\
y_{13}\, \overline{u}_i\,Q_3 \overline{L}_k &: I_{3u}\;.
\label{eq:model1yukawaparams}
\end{aligned}
\end{align}
We again label couplings as indicated in  footnote ${}^{\ref{foot:4})}$.
To avoid non-linear nullclines we also made choices as in the analysis of the RPV MSSM (Sec.~\ref{sec:RPV}).
Let us  explain the construction principle leading to  \eq{eq:model1yukawaparams}: 
\begin{itemize}
\item[$(i)$]The bottom and top quarks $\overline{d}_3$ and $\overline{u}_3$ are only allowed in  the bottom or top Yukawa terms already present in the MSSM, see \eq{eq:superpotentialMSSM2}. They can be switched on and off individually with the parameters  $x_b,x_t\in\{1,0\}$. 
\item[$(ii)$] Superfields  $\overline{d}_i$, $i \neq 3$  that is, any of them   but not the third generation may appear in exactly one superpotential term.
This can still be more than one term, one for each $i \neq 3$. The number of times any $\overline{d}_i$, $i \neq 3$, appears exactly once together with 
$Q_1$, $Q_2$, or $Q_3$ is given by $I_{1d}$, $I_{2d}$, and $I_{3d}$, respectively.
\item[$(iii)$]  The same as $(ii)$  but for up-type singlets: The number of times any $\overline{u}_i$, $i \neq 3$, appears exactly once together with 
$Q_1$, $Q_2$, or $Q_3$ is given by $I_{1u}$, $I_{2u}$, and $I_{3u}$, respectively.
\item[$(iv)$] Down-type quarks $\overline{d}_i$, $i \neq 3$  may be  present in two different Yukawa monomials. 
With $I_{12}$ ($I_{13})$ we count down quark superfields $\overline{d}_i, i 
\neq 3$ appearing exactly twice, once with  $Q_1$ and once with $Q_2$ ($Q_3$).
\item[$(v)$]  Each  lepton doublet $L$ and anti-lepton doublet $\overline{L}$ (both MSSM and BSM) is allowed   at most once in the superpotential. 
\end{itemize}
A concrete benchmark model where this machinery can be seen at work is given in Sec.~\ref{sec:Benchmark}.

The  reduced set of Yukawa interactions \eq{eq:model1yukawaparams}
is the result of an extensive trial and error search.
In fact, we have initially performed scans within the much wider  set of superpotentials \eq{eq:model1superpotential}, but  noticed that viable fixed point candidates do not arise without down-quarks $\bar d_i$, $i \neq 3$ appearing twice in the superpotential. 
Moreover, we also noticed that ultraviolet fixed points cannot be found if we allow for lepton doublets to appear twice in the superpotential with each quark singlet appearing at most once. 
We believe that our selection of Yukawa structures are the simplest ones to enable viable fixed points.
 
 As a result, in terms of \eq{eq:model1integers} the number of independent Yukawa couplings retained in our investigations reads
 \begin{equation}
 N_Y=2(I_{12}+I_{13})+\sum\limits_{i=1}^3(I_{id}+I_{iu})+x_b+x_t\;.
 \label{eq:model1AnzYukawas}
 \end{equation}
This is only a small subset of  the formally allowed  superpotential terms \eq{eq:model1AnzYukawasgeneral}, yet, suffices to identify 
 interacting gauge-Yukawa fixed points.

By construction, the models are described by three gauge and $N_Y$ independent Yukawa couplings.  Due to remaining flavor  symmetries acting on quark singlets and on lepton doublets appearing in terms of the same Yukawa coupling type, we  encounter at most 12 different types of Yukawa beta functions  corresponding to those of the bottom and top Yukawas, and the 10 couplings $y_i$  introduced in \eq{eq:model1yukawaparams}, modulo copies thereof, see Sec.~\ref{sec:Benchmark}  for an example where symmetries reduce the number of independent beta functions. The beta functions for the Yukawa couplings are given in App.\ref{app:model1}.

Next, we detail the results of the fixed point search.
The  selection rules  {\it i) - \it v) } still allow for infinitely many  MSSM extensions. 
However, the number of new quark singlets is bounded from above ($N_{q,{\rm BSM}} =2 (n_u+n_d ) \le 4$,  see Sec.~\ref{sec:construction}) or else
weakly-interacting fixed points cannot arise in perturbation theory. Similarly, increasing  the number of 
new leptons makes the weak gauge coupling more infrared-free, and it becomes  increasingly challenging if not outright impossible
to find ultraviolet fixed points. For these reasons, we limit new matter field multiplicities  as follows
\begin{align}
\begin{aligned}
\label{eq:model1parameterscan}
&0\le n_d\le 2-n_u\;, \\&
0\le n_u\le 2-n_d\;, \\
&0\le n_L\le 11 \, . 
\end{aligned}
\end{align}
The set of Yukawa couplings is restricted by 
\begin{align}
\begin{aligned}
\label{eq:model1Yukawas}
&0\le   I_{1u}\le I_{2u}\,,
\\&0\le x_b\,,x_t\le 1\,,
\\&
0\le I_{12}\,,\,I_{13}\,,\,I_{1d}\,,\,I_{2d}\,,\,I_{3d}\,,\,I_{1u}\,,\,I_{2u}\,,\,I_{3u}\le 4\,.
\end{aligned}
\end{align}
Overall, the above choices cover  112.600 
different MSSM extensions with  up to $N_Y=12$ independent Yukawa couplings.
Amongst these, we find 109.926 settings with  viable IR fixed points.  
Further, 114 models also show  candidates for interacting UV fixed points.
Also, a small set of models do not show interacting fixed point despite the strong gauge sector remaining asymptotically free. The reason for this is that the Yukawa-induced corrections are so strong that the fixed point becomes unphysical in pertubation theory  $(\alpha_3<0)$. These settings would require a non-perturbative check.

\begin{figure}
\centering
\includegraphics[width = 0.45\textwidth]{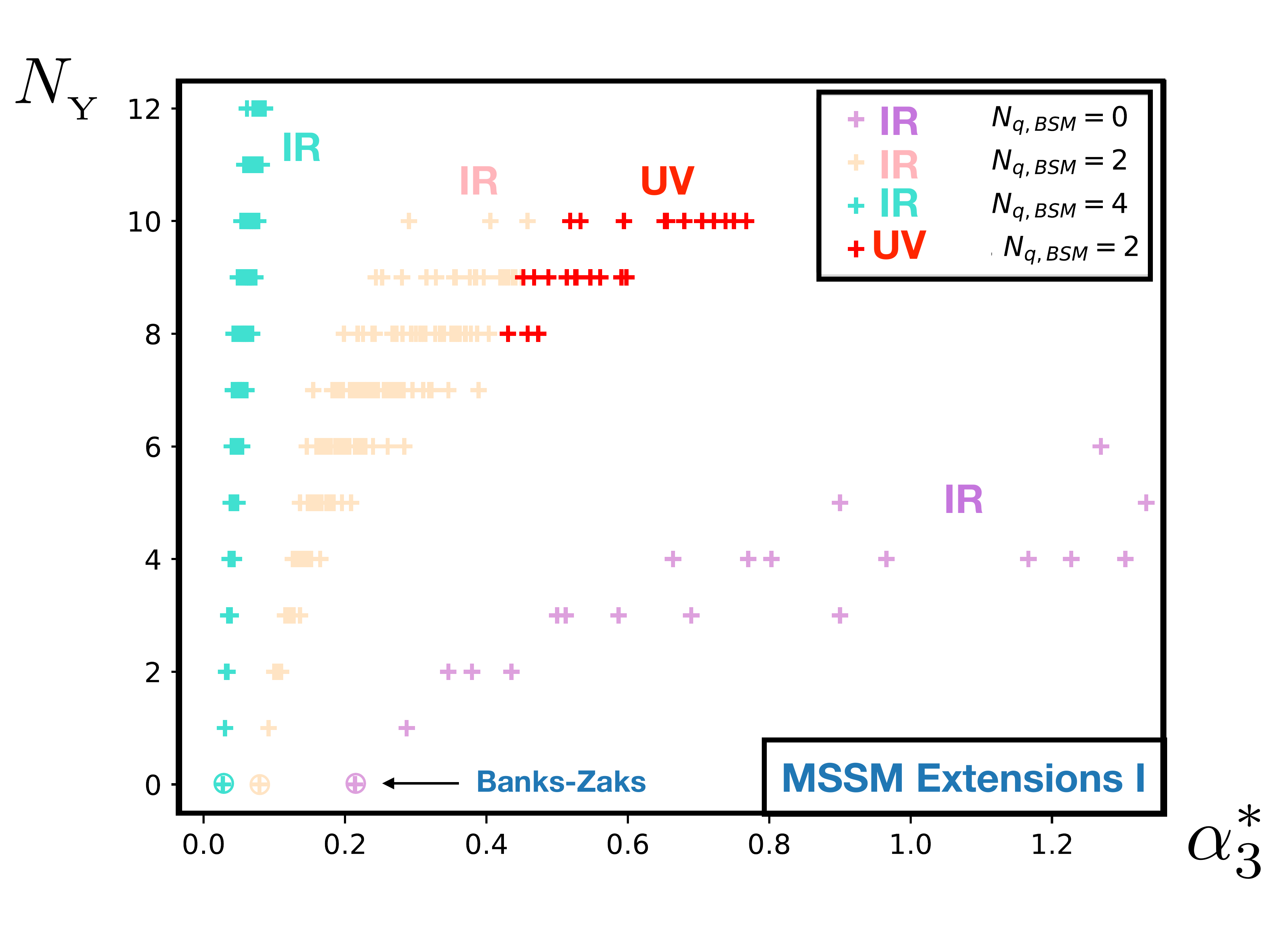}
\caption{Fixed points  of MSSM extensions (type I). Shown are the fixed point values  $\alpha_3^*$ against the number of Yukawa couplings $N_Y$, corresponding to 109.926  models, see text.  The color coding indicates  the number of additional quark singlets $N_{q,\text{BSM}}=2(n_u+n_d)$, and whether the fixed point is UV or IR.   114 models have an UV $\text{FP}_3$ (in red).}
\label{fig:branches}
\end{figure}

More specifically, in all models considered we find that  fixed points, if they arise,  remain interacting in the strong gauge sector with
\beq\label{bound-1}
0.027 \approx \frac{3}{110} \lesssim \alpha_3^*\big|_{\text{FP}_3}\,,
\eeq
in agreement with the analytical expression~\eqref{eq:alpha3bound}.
The weak and hypercharge gauge interactions are either switched off  
(in which case fixed points are  of the type $\text{FP}_3$), or the weak coupling remains non-zero as well 
(when fixed points are of the type $\text{FP}_{23}$), see Tab.~\ref{tab:FPnames}. Fixed points with vanishing strong gauge coupling, that is,  $\text{FP}_1$, $\text{FP}_2$, or $\text{FP}_{12}$, or fixed points with all gauge couplings non-zero  ($\text{FP}_{123}$)    do not arise (see 
App.\ref{sec:Expressions}). 

As an aside, we have verified explicitly  that  no interacting  fixed points arise once  $N_{q,{\rm BSM}}\ge 6$ by scanning  3.434.836 models including up to 4 pairs of singlet quarks, confirming the  reasoning put forward in Sec.~\ref{sec:construction}.

All fixed points with both non-abelian gauge couplings interacting, i.e.\ $\alpha_3^*>0$ and $\alpha_2^*>0$, and trivial or non-trivial superpotential couplings turn out to be infrared. In turn, the fixed points of the type $\text{FP}_3$ are found to be either infrared  or ultraviolet. If they are infrared, all gauge and non-trivial superpotential couplings are irrelevant. Most importantly,  there are no outgoing RG trajectories along which the weak gauge coupling can be switched on. Moreover, none of the models with infrared $\text{FP}_{3}$ have a simultaneous fixed point of the type $\text{FP}_{23}$.

In Fig.~\ref{fig:branches}, we show the strong gauge coupling for all fixed points 
of the type $\text{FP}_3$. Also displayed are the number of non-trivial Yukawa couplings $N_Y$. Different numbers of new quark singlets $N_{q,\text{BSM}}$ lead to different branches of fixed points. Their color-coding relates to $N_{q,\text{BSM}}$ 
 and whether the fixed point is infrared (magenta: $N_{q,{\rm BSM}}=0$, yellow: $N_{q,{\rm BSM}}=2$, green: $N_{q,{\rm BSM}}=4$) or ultraviolet (red: $N_{q,{\rm BSM}}=2$).

For $N_Y=0$, and for any $0\le N_{q,\text{BSM}}\le 4$ we find an infrared Banks-Zaks fixed point. For $N_Y>0$, fixed points are of the gauge-Yukawa type and can be IR or UV. For fixed $N_Y$, we observe that the strong coupling becomes larger with increasing $N_{q,{\rm BSM}}$.
For fixed  $N_{q,\text{BSM}}$, we also observe that  the strong gauge coupling tends  to increase with increasing $N_Y$. 

For each strand of models, Fig.~\ref{fig:branches}  indicates that the Banks-Zaks fixed point provides a lower bound on the strong coupling. The reason for this is that non-trivial Yukawa couplings reduce the effective two-loop coefficient and enhance $\alpha_3^*$. To the leading orders in perturbation theory, the lower bounds  are   $\alpha_3^*\ge \s03{14}\approx 0.214$ for $N_{q,\rm BSM}=0$, $\alpha_3^*\ge \s03{38}\approx 0.089$  for $N_{q,\rm BSM}=2$, and $\alpha_3^*\ge \s0{3}{110}\approx 0.027$ as at the minimum (\ref{bound-1})  for $N_{q,\rm BSM}=4$.   
Moreover, the models with $N_{q,\text{BSM}}=4$ (green points) lead to weakly interacting IR fixed points, and the threshold towards UV fixed points is not crossed. 
For models with $N_{q,\text{BSM}}=0$ (magenta points), the fixed point is more strongly interacting, and once more a regime with UV fixed points is not reached.  Inbetween, the models with $N_{q,\text{BSM}}=2$ lead to weakly interacting IR fixed points for any $N_Y\le 7$, and for some $N_Y>7$ (yellow points). Overall fixed points are mostly perturbative ($\alpha_3^*\ll 1$) 
though with increasing $N_Y$  some of the fixed points become borderline perturbative  $(\alpha_3^*<1)$ 
or even strongly coupled $(\alpha_3^*\approx 1)$ such as in the $N_{q,\text{BSM}}=0$ strand.

Finally, provided that $N_{q,\text{BSM}}=2$, that is, either a pair of additional up-singlets $(u, \bar u)$, or a pair of down-type ones $(d, \bar d)$,  and $N_Y= 8,9$ or $10$, we  also find models  where the fixed point is UV
with $\alpha_2$ becoming marginally relevant  due to quantum effects (red points). 	
\begin{figure}[t]
\vskip-.2cm
		\includegraphics[width=.85\linewidth]{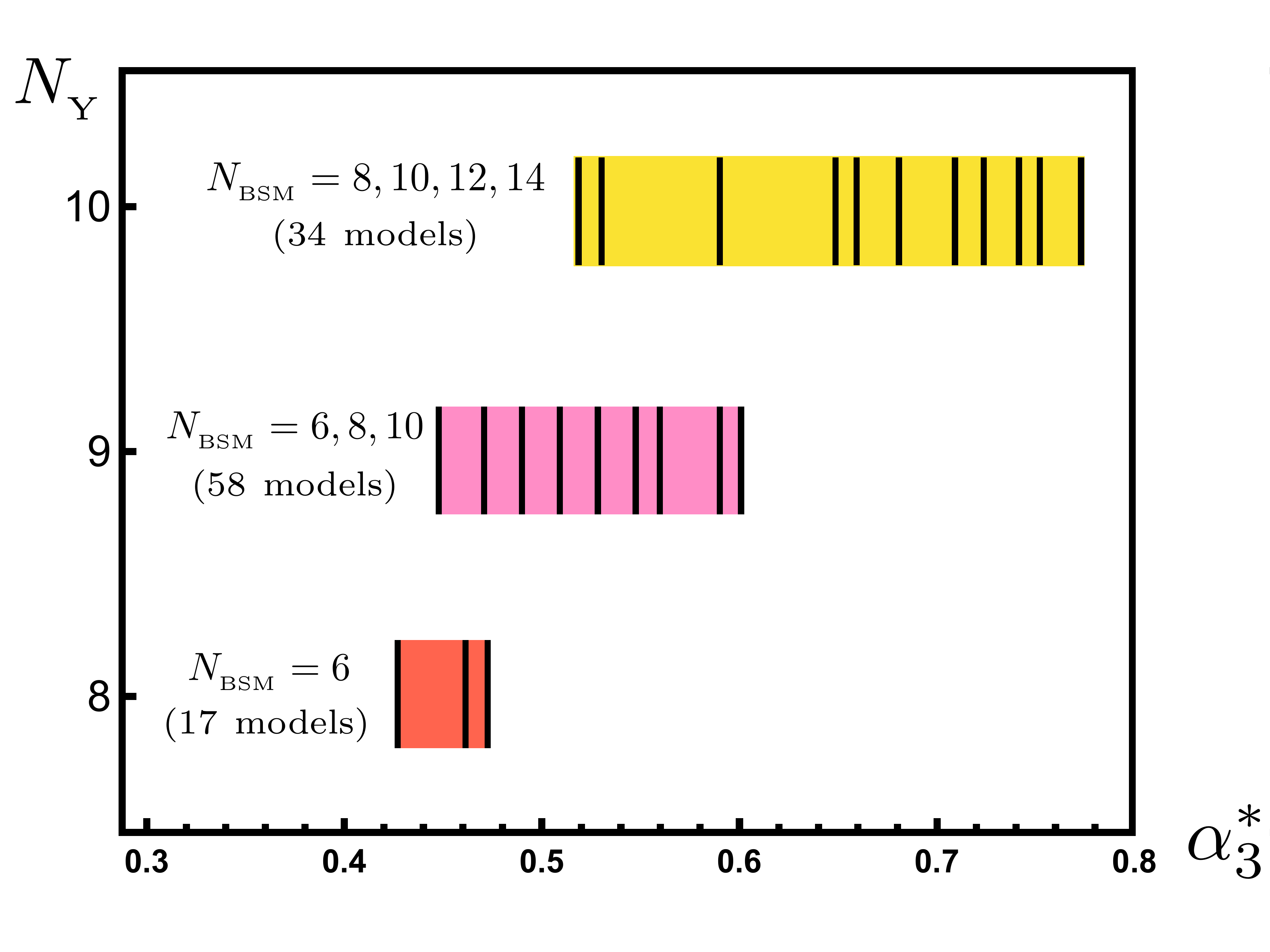}
\vskip-.3cm
		\caption{Spectroscopy of ultraviolet fixed points in type I models. Shown are the ranges of  $\alpha_3^*$,  
				sorted according to
		 numbers of BSM Yukawa couplings $(N_{{}_{\rm Y}})$, and the number of BSM superfields $(N_{{}_{\rm BSM}})$ .}
		\label{p109compare}
\end{figure}
In Fig.~\ref{p109compare} we show  $\alpha_3^*$ at the UV fixed point 
as a function of the number of BSM Yukawa couplings $(N_{{}_{\rm Y}})$, and the number of BSM 
superfields $(N_{{}_{\rm BSM}})$. Evidently, the fixed point tends to become more strongly interacting the 
more independent Yukawa couplings are present.
The settings with UV fixed points are further discussed  in Sec.~\ref{sec:ASanalysis}.

\subsection{New Quark Doublets and Leptons}
\label{sec:model2}

For the second type of MSSM extensions, we introduce a quark doublet $Q_4$ and  an anti-quark doublet $\overline{Q}_1$ as the new colored field content beyond the MSSM. Furthermore, we allow for $n_L$ pairs of BSM lepton and anti-lepton doublets $L$ and $\overline{L}$. We also include $n_S$ gauge singlet superfields $S$. 
The resulting superfield content (type II models) is summarized in Tab.~\ref{tab:model123particles}.
We study the superpotential 
\begin{align}
\begin{aligned}
W_2&=Y^{ijk}\overline{d}_iQ_jL_k+\overline{Y}^{ijk}\overline{u}_iQ_j\overline{L}_k+Y^{i}_SS_i\overline{Q}_1Q_4\\&+x_by_b\overline{d}_3Q_3H_d+x_ty_t\overline{u}_3Q_3H_u\;,
\label{eq:superpotentialmodel2}
\end{aligned}
\end{align}
with $i,j,k$ summing over all flavor indices and $x_b,x_t\in\{0,1\}$ are parameters which switch on and off the bottom and top Yukawa couplings.
The first few terms of~\eqref{eq:superpotentialmodel2} resemble the non-MSSM terms of the superpotential~\eqref{eq:model1superpotential} of model type I, with the difference  that here $i$ and $j$ run over different numbers of flavors. We compensate for the smaller amount of quark singlets, present in the Yukawa terms of $Y^{ijk}$ and $\overline{Y}^{ijk}$, by including terms with Yukawa couplings $Y_S^{i}$ involving the new anti-quark doublet $\overline{Q}_1$. The number of generally possible non-zero Yukawa couplings in the superpotential~\eqref{eq:superpotentialmodel2} is
\begin{align}
\begin{aligned}
N_Y^\text{general}&=3\cdot4\cdot(n_L+3)+3\cdot4\cdot n_L+n_S+2\\ 
&=24n_L+38+n_S\;,
\label{eq:generalYukawasmodel2}
\end{aligned}
\end{align}
with each term of the first line counts the number of Yukawa couplings in the respective  term of the superpotential~\eqref{eq:superpotentialmodel2}.

To parametrize different models efficiently, we introduce
$n_L$ and $n_S$
to count leptons and singlets.
Further, the non-zero Yukawa couplings in \eqref{eq:superpotentialmodel2}
are parametrised by the  integers
\begin{equation}
I_Q,\;I_d\;,\;I_u\;,\;x_3\;,\;\overline{x}_3\;,\;x_4\;,\;\overline{x}_4\;,\;x_S\;,\;x_b\;,\;x_t\;,
 \label{eq:model2Iparams}
\end{equation}
which count the different types of monomials appearing in the superpotential according to
 \begin{align}
\begin{aligned}
\text{\underline{type of monomial}}& : \text{\underline{how many of them}}\\
y_b\,\overline{d}_3\,Q_3\,H_d&:x_b\;,\\
y_t\,\overline{u}_3\,Q_3\,H_u&:x_t\;,\\
y_4\,\overline{d}_{1,2}\,Q_{1,2}\,L_k&: I_d\cdot I_Q\;,\\
y_5\,\overline{u}_{1,2}\,Q_{1,2}\,\overline{L}_k&:I_u\cdot I_Q\;,\\
y_6\,\overline{d}_{3}\,Q_{1,2}\,L_k&:x_3\cdot I_Q\;,\\
y_7\,\overline{u}_{3}\,Q_{1,2}\,\overline{L}_k&:\overline{x}_3\cdot I_Q\;,\\
y_8\,\overline{d}_{1,2}\,Q_{4}\,L_k&:x_4\cdot I_d\;,\\
y_9\,\overline{u}_{1,2}\,Q_{4}\,\overline{L}_k&:\overline{x}_4\cdot I_u\;,\\
y_{10}\,S_i\,\overline{Q}_{1}\,Q_4&:x_\text{S}\cdot n_\text{S}\;.
\label{eq:model2yukawaparams}
\end{aligned}
\end{align}
The  selection of  superpotentials with \eq{eq:model2Iparams}, \eq{eq:model2yukawaparams} 
from general superpotentials \eq{eq:superpotentialmodel2} is
similar in spirit to the choices we made previously for the MSSM with quark singlet extensions (type I).
Further details including  all RG beta functions are detailed in App.\ref{app:model2}.

To illustrate the construction principle, we consider a subset of terms from \eqref{eq:superpotentialmodel2}
\begin{align}
\begin{aligned}
W_2 \supset &\ Y^{111}\overline{d}_1Q_1L_1+Y^{142}\overline{d}_1Q_4L_2+\overline{Y}^{111}\overline{u}_1Q_1\overline{L}_1\\&
+Y^{313}\overline{d}_3Q_1L_3+\overline{Y}^{312}\overline{u}_3Q_1\overline{L}_2
\\&+Y_S^{1}S_1\overline{Q}_1Q_4
+y_b\overline{d}_3Q_3H_d+y_t\overline{u}_3Q_3H_u\;.
\label{eq:model2examplesuperpotential}
\end{aligned}
\end{align}
It corresponds to the  parameters
\begin{align}\nonumber
\begin{aligned}
&x_3=\overline{x}_3=x_4=x_S=x_b=x_t=1, \\
& \overline{x}_4=0\;, \ I_d=I_u=I_Q=n_S=1\;,
 \end{aligned}
\end{align}
with $n_L\ge 2$.
The map from \eqref{eq:model2examplesuperpotential} to  \eqref{eq:model2yukawaparams} is given by
\begin{align}\nonumber
\begin{aligned}
&Y^{111} \leftrightarrow y_4\;,\;Y^{142} \leftrightarrow y_8\;,\;Y^{313} \leftrightarrow y_6\;,\;\\ &\overline{Y}^{111} \leftrightarrow y_5\;,\;
\overline{Y}^{312} \leftrightarrow y_7\;,\;Y_S^{1} \leftrightarrow y_{10}
\end{aligned}
\end{align}
and the expressions for the Yukawa and gauge beta functions \eq{generalYukawabetas}, \eq{gauge2} can be found in App.\ref{app:model2}. 

Within the general model setup  (\ref{eq:model2yukawaparams})  we scanned  79.920 models in the parameter ranges
\begin{align}
\begin{aligned}
& 1\le n_S \le 5\;,\qquad\;\; 0\le n_L\le 8\;,\\
&0\le I_d\;,\;I_u\le 2\;,\quad 1\le I_Q\le 2\;,\quad\\
&0\le x_3\;,\;\overline{x}_3\;,\;x_4\;,\;\overline{x}_4\;,\;x_S\;,\;x_b\;,\;x_t\le 1\;.
\label{eq:scannedparamsmodel2}
\end{aligned}
\end{align}
We find that amongst all possible interacting fixed points  (see Tab.~\ref{tab:FPnames}) only those
of the type $\text{FP}_3$ where $\alpha_{3}^*\neq 0$ are realised. $\text{FP}_3$ can be either of the Banks-Zaks type or of the gauge-Yukawa type \eq{FP3}.
It exists for all  models and is found to be IR and perturbative, with the strong gauge coupling fixed point in the range
\begin{equation}\label{bound-2}
0.027\lesssim \alpha_{3}^*\big|_{\text{FP}_3}\lesssim 0.08\;.
\end{equation} 
The numerical lower bound is in agreement with the  bound dictated by the  leading order 
in perturbation theory, (\refeq{bound-1}).
We do not find instances where the fixed point $\text{FP}_3$ becomes ultraviolet.

\begin{figure}
\centering
\vskip-.3cm
\includegraphics[width = 0.45\textwidth]{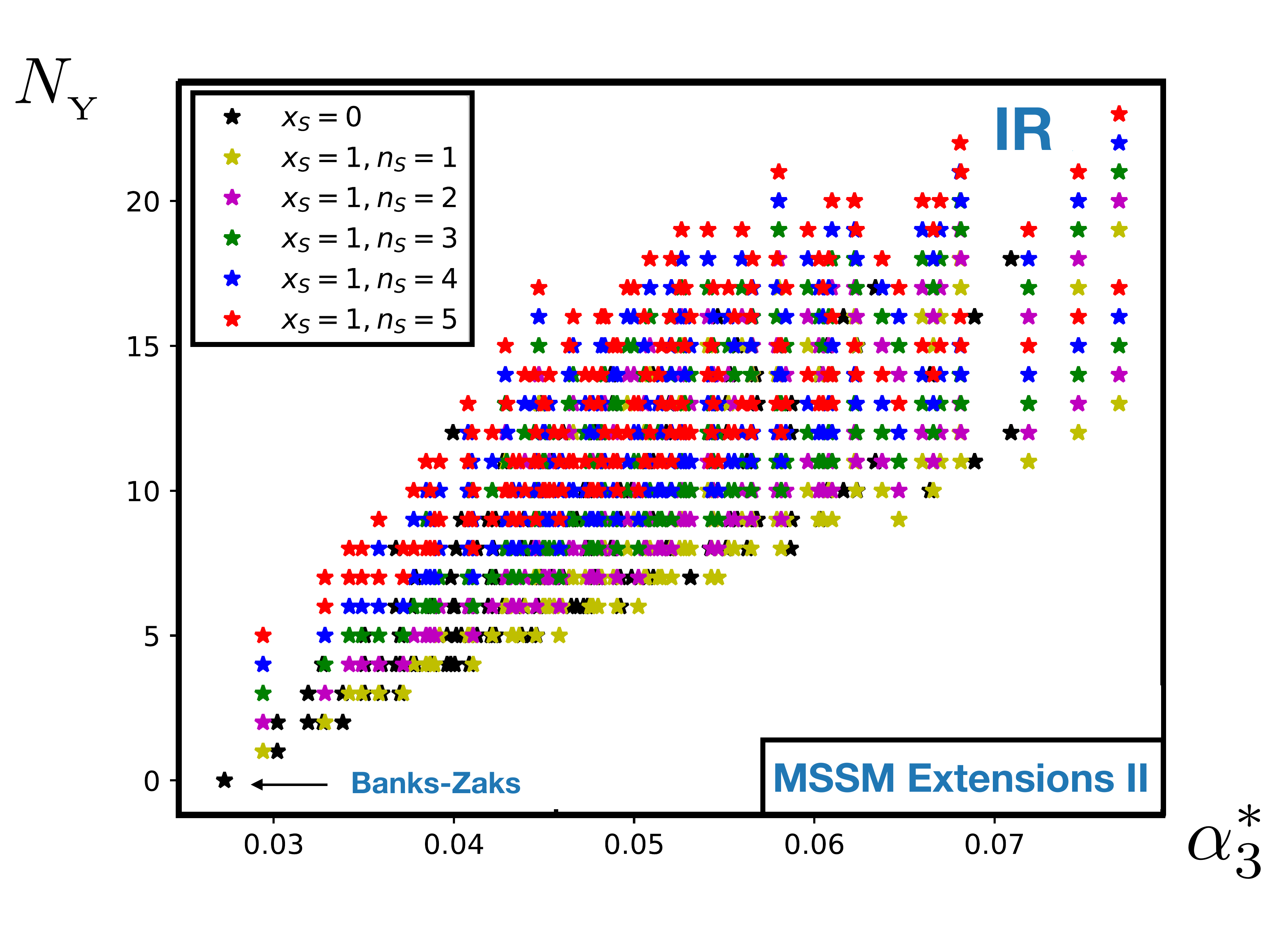}
\vskip-.3cm
\caption{Fixed points of MSSM extensions (type II). Shown are the values for the strong coupling constant $\alpha_3^*$ at the $\text{FP}_3$ fixed point for all  79.920 MSSM extensions of type II  given by the chiral superfields of Tab.~\ref{tab:model123particles} and superpotential~\eqref{eq:superpotentialmodel2}. Fixed points are IR throughout. Also shown is the number of Yukawa couplings $N_Y$~\eqref{eq:model2AnzYukawas}. Fixed points become slightly more strongly interacting with increasing number of Yukawa couplings.}
\label{fig:model2FPs}
\end{figure}

In Fig.~\ref{fig:model2FPs}, we compare the value of $\alpha_3$ at $\text{FP}_3$ against the number of Yukawa couplings $N_Y$ for all scanned models.
We observe that gauge-Yukawa fixed points are more strongly interacting than the Banks-Zaks fixed point.
Moreover, all fixed points are infrared and do not qualify as UV completions for the theory. 
Notice that our setup retains up to
\begin{align}
\begin{aligned}
N_{Y}&=(I_d+x_3+I_u+\overline{x}_3)I_Q\\
&+x_4I_d+\overline{x}_4I_u+x_Sn_S+x_b+x_t\;,
\label{eq:model2AnzYukawas}
\end{aligned}
\end{align}
different Yukawa couplings. 
Of these, the scan~\eqref{eq:scannedparamsmodel2} covered models with up to $N_{Y}=23$. 
From  Fig.~\ref{fig:model2FPs}, we learn that models tend to become more strongly interacting the more Yukawa couplings are switched on. 
Hence, although our scan only covered 
a small fraction of the $N_{Y,\text{scan}}^\text{general}=307$ different Yukawa couplings that could have been retained according to \eqref{eq:generalYukawasmodel2}, \eqref{eq:scannedparamsmodel2}, we do not expect that they would have opened a window for weakly coupled UV fixed points.

\subsection{Fourth Generation and New Leptons}
\label{sec:model3}

Here, we turn to MSSM extensions involving fourth generation quark doublet $Q_4$, and quark singlets $\overline{d}_4$ and $\overline{u}_4$. To avoid gauge anomalies, a fourth lepton generation consisting of a lepton doublet $L_4$ and a lepton singlet $\overline{e}_4$ are added as well. In addition, we allow for $n_L$ pairs of leptons and anti-leptons $(L,\overline{L})$ (see Tab.~\ref{tab:model123particles}).
The  superpotential reads 
\begin{align}
\begin{aligned}
W_3=&\ Y^{ijk}\overline{d}_iQ_jL_k+\overline{Y}^{ijk}\overline{u}_iQ_j\overline{L}_k\\&+y_b\overline{d}_3Q_3H_d+y_t\overline{u}_3Q_3H_u\;,
\label{eq:model3superpotential}
\end{aligned}
\end{align}
which looks similar to~\eqref{eq:model1superpotential} of type I models.
The main difference  is the presence of  $Q_4$  in~\eqref{eq:model3superpotential}, and that 
 non-trivial  bottom- and top Yukawa interactions are considered from the outset.
Note, possible terms  $\bar u_4 Q_4 H_u$ and $\bar d_4 Q_4 H_d$ have not been included  to avoid multiple appearances of the fourth generation, and corresponding challenges, see  App.\ref{sec:appendixunnaturalYukawas}.
The maximal number of non-zero Yukawa couplings in $W_3$  is given by
\begin{align}
\begin{aligned}
N_Y^\text{general}&=4\cdot4\cdot(n_L+4)+4\cdot4\cdot n_L+2\\
&=32n_L+66\;.
\label{eq:generalYukawasmodel3}
\end{aligned}
\end{align}
Our models are characterised by the number $n_L$ of BSM lepton pairs,
and integers 
\begin{equation}
I_{1d},\; I_{3d},\; I_{12},\;I_{13},\; I_{14},\; I_{1u},\; I_{4u}\;
\label{eq:model3parameters}
\end{equation}
 characterising the Yukawa interactions in \eqref{eq:model3superpotential}  as
\begin{align}
\begin{aligned}\label{model3}
\text{\underline{type of monomial}}& : \text{\underline{how many of them}}\\
y_b\overline{d}_3Q_3H_d& : 1\,,\\
y_t\overline{u}_3Q_3H_u& : 1\,,\\
y_4\overline{d}_iQ_1L_k+y_5\overline{d}_iQ_2L_{k'}& : I_{12}\;,\\
y_6\overline{d}_iQ_1L_k+y_7\overline{d}_iQ_3L_{k'}& : I_{13}\;,\\
y_8\overline{d}_iQ_1L_k+y_9\overline{d}_iQ_4L_{k'}& : I_{14}\;,\\
y_{10}\overline{d}_iQ_1L_k& : I_{1d}\;,\\
y_{11}\overline{d}_iQ_3L_k& : I_{3d}\;,\\
y_{12}\overline{u}_iQ_1\overline{L}_k& : I_{1u}\;,\\
y_{13}\overline{u}_iQ_4\overline{L}_k& : I_{4u}\;.
\end{aligned}
\end{align}
Flavor symmetries limit the number of different BSM beta functions in \eq{model3} to be at most 10 ($\beta_4,...,\beta_{13}$). 
All beta functions, and further details are given in App.\ref{app:model3}. 

Based on this ansatz, we have scanned 3.868 models within the ranges
\begin{align}
\begin{aligned}
&0\le I_{12},I_{13},I_{14},I_{1d},I_{3d},I_{1u},I_{4u}\le 3\;,\\& 0\le n_L\le 8\;.
\label{eq:scannedparamsmodel3}
\end{aligned}
\end{align}
Once more, we find that only $\text{FP}_3$ arises, with any of the other fixed point candidates coming out as unphysical. 
Moreover,  whenever it arises, $\text{FP}_3$ is numerically small with couplings in the range
\begin{equation}\label{bound-3}
0.027\lesssim \alpha_{3}^{*} |_{\text{FP}_3}\lesssim 0.10\;,
\end{equation} 
and in accord with the  lower bound  on the gauge coupling fixed point found to the leading orders in perturbation theory, (\refeq{bound-1}).  In total, we find that all  3.868 different models show conformal fixed points of the type $\text{FP}_3$, all of which are infrared. We neither find any  other types of fixed points, nor candidates for ultraviolet fixed points. 
\begin{figure}[t]
\vskip-.3cm
\includegraphics[width = 0.45\textwidth]{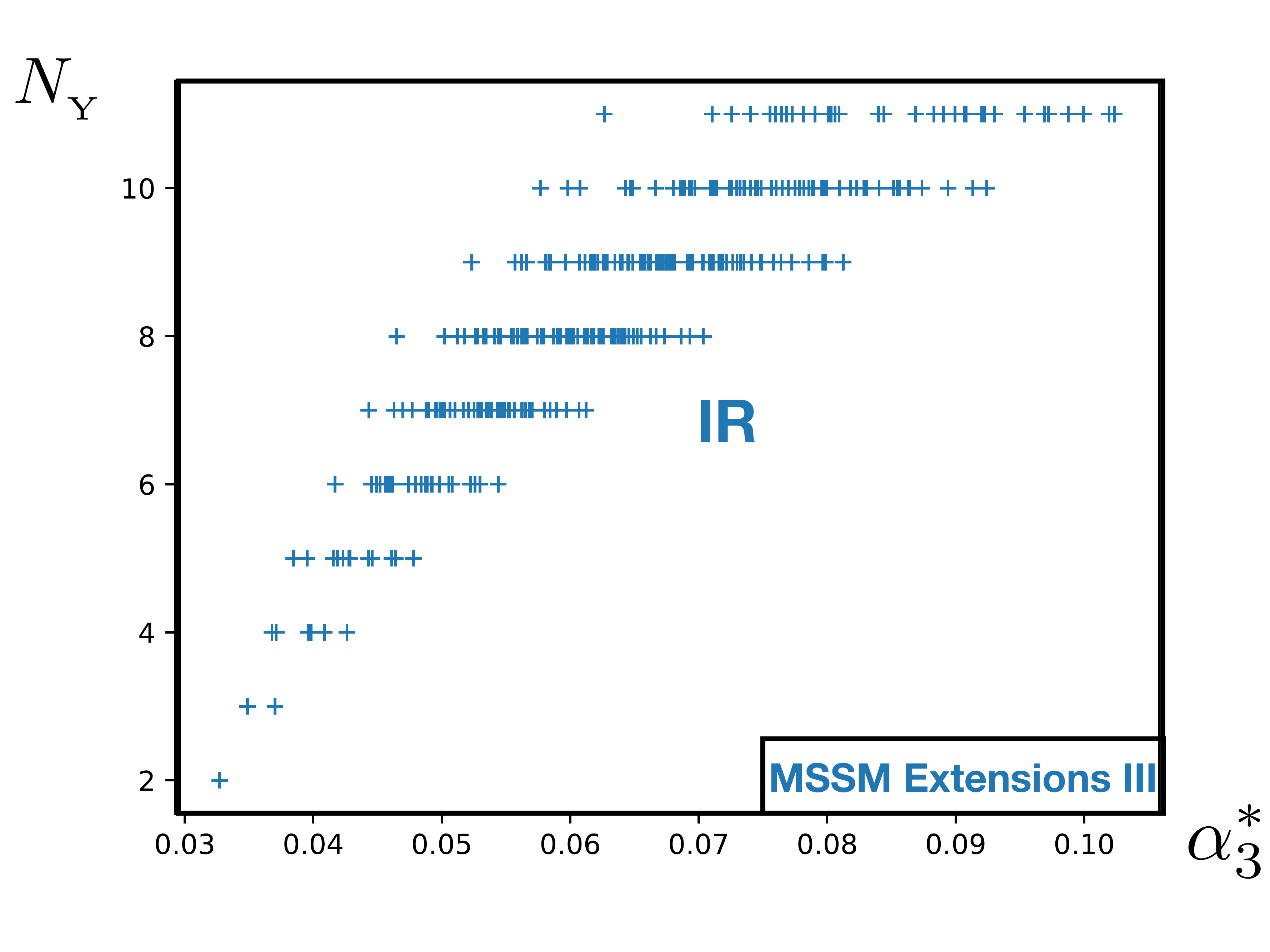}
\vskip-.3cm
\caption{Fixed points of MSSM extensions (type III). Shown is $\alpha_3^*$ 
at the  fixed point   $\text{FP}_3$ for  models 
with superpotential~\eqref{eq:model3superpotential}, corresponding to  the 3.868  different models.
All fixed points are IR and perturbative. Also shown is the number of Yukawa couplings $N_Y$~\eqref{eq:model3AnzYukawas}. The fixed point tends to become more strongly interacting 
with larger $N_Y$.}
\label{fig:model3FPs}
\end{figure}

\begin{table*}[t]
  \centering
 \sisetup {
    per-mode = fraction
  }
  \scalebox{0.75}{
  \begin{tabular}{`R`ccccccccccc`GcG`}
\toprule
\rowcolor{Yellow}
\cellcolor{white}
&&&&&&&&&&&&&&\\[-2.2ex]
\rowcolor{Yellow}
\cellcolor{white}
{\rm No.}
&$n_u$
&$n_d$
&$n_L$
&$I_{12}$
&$I_{13}$
&$I_{1d}$
&$I_{2d}$
&$I_{3d}$
&$I_{1u}$
&$I_{2u}$
&$I_{3u}$
&$\bm{N}_{{}_{\rm\bf BSM}}$
&$\bm{N}_{{}_{\rm\bf Y}}$
&$\bm{\al{3}^*}$
\\[.5ex]
\midrule
1  & 0 & 1 & 2 & 0 & 1 & 1 & 1 & 0 & 0 & 1 & 1 & 6 & 8 & \ 0.431\ \  \\
2 & 0 & 1 & 2 & 1 & 0 & 0 & 2 & 0 & 1 & 1 & 0 & 6 & 8 & 0.431 \\
3 & 0 & 1 & 2 & 1 & 0 & 1 & 1 & 0 & 0 & 2 & 0 & 6 & 8 & 0.431 \\
4 & 0 & 1 & 2 & 0 & 1 & 0 & 1 & 1 & 1 & 1 & 0 & 6 & 8 & 0.431 \\
5 & 0 & 1 & 2 & 0 & 1 & 1 & 0 & 1 & 0 & 2 & 0 & 6 & 8 & 0.431 \\
6
& 0 & 1 & 2 & 1 & 0 & 2 & 0 & 0 & 1 & 1 & 0 & 6 & 8 & 0.431 \\
\rowcolor{LightBlue}\cellcolor{LightRed}
7 \ & 0 & 1 & 2 & 0 & 1 & 0 & 2 & 0 & 1 & 1 & 0 & 6 & 8 & 0.458 \\
8 & 0 & 1 & 2 & 0 & 1 & 1 & 1 & 0 & 0 & 2 & 0 & 6 & 8 & 0.458 \\
9 & 0 & 1 & 2 & 1 & 0 & 0 & 1 & 1 & 1 & 1 & 0 & 6 & 8 & 0.458 \\
10 & 0 & 1 & 2 & 1 & 0 & 1 & 0 & 1 & 0 & 2 & 0 & 6 & 8 & 0.458 \\
11 & 0 & 1 & 2 & 1 & 0 & 1 & 0 & 1 & 1 & 1 & 0 & 6 & 8 & 0.458 \\
12 & 0 & 1 & 2 & 1 & 0 & 1 & 1 & 0 & 0 & 1 & 1 & 6 & 8 & 0.458 \\
13 & 0 & 1 & 2 & 1 & 0 & 2 & 0 & 0 & 0 & 1 & 1 & 6 & 8 & 0.458 \\
14 & 0 & 1 & 2 & 0 & 1 & 1 & 1 & 0 & 1 & 1 & 0 & 6 & 8 & 0.473 \\
15 & 0 & 1 & 2 & 1 & 0 & 1 & 1 & 0 & 1 & 1 & 0 & 6 & 8 & 0.473 \\
16 & 0 & 1 & 2 & 1 & 0 & 2 & 0 & 0 & 0 & 2 & 0 & 6 & 8 & 0.473 \\
17 & 0 & 1 & 2 & 0 & 1 & 2 & 0 & 0 & 0 & 2 & 0 & 6 & 8 & 0.473 \\
\midrule
\ 18\ \  & 0 & 1 & 2 & 0 & 2 & 1 & 0 & 0 & 1 & 1 & 0 & 6 & 9 & 0.452 \\
19 & 0 & 1 & 2 & 2 & 0 & 0 & 1 & 0 & 0 & 2 & 0 & 6 & 9 & 0.468 \\
20 & 0 & 1 & 2 & 0 & 2 & 0 & 0 & 1 & 0 & 2 & 0 & 6 & 9 & 0.468 \\
21 & 0 & 1 & 2 & 0 & 2 & 0 & 1 & 0 & 0 & 1 & 1 & 6 & 9 & 0.468 \\
22 & 0 & 1 & 2 & 2 & 0 & 0 & 1 & 0 & 0 & 0 & 2 & 6 & 9 & 0.488 \\
23 & 0 & 1 & 2 & 2 & 0 & 1 & 0 & 0 & 0 & 0 & 2 & 6 & 9 & 0.488 \\
24 & 0 & 1 & 2 & 2 & 0 & 0 & 0 & 1 & 0 & 1 & 1 & 6 & 9 & 0.488 \\
25 & 0 & 1 & 2 & 1 & 1 & 0 & 0 & 1 & 0 & 2 & 0 & 6 & 9 & 0.514 \\
26 & 0 & 1 & 2 & 1 & 1 & 0 & 1 & 0 & 0 & 1 & 1 & 6 & 9 & 0.514 \\
27 & 0 & 1 & 2 & 1 & 1 & 0 & 1 & 0 & 0 & 2 & 0 & 6 & 9 & 0.514 \\
28 & 1 & 0 & 3 & 1 & 1 & 0 & 0 & 0 & 0 & 2 & 1 & 8 & 9 & 0.514 \\
29 & 1 & 0 & 3 & 1 & 1 & 0 & 0 & 0 & 0 & 3 & 0 & 8 & 9 & 0.514 \\
30 & 0 & 1 & 3 & 1 & 1 & 0 & 0 & 1 & 0 & 2 & 0 & 8 & 9 & 0.514 \\
31 & 0 & 1 & 3 & 1 & 1 & 0 & 1 & 0 & 0 & 1 & 1 & 8 & 9 & 0.514 \\
32 & 0 & 1 & 3 & 1 & 1 & 0 & 1 & 0 & 0 & 2 & 0 & 8 & 9 & 0.514 \\
33 & 0 & 1 & 2 & 0 & 2 & 0 & 1 & 0 & 0 & 2 & 0 & 6 & 9 & 0.526 \\
34 & 0 & 1 & 2 & 2 & 0 & 0 & 0 & 1 & 0 & 2 & 0 & 6 & 9 & 0.526 \\
35 & 0 & 1 & 2 & 2 & 0 & 0 & 1 & 0 & 0 & 1 & 1 & 6 & 9 & 0.526 \\
36 & 1 & 0 & 3 & 0 & 2 & 0 & 0 & 0 & 0 & 3 & 0 & 8 & 9 & 0.526 \\
37 & 0 & 1 & 3 & 0 & 2 & 0 & 1 & 0 & 0 & 2 & 0 & 8 & 9 & 0.526 \\
38 & 1 & 0 & 3 & 2 & 0 & 0 & 0 & 0 & 0 & 2 & 1 & 8 & 9 & 0.526 \\
39 & 0 & 1 & 3 & 2 & 0 & 0 & 0 & 1 & 0 & 2 & 0 & 8 & 9 & 0.526 \\
40 & 0 & 1 & 3 & 2 & 0 & 0 & 1 & 0 & 0 & 1 & 1 & 8 & 9 & 0.526 \\
41 & 0 & 1 & 2 & 1 & 1 & 0 & 0 & 1 & 1 & 1 & 0 & 6 & 9 & 0.528 \\
42 & 0 & 1 & 2 & 1 & 1 & 1 & 0 & 0 & 0 & 1 & 1 & 6 & 9 & 0.528 \\
43 & 1 & 0 & 3 & 1 & 1 & 0 & 0 & 0 & 1 & 1 & 1 & 8 & 9 & 0.528 \\
44 & 0 & 1 & 3 & 1 & 1 & 0 & 0 & 1 & 1 & 1 & 0 & 8 & 9 & 0.528 \\
45 & 0 & 1 & 3 & 1 & 1 & 1 & 0 & 0 & 0 & 1 & 1 & 8 & 9 & 0.528 \\
46 & 0 & 1 & 2 & 1 & 1 & 1 & 0 & 0 & 1 & 1 & 0 & 6 & 9 & 0.547 \\
47 & 0 & 1 & 3 & 1 & 1 & 1 & 0 & 0 & 1 & 1 & 0 & 8 & 9 & 0.547 \\
48 & 0 & 1 & 2 & 2 & 0 & 1 & 0 & 0 & 1 & 1 & 0 & 6 & 9 & 0.561 \\
49 & 0 & 1 & 3 & 2 & 0 & 1 & 0 & 0 & 1 & 1 & 0 & 8 & 9 & 0.561 \\
50 & 0 & 1 & 2 & 0 & 2 & 0 & 1 & 0 & 1 & 1 & 0 & 6 & 9 & 0.561 \\
51 & 0 & 1 & 2 & 0 & 2 & 1 & 0 & 0 & 0 & 2 & 0 & 6 & 9 & 0.561 \\
52 & 0 & 1 & 2 & 2 & 0 & 0 & 1 & 0 & 1 & 1 & 0 & 6 & 9 & 0.561 \\
53 & 0 & 1 & 2 & 2 & 0 & 1 & 0 & 0 & 0 & 2 & 0 & 6 & 9 & 0.561 \\
54 & 1 & 0 & 3 & 0 & 2 & 0 & 0 & 0 & 1 & 2 & 0 & 8 & 9 & 0.561 \\
55 & 0 & 1 & 3 & 0 & 2 & 0 & 1 & 0 & 1 & 1 & 0 & 8 & 9 & 0.561 \\
56 & 0 & 1 & 3 & 0 & 2 & 1 & 0 & 0 & 0 & 2 & 0 & 8 & 9 & 0.561 \\
57 & 1 & 0 & 3 & 2 & 0 & 0 & 0 & 0 & 1 & 2 & 0 & 8 & 9 & 0.561 \\
\bottomrule
  \end{tabular}
\begin{tabular}{`R`ccccccccccc`GcG`}
\toprule
\rowcolor{Yellow}
\cellcolor{white}
&&&&&&&&&&&&&&\\[-2.2ex]
\rowcolor{Yellow}
\cellcolor{white}
{\rm No.}
&$n_u$
&$n_d$
&$n_L$
&$I_{12}$
&$I_{13}$
&$I_{1d}$
&$I_{2d}$
&$I_{3d}$
&$I_{1u}$
&$I_{2u}$
&$I_{3u}$
&$\bm{N}_{{}_{\rm\bf BSM}}$
& $\bm{N}_{{}_{\rm\bf Y}}$
&$\bm{\al3^*}$
\\[.5ex]
\midrule
\ 58\ \  & 0 & 1 & 3 & 2 & 0 & 0 & 1 & 0 & 1 & 1 & 0 & 8 & 9 & \ 0.561\ \  \\
59 & 0 & 1 & 3 & 2 & 0 & 1 & 0 & 0 & 0 & 2 & 0 & 8 & 9 & 0.561 \\
60 & 0 & 1 & 2 & 2 & 0 & 0 & 0 & 1 & 1 & 1 & 0 & 6 & 9 & 0.591 \\
61 & 0 & 1 & 2 & 2 & 0 & 1 & 0 & 0 & 0 & 1 & 1 & 6 & 9 & 0.591 \\
62 & 0 & 1 & 3 & 2 & 0 & 1 & 0 & 0 & 0 & 1 & 1 & 8 & 9 & 0.591 \\
63 & 0 & 1 & 3 & 2 & 0 & 0 & 0 & 1 & 1 & 1 & 0 & 8 & 9 & 0.591 \\
64 & 1 & 0 & 3 & 2 & 0 & 0 & 0 & 0 & 1 & 1 & 1 & 8 & 9 & 0.591 \\
65 & 1 & 0 & 4 & 2 & 0 & 0 & 0 & 0 & 1 & 1 & 1 & 10 & 9 & 0.591 \\
66 & 0 & 1 & 4 & 2 & 0 & 0 & 0 & 1 & 1 & 1 & 0 & 10 & 9 & 0.591 \\
67 & 0 & 1 & 4 & 2 & 0 & 1 & 0 & 0 & 0 & 1 & 1 & 10 & 9 & 0.591 \\
68 & 0 & 1 & 2 & 1 & 1 & 0 & 1 & 0 & 1 & 1 & 0 & 6 & 9 & 0.598 \\
69 & 0 & 1 & 2 & 1 & 1 & 1 & 0 & 0 & 0 & 2 & 0 & 6 & 9 & 0.598 \\
70 & 1 & 0 & 3 & 1 & 1 & 0 & 0 & 0 & 1 & 2 & 0 & 8 & 9 & 0.598 \\
71 & 0 & 1 & 3 & 1 & 1 & 0 & 1 & 0 & 1 & 1 & 0 & 8 & 9 & 0.598 \\
72 & 0 & 1 & 3 & 1 & 1 & 1 & 0 & 0 & 0 & 2 & 0 & 8 & 9 & 0.598 \\
73 & 1 & 0 & 4 & 1 & 1 & 0 & 0 & 0 & 1 & 2 & 0 & 10 & 9 & 0.598 \\
74 & 0 & 1 & 4 & 1 & 1 & 0 & 1 & 0 & 1 & 1 & 0 & 10 & 9 & 0.598 \\
75 & 0 & 1 & 4 & 1 & 1 & 1 & 0 & 0 & 0 & 2 & 0 & 10 & 9 & 0.598 \\
\midrule
76 & 0 & 1 & 3 & 0 & 3 & 0 & 0 & 0 & 1 & 1 & 0 & 8 & 10 & 0.519 \\
77 & 0 & 1 & 3 & 2 & 1 & 0 & 0 & 0 & 0 & 0 & 2 & 8 & 10 & 0.533 \\
78 & 0 & 1 & 3 & 1 & 2 & 0 & 0 & 0 & 0 & 1 & 1 & 8 & 10 & 0.594 \\
79 & 0 & 1 & 4 & 1 & 2 & 0 & 0 & 0 & 0 & 1 & 1 & 10 & 10 & 0.594 \\
80 & 0 & 1 & 3 & 0 & 3 & 0 & 0 & 0 & 0 & 2 & 0 & 8 & 10 & 0.652 \\
81 & 0 & 1 & 3 & 3 & 0 & 0 & 0 & 0 & 0 & 2 & 0 & 8 & 10 & 0.652 \\
82 & 0 & 1 & 4 & 0 & 3 & 0 & 0 & 0 & 0 & 2 & 0 & 10 & 10 & 0.652 \\
83 & 0 & 1 & 4 & 3 & 0 & 0 & 0 & 0 & 0 & 2 & 0 & 10 & 10 & 0.652 \\
84 & 0 & 1 & 5 & 0 & 3 & 0 & 0 & 0 & 0 & 2 & 0 & 12 & 10 & 0.652 \\
85 & 0 & 1 & 5 & 3 & 0 & 0 & 0 & 0 & 0 & 2 & 0 & 12 & 10 & 0.652 \\
86 & 0 & 1 & 3 & 3 & 0 & 0 & 0 & 0 & 0 & 0 & 2 & 8 & 10 & 0.655 \\
87 & 0 & 1 & 4 & 3 & 0 & 0 & 0 & 0 & 0 & 0 & 2 & 10 & 10 & 0.655 \\
88 & 0 & 1 & 5 & 3 & 0 & 0 & 0 & 0 & 0 & 0 & 2 & 12 & 10 & 0.655 \\
89 & 0 & 1 & 3 & 1 & 2 & 0 & 0 & 0 & 1 & 1 & 0 & 8 & 10 & 0.680 \\
90 & 0 & 1 & 4 & 1 & 2 & 0 & 0 & 0 & 1 & 1 & 0 & 10 & 10 & 0.680 \\
91 & 0 & 1 & 5 & 1 & 2 & 0 & 0 & 0 & 1 & 1 & 0 & 12 & 10 & 0.680 \\
92 & 0 & 1 & 3 & 2 & 1 & 0 & 0 & 0 & 0 & 1 & 1 & 8 & 10 & 0.705 \\
93 & 0 & 1 & 4 & 2 & 1 & 0 & 0 & 0 & 0 & 1 & 1 & 10 & 10 & 0.705 \\
94 & 0 & 1 & 5 & 2 & 1 & 0 & 0 & 0 & 0 & 1 & 1 & 12 & 10 & 0.705\\
95 & 0 & 1 & 3 & 3 & 0 & 0 & 0 & 0 & 1 & 1 & 0 & 8 & 10 & 0.722 \\
96 & 0 & 1 & 4 & 3 & 0 & 0 & 0 & 0 & 1 & 1 & 0 & 10 & 10 & 0.722 \\
97 & 0 & 1 & 5 & 3 & 0 & 0 & 0 & 0 & 1 & 1 & 0 & 12 & 10 & 0.722 \\
98 & 0 & 1 & 6 & 3 & 0 & 0 & 0 & 0 & 1 & 1 & 0 & 14 & 10 & 0.722 \\
99 & 0 & 1 & 3 & 2 & 1 & 0 & 0 & 0 & 0 & 2 & 0 & 8 & 10 & 0.738 \\
100 & 0 & 1 & 4 & 2 & 1 & 0 & 0 & 0 & 0 & 2 & 0 & 10 & 10 & 0.738 \\
101 & 0 & 1 & 5 & 2 & 1 & 0 & 0 & 0 & 0 & 2 & 0 & 12 & 10 & 0.738 \\
102 & 0 & 1 & 6 & 2 & 1 & 0 & 0 & 0 & 0 & 2 & 0 & 14 & 10 & 0.738 \\
103 & 0 & 1 & 3 & 1 & 2 & 0 & 0 & 0 & 0 & 2 & 0 & 8 & 10 & 0.738 \\
104 & 0 & 1 & 4 & 1 & 2 & 0 & 0 & 0 & 0 & 2 & 0 & 10 & 10 & 0.738 \\
105 & 0 & 1 & 5 & 1 & 2 & 0 & 0 & 0 & 0 & 2 & 0 & 12 & 10 & 0.738 \\
106 & 0 & 1 & 6 & 1 & 2 & 0 & 0 & 0 & 0 & 2 & 0 & 14 & 10 & 0.738 \\
107 & 0 & 1 & 3 & 3 & 0 & 0 & 0 & 0 & 0 & 1 & 1 & 8 & 10 & 0.750 \\
108 & 0 & 1 & 4 & 3 & 0 & 0 & 0 & 0 & 0 & 1 & 1 & 10 & 10 & 0.750 \\
109 & 0 & 1 & 5 & 3 & 0 & 0 & 0 & 0 & 0 & 1 & 1 & 12 & 10 & 0.750 \\
110 & 0 & 1 & 6 & 3 & 0 & 0 & 0 & 0 & 0 & 1 & 1 & 14 & 10 & 0.750 \\
111 & 0 & 1 & 3 & 2 & 1 & 0 & 0 & 0 & 1 & 1 & 0 & 8 & 10 & 0.767 \\
112 & 0 & 1 & 4 & 2 & 1 & 0 & 0 & 0 & 1 & 1 & 0 & 10 & 10 & 0.767 \\
113 & 0 & 1 & 5 & 2 & 1 & 0 & 0 & 0 & 1 & 1 & 0 & 12 & 10 & 0.767 \\
114 & 0 & 1 & 6 & 2 & 1 & 0 & 0 & 0 & 1 & 1 & 0 & 14 & 10 & 0.767 \\
\bottomrule
  \end{tabular}
  }
  \caption{Overview of the  UV fixed points  candidates  in type I models obtained in Sec.~\ref{sec:model1}. Shown is the number of left-handed up-type quark singlets $(n_u)$, down-type quark singlets $(n_d)$, and lepton  $(n_L)$ chiral superfields, the parameters~\eqref{eq:model1integers} characterising  the superpotential, the total number of  superfields ${N}_{{}_{\rm\bf BSM}}$ beyond the MSSM, the total number of  non-trivial Yukawa couplings ${N}_{{}_{\rm\bf Y}}$, and the fixed point values ${\al3^*}$ (see main text). Models are ordered according to increasing ${N}_{{}_{\rm\bf Y}}$,  ${\al3^*}$, and  ${N}_{{}_{\rm\bf BSM}}$. A benchmark, model 7, discussed further  in Sec.~\ref{sec:Benchmark} is highlighted in blue.}
  \label{tab:model1AScandidates}
\end{table*}

In Fig.~\ref{fig:model3FPs}, we show  $\alpha_3^*$ at $\text{FP}_3$  for all scanned type III models versus the number of  Yukawa couplings $N_Y$. Again, we observe 
that $\alpha_3^*$ tends to grow for a larger numbers of Yukawa couplings.
In our scans, the number of Yukawa couplings  equals
\begin{equation}
N_Y=2\sum\limits_{i=2}^4 I_{1i}+I_{1d}+I_{3d}+I_{1u}+I_{4u}+2\;.
\label{eq:model3AnzYukawas}
\end{equation}
Hence the scanned parameter space~\eqref{eq:scannedparamsmodel3} covers models with up to $11$ interacting Yukawa couplings.
Based on the structure of results, and also in comparison with the previous two models, we do not expect to find UV  fixed points by increasing the number of independent Yukawas.

\section{\bf Ultraviolet Completions}
\label{sec:ASanalysis}
In this section, we focus on   MSSM extensions with ultraviolet fixed points and the prospects for  matching  them to the Standard Model at low energies.

\subsection{Main Features and Benchmark}\label{sec:Benchmark}

In Sec.~\ref{sec:model1}, we obtained  MSSM extensions with ultraviolet fixed points (type I models). 
They are summarised in Tab.~\ref{tab:model1AScandidates} showing  for each model 
the number of left-handed up-type quark singlets $(n_u)$, down-type quark singlets $(n_d)$, and lepton  $(n_L)$ chiral superfields, the parameters~\eqref{eq:model1integers} characterising  the superpotential~\eq{eq:model1yukawaparams}, the total number of  superfields ${N}_{{}_{\rm\bf BSM}}$ beyond the MSSM, the total number of  non-trivial Yukawa couplings ${N}_{{}_{\rm\bf Y}}$, and the fixed point value of the strong coupling ${\al3^*}$. The  models are sorted according to increasing ${N}_{{}_{\rm\bf Y}}$,  ${\al3^*}$, and  ${N}_{{}_{\rm\bf BSM}}$, in this order.
For all models, we observe that the superpotential parameters obey  $I_{12}+I_{13}+I_{1d}+I_{2d}+I_{3d}+I_{1u}+I_{2u}+I_{3u}=5$, and   $I_{12}+I_{13}$  is either 1, 2 or 3. 
Furthermore, we always find 
that the MSSM bottom and top Yukawa couplings are  interacting in the UV, and that $n_u+n_d=1$.

\begin{figure}
\hskip-.4cm
\includegraphics[width = 0.4\textwidth]{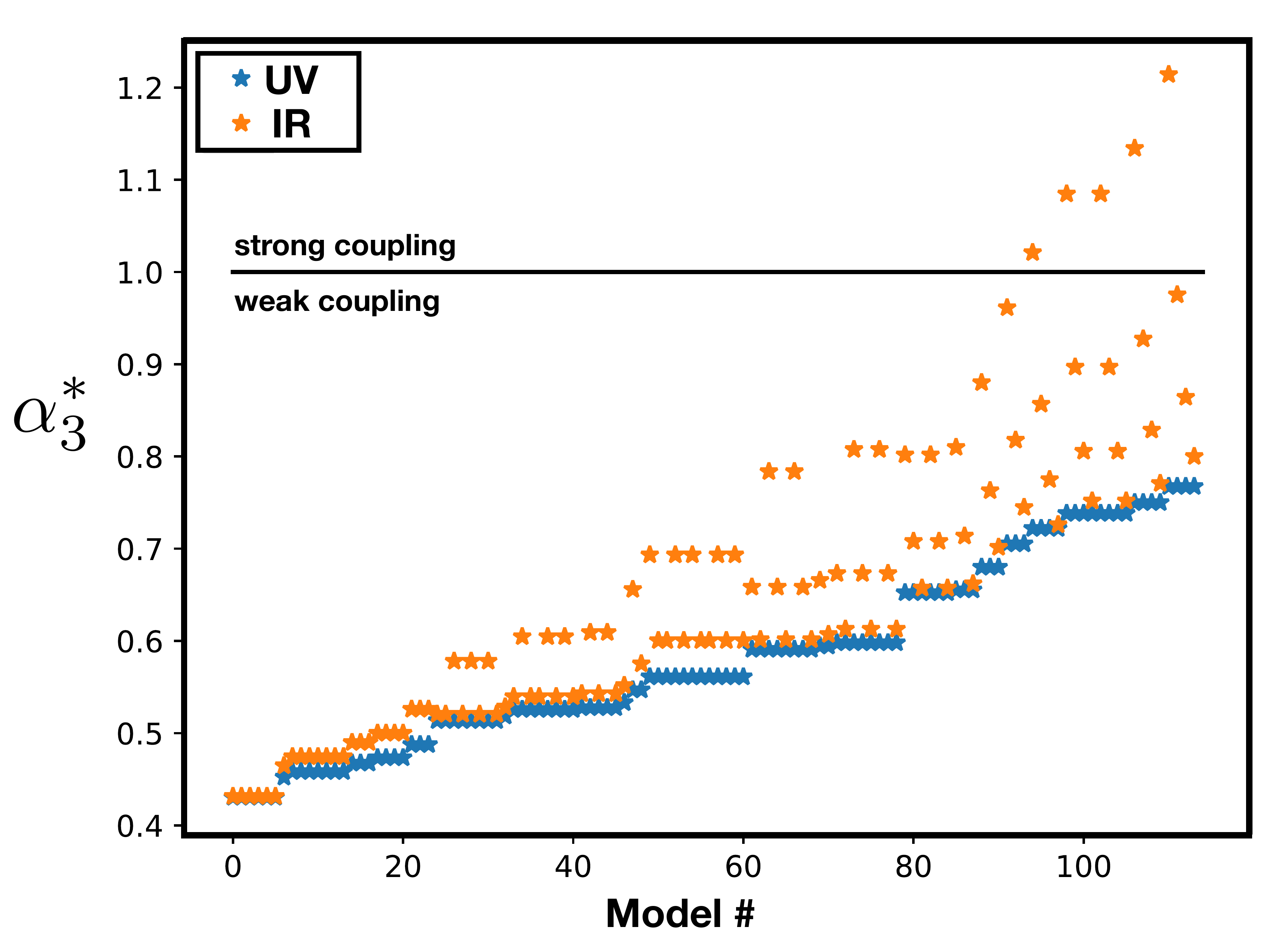}
\vskip-.3cm
\caption{The strong gauge coupling $\alpha_3^*$ at the UV fixed point ($\text{FP}_3$) (blue stars) and the associated IR fixed point ($\text{FP}_{23}$)  (orange stars) for  type I models  in Tab.~\ref{tab:model1AScandidates}. 
Models have  two quark singlets beyond the MSSM, plus leptons. }
\label{fig:FP3andFP23}
\end{figure}

Models with UV fixed points and  vanishing $\alpha_2^*$  always come with an associated IR fixed point  where $\alpha_2^*$ remains non-zero.
In Fig.~\ref{fig:FP3andFP23}, we compare the corresponding values of the strong gauge coupling.
They are in the range 
\begin{equation}
0.43 \lesssim \alpha_3^*\big|_{\text{UV}}< \alpha_3^*\big|_{\text{IR}}\,.
\label{eq:model1alpha3bound}
\end{equation}
Note that fixed points are
borderline perturbative. As such, they must be taken with a  grain of salt as higher loop corrections  \cite{Novikov:1983uc,Novikov:1985rd}  or  non-perturbative effects  \cite{Intriligator:2003jj,HLM22} may well be of a  similar magnitude.
In Fig.~\ref{fig:FP3andFP23}, the black horizontal line indicates the onset of strong coupling $(\alpha_3\ge 1)$, which is the case for a few IR fixed points.

Our results are in accord with more formal constraints such as the $a$-theorem, which states that  the central
charge $a=\frac{3}{32}\left[2d_G+\sum_i(1-R_i)(1-3(1-R_i)^2)\right]$must be a decreasing function along RG trajectories in any $4d$ quantum field theory  \cite{Anselmi:1997am}.  Here, $d_G$ denotes the dimension of the gauge groups,  $i$ runs over all chiral superfields, and $\gamma_i$ and $R_i=\s023(1+\gamma_i)$  the corresponding anomalous dimensions and $R$-charges, respectively.
We find
\begin{equation}
\Delta a =a_{\text{UV}}-a_{\text{IR}}>0
\end{equation}
on any of the UV-IR connecting trajectories. Had the IR limit been the Gaussian, validity of the $a$-theorem would imply strong coupling and non-perturbatively large $R$-charges in the UV, at least for some of the fields. In our models, this cannot arise because the Gaussian is a saddle  and the IR is not free. Hence, no trajectories connect the  UV to the Gaussian, which supports the weak form of the $a$-theorem as $a_{\text{UV}}-a_{\text{G}}<0$. 
We have also checked that fixed points are in accord with the positivity of central charges, the conformal collider bound, and constraints from unitarity \cite{Cardy:1988cwa,Osborn:1989td,Anselmi:1997am,Hofman:2008ar,Komargodski:2011vj}.

Next, we focus on a benchmark, model 7 from Tab.~\ref{tab:model1AScandidates}  with blue background color, and  matter content summarised in Tab.~\ref{tab:ASmodelparticles},
and with the superpotential 
\begin{align}
\begin{aligned}
W_1 \supset &\,Y^{411}\overline{d}_4Q_1L_1+Y^{432}\overline{d}_4Q_3L_2+Y^{124}\overline{d}_1Q_2L_{4}\\&
+Y^{225}\overline{d}_2Q_2L_{5}+\overline{Y}^{211}\overline{u}_2Q_1\overline{L}_1
\\&+\overline{Y}^{122}\overline{u}_1Q_2\overline{L}_2
+y_b\overline{d}_3Q_3H_d+y_t\overline{u}_3Q_3H_u\;.
\label{eq:ASmodelsuperpotential}
\end{aligned}
\end{align}
The model features the parameters
\begin{align}
\begin{aligned}
&I_{12}=I_{1d}=I_{3d}=I_{3u}=0\;,\\&
I_{13}=I_{1u}=I_{2u}=x_b=x_t=1\;, \\
&I_{2d}=2\;, n_u=0\;,\ n_d=1\;,\ n_L=2\;,
\end{aligned}
\end{align}
see   \eq{eq:model1yukawaparams}. Every term of the superpotential~\eqref{eq:ASmodelsuperpotential} contains exactly one superfield beyond the MSSM. Hence even though $R$-parity violation is a crucial feature of the superpotential~\eqref{eq:model1superpotential}, we can stay within experimental bounds in our benchmark model if the masses of these fields beyond the MSSM are large enough~\cite{Dawson:1985vr,Barbieri:1985ty,Barger:1989rk,Godbole:1992fb,Bhattacharyya:1995pq,Dreiner:1997uz,Domingo:2018qfg}. Further, the Yukawa couplings of~\eqref{eq:model1yukawaparams} and~\eqref{eq:ASmodelsuperpotential} are related as
\begin{align}
\begin{aligned}
Y^{411}\rightarrow y_{8},\; \ Y^{432}\rightarrow y_{9},\;\ Y^{124}\rightarrow y_{7},\;\\
\overline{Y}^{211}\rightarrow y_{11},\; \overline{Y}^{122}\rightarrow y_{12},\; Y^{225}\rightarrow y_{7}\,,
\end{aligned}
\end{align}
and  $y_4,y_5, y_6, y_{10}, y_{13}=0$.
Notice that the permutation flavor symmetry
\begin{equation}
(\overline{d}_1,L_4)\leftrightarrow (\overline{d}_2,L_5)\;.
\end{equation}
implies that the RG beta functions for the couplings $Y^{124}$ and $Y^{225}$ are equivalent, and mapped onto the same type of
beta function. 
The benchmark data is given in Tab.~\ref{FPmod3}.
All non-zero components of $\text{FP}_{23}$ are slightly larger than the corresponding couplings at $\text{FP}_3$.

\begin{table}
  		\addtolength{\tabcolsep}{3pt}
		\setlength{\extrarowheight}{3pt}			
 \centering
  \sisetup {
    per-mode = fraction
  }
  \scalebox{0.7}{
    \begin{tabular}{`ccccc`}
\toprule
\rowcolor{Yellow}
\bf Superfield&$\bm{SU(3)_C}$&$\bm{SU(2)_L}$&$\bm{U(1)_Y}$& \bf Multiplicity\\
\midrule
\underline{MSSM:}\ \  quark doublet $Q$  &\textbf{3}&\textbf{2}&$+\s016$&$3$\\
\rowcolor{Gray}
up-quark singlet $\overline{u}$&$\overline{\textbf{3}}$&\textbf{1}&$-\s023$&$3$\\
down-quark singlet  $\overline{d}$&$\overline{\textbf{3}}$&\textbf{1}&$+\s013$&$3$\\
\rowcolor{Gray}
lepton doublet $L$&\textbf{1}&\textbf{2}&$-\s012$&$3$\\
lepton singlet $\overline{e}$&\textbf{1}&\textbf{1}&$+1$&$3$\\
\rowcolor{Gray}
up-Higgs $H_u$&\textbf{1}&\textbf{2}&$+\s012$&$1$\\
down-Higgs $H_d$&\textbf{1}&\textbf{2}&$-\s012$&$1$\\
\midrule
\rowcolor{Gray}
\underline{BSM:}\ \ \ quark singlet $\overline{d}_4$\quad \ 
&$\overline{\textbf{3}}$&\textbf{1}&$+\s013$&$1$\\
 anti-quark singlet $d_1$&$\textbf{3}$&\textbf{1}&$-\s013$&$1$\\
\rowcolor{Gray}
 lepton doublets $L_{4,5}$&\textbf{1}&\textbf{2}&$-\s012$&$2$\\
 anti-lepton doublets $\overline{L}_{1,2}$&\textbf{1}&$\overline{\textbf{2}}$&$+\s012$&$2$\\
\bottomrule
  \end{tabular}
}
  \caption{Summary of left-handed superfields in the  benchmark model, also showing their gauge charges and multiplicity  (model 7 of Tab.~\ref{tab:model1AScandidates}). 
  The four bottom rows show the  superfield content beyond the MSSM.
  }
  \label{tab:ASmodelparticles}
\end{table}

\begin{table}[b]
  		\addtolength{\tabcolsep}{3pt}
		\setlength{\extrarowheight}{2pt}			
 \centering
  \scalebox{0.6}{  \begin{tabular}{`ccccccccccc`}
\toprule
\rowcolor{Yellow}
&$\bm{\alpha_3} $
&$ \bm{\alpha_2}$
&$\bm{\alpha_{Y^{411}}}$
&$\bm{\alpha_{Y^{432}}}$
&$\bm{\alpha_{Y^{124}}}$
&$\bm{\alpha_{Y^{225}}}$
&$\bm{\alpha_{\overline{Y}^{211}}}$
&$\bm{\alpha_{\overline{Y}^{122}}}$
&$\bm{\alpha_{y_t}}$
&$\bm{\alpha_{y_b}}$ \\
\midrule
\rowcolor{Gray}
$\bm{\text{\bf FP}_3}$&0.458&0
&0.278&0.208&0.306&0.306&0.361&0.306&0.320&0.320\\
$\bm{\text{\bf FP}_{23}}$&0.474&0.025
&0.296&0.222&0.326&0.326&0.385&0.326&0.341&0.341\\
\bottomrule
  \end{tabular}
}
  \caption{Coordinates of the UV and IR fixed points of the benchmark model (model 7 of Tab.~\ref{tab:model1AScandidates}).}
  \label{FPmod3}
\end{table}

Finally, we  compare the benchmark superpotential~\eqref{eq:ASmodelsuperpotential} with a sample superpotential~\eqref{eq:model2examplesuperpotential} which arises in models with additional quark doublets (see Sec.~\ref{sec:model2}).
Neglecting hypercharge (so that $\overline{d}$ and $\overline{u}$ have the same gauge representations), we see that these two superpotentials differ in two aspects. Firstly, in~\eqref{eq:model2examplesuperpotential}, $\overline{u}_3$ appears once outside of $W_\text{MSSM}$, inducing a mixing with $y_t$ ($E_{t7}{\neq} 0$ in App.~\ref{app:model2}). Secondly,  the term involving $\overline{Q}_1$ in~\eqref{eq:model2examplesuperpotential} has its gauge indices contracted in a different manner than $Q_2$ in the superpotential \eqref{eq:ASmodelsuperpotential}, yielding a Yukawa self-coupling term of $E_{10,10}=16$ (see App.~\ref{app:model2}) instead of $E_{12,12}=12$ (see App.~\ref{app:model1}). While these differences are small in that they lead to  only small differences in the  beta functions, they suffice to alter the nature of the fixed point from UV to IR.

\begin{figure*}
\vskip-.5cm
\hskip-6mm
\includegraphics[width=0.75\textwidth]{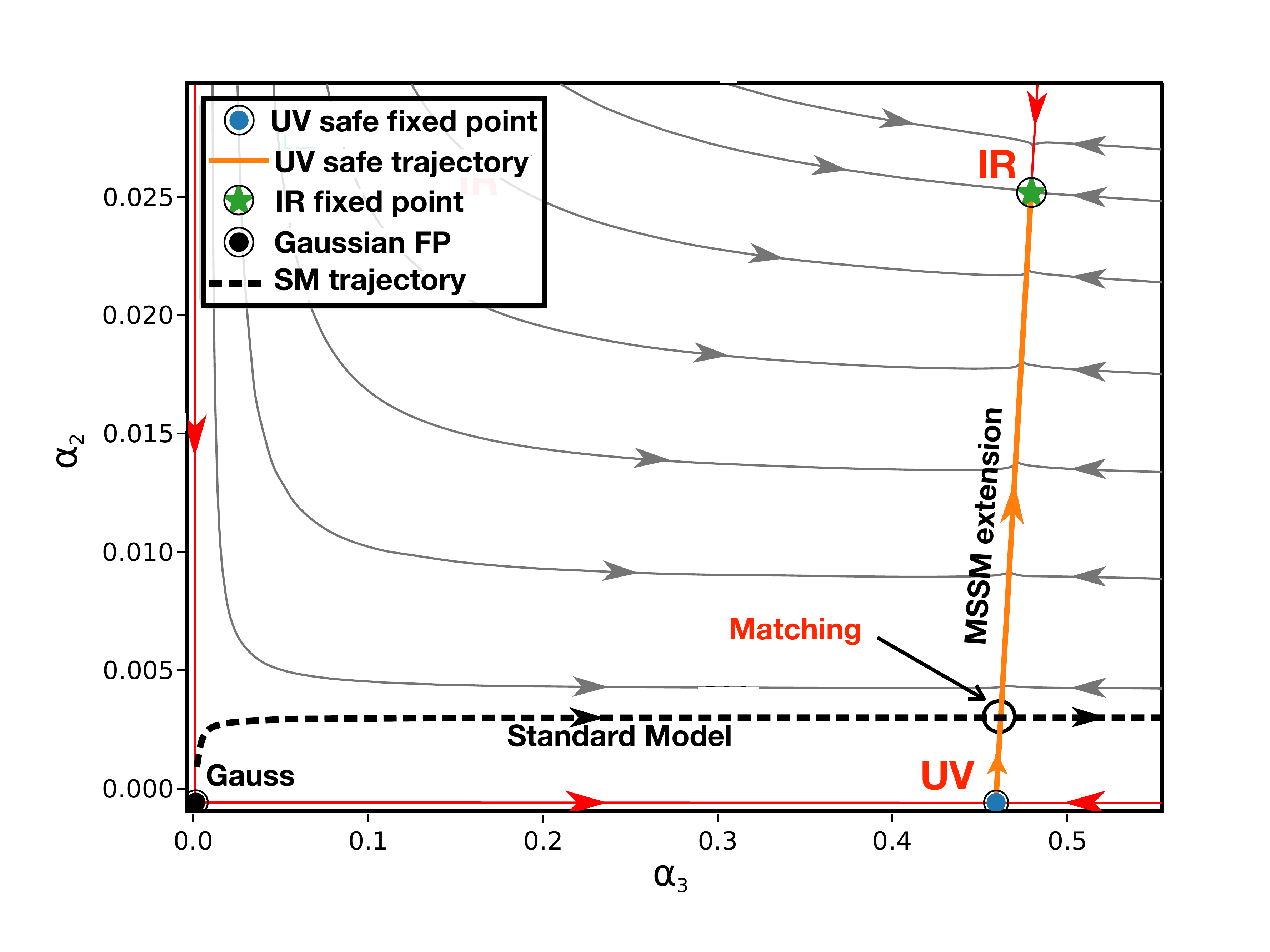}
\vskip-.5cm
\caption{
Shown are the RG flows of the benchmark model with  particle content as in Tab.~\ref{tab:ASmodelparticles} and  superpotential~\eqref{eq:ASmodelsuperpotential}, projected onto the $(\alpha_2,\alpha_3)$ plane, and all Yukawas  on their nullclines. 
Full dots show the Gaussian, the UV ($\text{FP}_3$), and the IR fixed point ($\text{FP}_{23}$), and 
arrows on trajectories point from the UV to the IR.
We also indicate  the SM running of gauge couplings (dashed black line), and the  trajectory emanating from the UV fixed point (orange).}
\label{fig:general_AS_Nullcline}
\end{figure*}

\subsection{Asymptotic Safety with Logarithmic Scaling and UV Critical Surface}
In four dimensions, the free Gaussian fixed point of a gauge coupling corresponds to a double-zero of its beta function \eqref{gaugeYukawa2}. This implies that the scaling dimension $\sim \partial_\alpha\beta(\alpha)|_{\alpha^*=0}$ vanishes, meaning that  the running of asymptotically free gauge couplings becomes logarithmically slow close to the Gaussian.
Conversely,  interacting fixed points generically correspond to single zeros with $\partial_\alpha\beta(\alpha)|_{\alpha^*\neq 0}\neq 0$, which implies 
that the running of couplings becomes power law, and much faster. 

Perhaps unexpectedly, however, it turns out that the RG scaling out of an interacting  fixed point  may still only be logarithmic  in some cases. The reason for this is that  fixed points may be partially interacting in gauge theories with product gauge groups, meaning that some of the gauge couplings are switched off at the fixed point. If so, the gauge couplings which vanish at the fixed point only run logarithmically, even if the other couplings achieve an interacting fixed point, $e.g.$~\eqref{B2-1}.

Here, this scenario is realised for all UV fixed points.
Specifically, the weak gauge coupling vanishes in the UV, where it represents a marginally relevant interaction as in \eqref{B2-1} with \eqref{B2eff} and \eqref{B2eff>0>B2}. In consequence, its RG running out of the fixed point is given by 
\beq\label{relevant}
\alpha_2(\mu)=\frac{\delta\alpha_2(\Lambda)}{1+B_{2,\rm eff}\,\delta\alpha_2\,\ln(\mu/\Lambda)}\,,
\eeq
with $\delta\alpha_2(\Lambda)$ the sole  free parameter of the theory at the high scale $\Lambda$, and $B_{2,\rm eff}>0$
the interaction-induced one-loop coefficient. 
Hence, despite of the theory being asymptotically safe with residual interactions in the UV, we find that the marginally relevant coupling $\alpha_2$ runs logarithmically  as in asymptotic freedom. Further, dimensional transmutation leads to a RG invariant mass scale
\beq\label{mutr}
\mu_{\rm tr}
=\Lambda\exp\big[-{B_{2,\rm eff}\,\delta\alpha_2(\Lambda)}\big]^{-1}\,,
\eeq
which is the analogue of the scale $\Lambda_{\rm QCD}$ in QCD, and independent of the high scale $\Lambda\gg \mu_{\rm tr}$ where the RG flow is started.

As an aside, we note that a power-law running of relevant perturbations out of a UV fixed point with supersymmetry can only arise if one or several of the asymptotically nonfree gauge couplings remain interacting in the UV. Here, this minimally requires  an interacting fixed point of the type FP${}_{23}$ with  both $\alpha_2^*$ and $\alpha_3^*$ non-zero, and outgoing trajectories. Although this scenario does not arise in the models studied here (nor in the models of \cite{Bond:2017suy}) it would be useful to establish conditions under which power-law scaling  becomes available.

Another  feature of the fixed points is that the strong gauge coupling $\alpha_3$ and the non-trivial Yukawa  couplings have become marginally   irrelevant interactions in the UV.  Their running is   fully determined by the one of $\alpha_2$ along the  outgoing trajectory, $\alpha_3(\mu)=F_3(\alpha_2(\mu))$ for the strong gauge coupling and $\alpha_i(\mu)=G_i(\alpha_2(\mu))$ for the non-trivial Yukawas,  with  $F_3(x)$ and $G_i(x)$  model-dependent functions. Close to the fixed point, this becomes
\beq\label{irrelevant}
\begin{aligned}
\alpha_3(\mu)=\alpha_3^*+A_3 \,\alpha_2(\mu)\\
\alpha_i(\mu)=\alpha_i^*+B_i \,\alpha_2(\mu)
\end{aligned}
\eeq
with $A_3$ and $B_i$ model-dependent parameters. Hence, all gauge and Yukawa couplings run logarithmically rather than power-law close to the partially interacting UV fixed point, which also percolates to  other  parameters including soft supersymmetry breaking terms or gaugino masses \cite{Martin:2000cr}. Most notably, the UV critical surface has only one free parameter  \cite{Bond:2017suy}. This should be contrasted with  asymptotic freedom where, instead, non-abelian gauge and Yukawa couplings are all marginally relevant. Hence, interacting UV fixed points with supersymmetry  enhance the predictive power over asymptotically free models and over fixed point theories without supersymmetry  \cite{Bond:2017suy}.

\subsection{Matching to the Standard Model}
We now discuss the phase diagram of the benchmark model, shown in Fig.~\ref{fig:general_AS_Nullcline}, which is of the same form as anticipated in Fig.~\ref{fig:templateRG}. Relevant perturbations such as $\delta\alpha_2$ can trigger outgoing RG flows. 
Specifically, Fig.~\ref{fig:general_AS_Nullcline} shows  RG trajectories in the $(\alpha_3,\alpha_2)$ plane with Yukawas projected onto their nullcline values, and the various fixed points, which are the Gaussian, the UV ($\text{FP}_3$), and the IR fixed point ($\text{FP}_{23}$). 
Arrows on trajectories point from the UV to the IR, and coloured trajectories indicate separatrices between the various fixed points.
The dashed black line indicates the SM running of gauge couplings, covering the range from MeV to Planckian energies. The UV safe trajectory emanating from the UV fixed point is depicted in orange. It would cross over into the IR fixed point provided all fields remain massless. Further,  we  note a bound for the IR fixed point value of the weak gauge coupling,
\beq\label{lower}
0.003 < \alpha_2^*\big|_{\rm IR}\,,
\eeq
or else the UV-IR connecting separatrix terminates at the IR fixed point before the SM line is ever reached. Then, to match the theory to the standard model, some fields need to decouple and become massive. If this happens at the appropriate energy scale, the UV safe trajectory can be matched to the SM, as indicated in Fig.~\ref{fig:general_AS_Nullcline}.

Next, we determine the matching scale $\mu=\mu_{\rm SM}$. 
Recall that since the UV safe theory only has a single free parameter \eqref{relevant}, $\alpha_3(\mu)$ is  uniquely determined by $\alpha_2(\mu)$. Hence, the  UV-safe trajectory relates the gauge couplings as $\alpha^{\rm UV}_3(\mu)\equiv \alpha_3(\alpha_2(\mu))$. Similarly, for the Standard Model we may express the RG running of  the strong gauge coupling in terms of the weak gauge coupling and write $\alpha_3^{\rm SM}(\mu)\equiv \alpha_3^{\rm SM}(\alpha_2^{\rm SM}(\mu))$. The matching scale is then uniquely determined from the  condition $\alpha_3^{\rm UV}= \alpha_3^{\rm SM}$,
which has a unique solution for $\alpha_2(\mu_{\rm SM})$ (see Fig.~\ref{fig:general_AS_Nullcline}). We find 
\begin{equation}\label{match}
\mu_{\rm SM}\lesssim \mathcal{O}(\SI{1}{\giga\electronvolt})
\end{equation} 
for the benchmark model of Tab.~\ref{tab:ASmodelparticles}, and, for that matter, for any of the models in Tab.~\ref{tab:model1AScandidates}.
Hence, despite of the remarkable fact that the MSSM extension can be matched to the SM, the
matching scale comes out too low to be in accord with observation. We conclude that the models cannot be taken 
as viable UV completions of the SM.

The result \eqref{match} can be understood from \eq{eq:model1alpha3bound}, which provides a lower bound
on $\alpha_3$, and the fact that the corresponding IR fixed point is more strongly coupled. The latter implies that
$\alpha_3(\mu)$ remains larger than its UV fixed point value along the trajectory down to the matching scale.
Therefore, we conclude that $\alpha_3^*$ being numerically too large at the UV fixed point is the culprit for disallowing 
a successful matching of $\delta\alpha_2$ perturbations.

\begin{figure}
\centering
\vskip-.5cm
\includegraphics[width=0.45\textwidth]{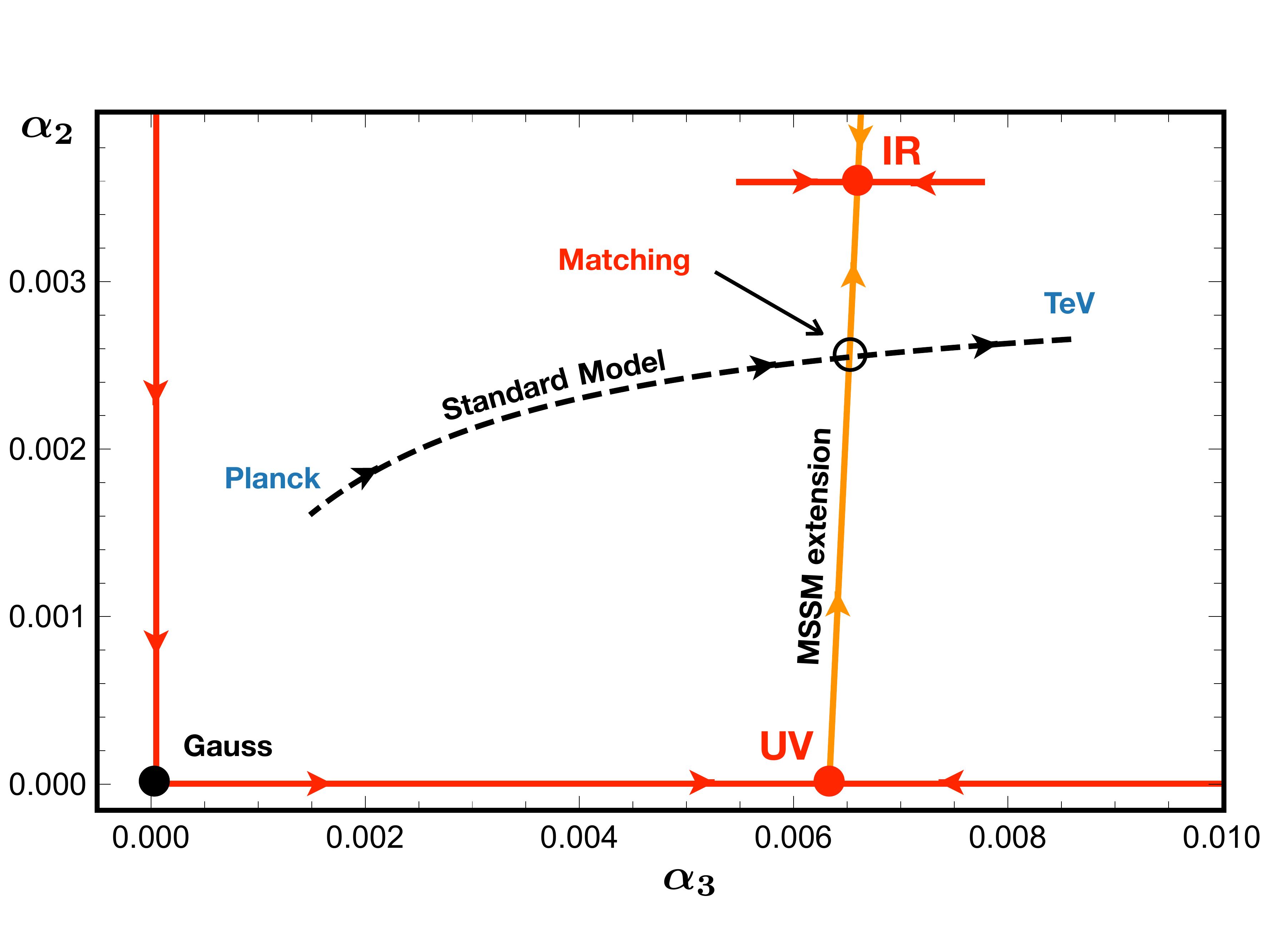}
\vskip-.5cm
\caption{Template for the matching of an asymptotically safe 
MSSM extension with $10^{-2}\lesssim \alpha_3^*|_\text{UV}\lesssim \alpha_3^*|_\text{IR}$ to the SM  at low energies (layout as in Fig.~\ref{fig:general_AS_Nullcline}). Notice that the SM running dictates the bound \eqref{limits} for the strong gauge coupling in the UV.}
\label{fig:TemplateUVsafe}
\end{figure}

\subsection{Standard Hierarchy}

Given the result \eq{match}, we now discuss prospects for MSSM extensions with UV fixed points where 
$\delta\alpha_2$ perturbations can be matched to the SM.
It either requires lower values for $\alpha_3$ in the UV, or a tilt of the UV-IR connecting separatrix, or a combination of both.

The first  scenario is illustrated in Fig.~\ref{fig:TemplateUVsafe}, where the black dashed line shows the SM running of couplings between the TeV and the Planck scale. Here,  we assume that the standard hierarchy 
\beq \label{hierarchy}
\alpha^*_3|_{\rm UV} < \alpha^*_3|_{\rm IR}\eeq
is observed.
Unlike in the benchmark model, however, we speculate that the fixed point coupling $\alpha^*_3$ is small enough to allow for a matching at TeV energies or above. More specifically, this would require that the gauge coupling fixed point sits within the range
\begin{equation}\label{limits}
0.001\lesssim \alpha_3^*\big|_{\rm UV}\lesssim 0.01\,,
\end{equation}
and is  smaller by at least one order of magnitude than what has been found in our models, see \eq{eq:model1alpha3bound}. 
To leading order in perturbation theory, we have observed the bound  (\refeq{bound-1}) for all our models.
This technical constraint may be overcome at higher loop order, or  non-perturbatively.

\subsection{Inverted Hierarchy}
\label{sec:Inverted}

\begin{figure}
\centering
\vskip-.5cm
\includegraphics[width=0.45\textwidth]{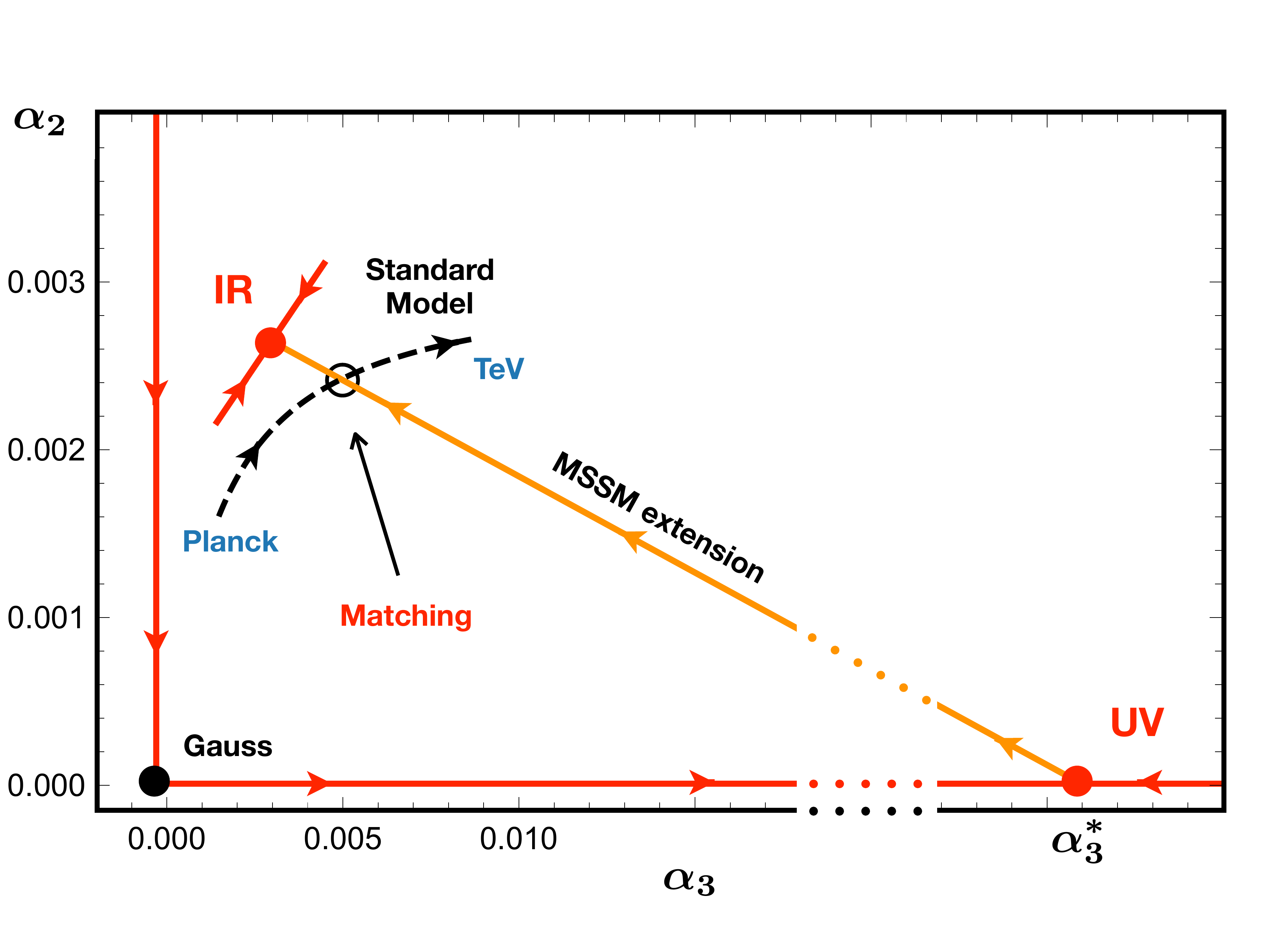}
\vskip-.5cm
\caption{Template for the matching of an asymptotically safe MSSM extension with $10^{-2}\ll \alpha_3^*|_\text{UV}$ to the SM at low energies  (layout as in Fig.~\ref{fig:general_AS_Nullcline}). A successful matching requires  $\alpha_3^*|_\text{IR}\lesssim \alpha_3^*|_\text{UV}$ and a  tilted separatrix in comparison to Fig.~\ref{fig:TemplateUVsafe}.}
\label{fig:TemplateUVsafe2}
\end{figure}

The second   scenario  questions the robustness of the hierarchy \eqref{hierarchy}.
Assuming that MSSM extensions can be found where the converse holds true, 
\beq \alpha^*_3|_{\rm UV} > \alpha^*_3|_{\rm IR}\,,\eeq
a matching to the SM would become a possibility owing to   
a ``tilted'' separatrix  (Fig.~\ref{fig:TemplateUVsafe2}) . Consequently, the separatrix  may cross the SM line in the  energy range where  $\alpha_3$ is  small. 

To check  the feasibility of this in perturbation theory, we look into the general expressions for  fixed points \eq{FP3}, \eq{alpha3_23} and \eq{alpha2_23} in terms of loop coefficients. After re-arranging terms, we find  that the UV and IR fixed points are related as
\begin{equation}\label{relation}
\alpha_3^*\big|_\text{IR}=\alpha_3^*\big|_\text{UV} -\frac{C'_{32}}{C'_{33}}\,\alpha_2^*\big|_\text{IR}\,.
\end{equation}
Hence, an inverted hierarchy 
requires  $C'_{32}/C'_{33}>0$. For sufficiently small one-loop factor $B_a$, the diagonal entries $C'_{aa}$ are always positive  in any  QFT, though they may become negative for larger positive $B_a$, while the off-diagonal terms $C'_{ab}$ $(a\neq b)$ may have either sign for general $B_a$. 
Further, \eqref{B3>0>B2} together with the mandatory sign flip  \eqref{B2eff}, \eq{B2eff>0>B2}
requires a negative $C'_{23}<0$. 
Altogether, the necessary and sufficient conditions for an inverted hierarchy are
\beq\label{conditionsinvertedapp}
\begin{aligned}
B_2 &< 0\,,\\[1ex]
B_3\,C'_{23}&< B_2\, C'_{33}\,,\\[1ex] 
 B_2\,C_{22}'&< B_3\,  C'_{32}\,,\\[1ex]
 0&<B_3, C'_{32}, C'_{33}\,.
 \end{aligned}
\eeq
They imply that the Yukawa-shifted off-diagonal two-loop coefficients must have opposite signs,
\beq\label{2332}
C'_{23}<0<C'_{32}\,.
\eeq
Next, we check  the conditions \eqref{2332}, and hence \eqref{conditionsinvertedapp} for 
extensions of the  MSSM with superpotentials $W$ involving quark singlets $\overline{q}$, quark doublets $Q$, lepton doublets $L$, and gauge singlets $S$, which we write schematically as
\begin{equation}
W\supset y_i(\overline{q}QL),\; y_i(S\overline{Q}Q)\;,
\label{eq:generalsuperpotential1}
\end{equation}
and where the index $i$  counts the different Yukawa terms.\footnote{All MSSM extensions considered in Secs.~\ref{sec:MSSM} and~\ref{sec:ASanalysis} are of this type.}
The gauge beta functions \eqref{gauge2} are characterised by
 the two-loop matrix $C$,
\begin{align}\label{eq:coeffproof}
 C=\left(
\begin{array}{c|c}
\\
7\sum\limits_{\text{SU}(2)}d_3(R)-48\;\;&16\sum\limits_{\text{SU}(3)\otimes\text{SU}(2)}1\\[4ex]
\hline\\
6\sum\limits_{\text{SU}(3)\otimes\text{SU}(2)}1&\;\;\frac{34}{3}\sum\limits_{\text{SU}(3)}d_2(R)-108\\
\end{array}
\right)
\end{align}
where the  sums account for 
the MSSM and BSM fields charged under the SM gauge groups with dimension $d_{2,3}(R)$.
The off-diagonal entries, which  are positive as in any quantum field theory, obey 
\beq\label{C38}
 C_{32}=\s038\, C_{23} \,.
  \eeq 
 Turning to the Yukawas, they contribute to the running of the gauge couplings with coefficients
$D_{3i}=8$ and 
$D_{2i}=12$.
The  Yukawa beta functions \eqref{generalYukawabetas} are characterised by the 
one-loop matrix $E$ (which we do not need to specify explicitly), and by the  gauge contributions 
$F_{i3}=\frac{32}{3}$ and 
$F_{i2}=6$.
The latter are known explicitly because the superpotential~\eqref{eq:generalsuperpotential1} has exactly two superfields  in the fundamental representation of either $\text{SU}(3)_C$ and $\text{SU}(2)_L$.
From  \eqref{generalYukawabetas}, and in terms of \eqref{alphaa}  and \eqref{alphai} we find  the Yukawa nullclines
\begin{equation}
\alpha_i=(\s0{32}{3}\alpha_3+6\alpha_2)\sum\limits_{j}^{}(E^{-1})_{ij}\,. 
\label{eq:Yukawanullclinerelation}
\end{equation}
Note that while the matrix elements of $E$  are always positive, this does not need to hold true {\it a priori} for those of $E^{-1}$. 
However, for quantum field theories  with physical perturbative fixed points, the sums $\sum_{j}^{}(E^{-1})_{ij}$
{\it must} all be  positive
to ensure positivity for all squared Yukawa couplings $\alpha_i$ (we have checked explicitly that this is true for  all models studied here). 
Consequently, we  find  the Yukawa-shifted two-loop matrix  \eqref{eq:modcoeff} as
\beq\label{C'model}
C'=C - \left(
\begin{array}{ccc}
72& \ \  &128\\
48& &\frac{256}{3}
\end{array}
\right)\sum\limits_{ij}^{}(E^{-1})_{ij}\,.
\eeq
To test \eqref{2332} we  focus on the off-diagonal elements, which may take either sign. Most notably, the relation \eqref{C38} 
continues to hold true for the matrix $C'$ where Yukawa-induced shifts have been taken into account,
 \beq\label{C'38}
 C'_{32}=\s038\, C'_{23} \,.
  \eeq 
Thus, the condition \eqref{2332}, and hence \eqref{conditionsinvertedapp},  cannot be satisfied  for any of the models involving the MSSM with additional quark singlets $\overline{q}$, quark doublets $Q$, and lepton doublets $L$ and superpotential \eq{eq:generalsuperpotential1} (In App.\ref{sec:appendixrelationFP3FP23} we show that the result generalises for 
superfields in general representations.) 
We conclude that the  hierarchy \eqref{hierarchy}  is a rather robust feature of models and that scenarios with $ \mu_{\rm SM}\gtrsim {\cal O}(1\, {\rm TeV})$   require either superpotentials  different from those studied here,  higher order loop corrections, or non-perturbative effects. We plan to explore these possibilities elsewhere.

\section{\bf Discussion and Concluding Remarks}
\label{sec:conclusion}
Motivated by 
the recent discovery of  interacting ultraviolet fixed points 
in weakly-coupled supersymmetric 
theories  \cite{Bond:2017suy}, we have performed a comprehensive search for fixed points and asymptotic safety 
in  extensions of the minimally supersymmetric Standard Model involving
either new quark singlets, new quark doublets, or a fourth generation. 
We thereby have performed a  scan over about 200k different MSSM extensions, most of which show  infrared conformal fixed points (Figs.~\ref{fig:branches},~\ref{fig:model2FPs} and \ref{fig:model3FPs}), and  about a hundred  candidates with ultraviolet 
ones (Figs.~\ref{fig:branches},~\ref{p109compare}). 
All settings  predict low-scale supersymmetry-breaking and a violation of $R$-parity.

While interacting fixed points can arise prolifically in asymptotically free  gauge theories, here,  we observe that their occurrence  is much more constrained  due the loss of asymptotic freedom in the weak gauge sector. The latter
is an unavoidable  consequence of 
supersymmetry, following from  the already known charge carriers of the Standard Model.
We expect that the availability of interacting fixed points will be equally constrained in other supersymmetric extensions including string-inspired models with many vectorlike representations, or   supersymmetric  grand unified theories.

By and large,  in all models the fixed point couplings \eqref{bound-1}, \eqref{bound-2}, \eqref{bound-3} are found to be small. 
Enhancing the number of independent superpotential couplings 
tends to  enhance the values of fixed point couplings (Figs.~\ref{fig:branches},~\ref{p109compare},~\ref{fig:model2FPs} and ~\ref{fig:model3FPs}). 
Thus, a reduction of flavor symmetry effectively requires stronger gauge interactions to achieve conformality. 
In some settings  couplings may  become  borderline perturbative (Figs.~\ref{fig:branches}, \ref{fig:FP3andFP23}), which calls for higher loop studies  \cite{Novikov:1983uc,Novikov:1985rd}  or  non-perturbative checks  \cite{Intriligator:2003jj,HLM22}.
We further noticed that models with interacting UV fixed points always also display an interacting IR fixed point \cite{Bond:2017suy}, while the converse is   not   the case.
All our results are consistent with  formal constraints such as the $a$-theorem, positivity of central charges, the conformal collider bound, 
and bounds from unitarity \cite{Cardy:1988cwa,Osborn:1989td,Anselmi:1997am,Hofman:2008ar,Komargodski:2011vj}.

From a phenomenological perspective, the   fixed point candidates in Tab.~\ref{tab:model1AScandidates} are of  interest as they may lead to ultraviolet completions of the Standard Model. 
Supersymmetry enhances the predictive power of interacting fixed points over non-interacting ones, leading to a smaller number of fundamentally free parameters. 
However, although  fixed points can  be matched to the Standard Model  (Fig.~\ref{fig:general_AS_Nullcline}) and the intrinsic $R$-parity violation can be tuned to stay within experimental bounds~\cite{Dawson:1985vr,Barbieri:1985ty,Barger:1989rk,Godbole:1992fb,Bhattacharyya:1995pq,Dreiner:1997uz,Domingo:2018qfg}, 
the matching scale comes   out too low \eqref{match}. 
The reason for this null result is that the strong gauge coupling  $\alpha_3^*$ in the UV is simply not small enough, and 
that it grows  along trajectories leaving the fixed point.
 Settings with fixed point couplings in the range \eq{lower},  \eqref{limits} can alleviate this impasse (Fig.~\ref{fig:TemplateUVsafe}), 
as can extensions  where the UV-safe separatrix is tilted towards smaller couplings 
(Fig.~\ref{fig:TemplateUVsafe2}). 
Either of these options  require   higher loops, non-perturbative effects, or interactions beyond those considered here.
We have not been concerned with the $U(1)_Y$ sector which remains infrared free despite of interacting fixed points. 
This is viable  phenomenologically because   the $U(1)_Y$ Landau pole arises beyond the Planck scale for most of the fixed point scenarios discussed here.  Still, it will be worth investigating whether MSSM extensions can also stabilise $U(1)_Y$.

A perhaps unexpected aspect of partially interacting UV fixed points is that the running of relevant perturbations 
may still be only  logarithmic \eqref{relevant}, \eqref{irrelevant} as in asymptotic freedom   \cite{Bond:2017suy}, rather than power-law. This feature  percolates to the running of other  parameters such as soft supersymmetry breaking terms, or gaugino masses.
Therefore, and much unlike fully interacting UV fixed points in non-supersymmetric theories \cite{Litim:2014uca,Bond:2017lnq},  relevant supersymmetric perturbations  would only show power-law running  if at least one of the asymptotically nonfree gauge couplings remains interacting in the UV (meaning UV fixed points with $\alpha_2^*,\alpha_3^*>0$ in our models). If this scenario is realised, gaugino masses will exhibit power-law running with scale, and may offer a possible solution to the supersymmetric flavor problem \cite{Martin:2000cr}.
Future work should clarify if this  scenario can arise for semi-simple supersymmetric matter-gauge theories, because if it does,  it may open up yet another route to UV-complete  the Standard Model.
\\[2ex]

{\bf Acknowledgements.---}  This work is supported by the Science and Technology Facilities Council (STFC)  under the Consolidated Grant ST/T00102X/1 (DL). 

\setcounter{section}{0}
\section*{\bf Appendices}

\subsection{Yukawa nullclines}
\label{sec:appendixunnaturalYukawas}

In this appendix, we have a look into  Yukawa nullclines, motivated by the observation that  the one loop beta functions of certain Yukawa couplings may not come out proportional to the Yukawa couplings themself, but, instead, be driven by inhomogeneous terms.  Consider, for example, a model with the superpotential
\begin{equation}
W=y_1ABC+y_2ABD+y_3AEC+y_4AED,
\label{eq:YukawaproblemSuperpotential}
\end{equation}
involving chiral superfields $A,B,C,D$, and $E$. The one-loop Feynman diagram
\begin{align}
\begin{tikzpicture}[baseline=(b.north)]
\begin{feynman}
\vertex[label=$A$](i1) ;
\vertex[below right = of i1] (b);
\vertex[label=$D$, right=0.75cm of b](D);
\vertex[right = of b](c);
\vertex[label=$E$, above right=1cm of c](E);
\vertex[label=$A$, below=1.9cm of E](F);
\vertex[right = of c](d);
\vertex[label=$C$, right = of d](i2);
\vertex[label=$B$, below left = of b](i3);
\diagram* {(i1)--(b)--(c)--[ out=90,in=90](d)--(i2), (c)--[out=-90,in=-90](d), (i3) -- (b),
};
\end{feynman}
\end{tikzpicture}
\label{eq:YukawaproblemFeynman}
\end{align}
 contributes to  the running of  $y_1$ and yields a contribution $\propto y_2\,y_4^*\,y_3$. 
Hence, the coupling $y_1$ would seem unnatural  \cite{tHooft:1979rat} in that it can be switched on by fluctuations, as long as  
the other Yukawas are non-zero. 
Specifically, the system of Yukawa beta functions for the superpotential~\eqref{eq:YukawaproblemSuperpotential} reads
\begin{align}
\begin{aligned}
\beta_{y_1}&=y_1(3|y_1|^2+3|y_2|^2+|y_3|^2-A_1)+y_2y_4^*y_3\,,\\
\beta_{y_2}&=y_2(3|y_2|^2+3|y_1|^2+|y_4|^2 -A_2)+y_1y_3^*y_4\,,\\
\beta_{y_3}&=y_3(3|y_3|^2+3|y_4|^2+|y_1|^2-A_3)+y_4y_2^*y_1\,,\\
\beta_{y_4}&=y_4(3|y_4|^2+3|y_3|^2+|y_2|^2-A_4)+y_3y_1^*y_2\,,
\label{eq:Yukawaproblembetas}
\end{aligned}
\end{align}
where we absorbed  the loop factor  $4\pi$ into the couplings. The coefficients
$A_i=A_i(g^2)$ 
are positive and linear functions of the gauge coupling squares $g^2$~\cite{Bond:2016dvk}. 
The Yukawa nullclines are found by solving $\beta_{y_i}=0$ in~\eqref{eq:Yukawaproblembetas} for the Yukawas. In the absence of  inhomogeneous terms, the nullcline conditions are linear functions of $ |y_i|^2$. 
 In the presence of  inhomogeneous terms, the nullcline conditions become cubic functions of $y_i$.
Note that enhanced symmetry, for instance $y_1=y_2=y_3=y_4$ in ~\eqref{eq:Yukawaproblembetas}, also lead to linear nullcline conditions.

In this work, we limit ourselves to  superpotentials with linear nullcline conditions. 
In practice, this is achieved by permutation symmetries (as seen above), or by selecting  superpotentials  where any two trilinear terms have at most one  superfield in common.

\begin{widetext}
\subsection{Expressions for Fixed Points}\label{sec:Expressions}

General expressions for the partially interacting fixed points $\text{FP}_{3}$ and $\text{FP}_{23}$ have been given in the main text, see \eqref{alpha_3_3}, \eqref{FP3}, and \eqref{alpha3_23}, \eqref{alpha2_23}, respectively. 
Here, we provide formal expressions for the fixed point candidates $\text{FP}_{13}$ and $\text{FP}_{123}$ in terms of loop coefficients.
For the fixed point $\text{FP}_{13}$, we have
\begin{align}
\begin{aligned}  
\alpha_3^*\big|_{\text{FP}_{13}}&=\frac{B_3C_{11}'-B_1C_{31}' }{C_{11}'C_{33}'-C_{13}'C_{31}'}\;,\;\\
\alpha_2^*\big|_{\text{FP}_{13}}&=0\;,\;\\
\alpha_1^*\big|_{\text{FP}_{13}}&=\frac{B_1C_{33}' -B_3C_{13}'}{C_{11}'C_{33}'-C_{13}'C_{31}'}\;,
\end{aligned}  
\end{align}
while for $\text{FP}_{123}$, the expressions read
\begin{align}
\begin{aligned}  
\alpha_3^*\big|_{\text{FP}_{123}}&=\frac{B_3(C_{12}'C_{21}' - C_{11}'C_{22}')-B_1(C_{21}'C_{32}' - C_{22}'C_{31}') - B_2(C_{12}'C_{31}' - C_{11}'C_{32}')}{C_{12}'C_{21}'C_{33}' - C_{12}'C_{23}'C_{31}' - C_{13}'C_{21}'C_{32}' + C_{13}'C_{22}'C_{31}' + C_{11}'C_{23}'C_{32}' - C_{11}'C_{22}'C_{33}'}\;,\;\\
\alpha_2^*\big|_{\text{FP}_{123}}&=\frac{B_1(C_{21}'C_{33}' - C_{23}'C_{31}') + B_2(C_{13}'C_{31}' - C_{11}'C_{33}') - B_3(C_{13}'C_{21}' - C_{11}'C_{23}')}{C_{12}'C_{21}'C_{33}' - C_{12}'C_{23}'C_{31}' - C_{13}'C_{21}'C_{32}' + C_{13}'C_{22}'C_{31}' + C_{11}'C_{23}'C_{32}' - C_{11}'C_{22}'C_{33}'}\;,\;\\
\alpha_1^*\big|_{\text{FP}_{123}}&=\frac{B_1(C_{23}'C_{32}'- C_{22}'C_{33}') + B_2(C_{12}'C_{33}' - C_{13}'C_{32}') - B_3(C_{12}'C_{23}' - C_{13}'C_{22}')}{C_{12}'C_{21}'C_{33}' - C_{12}'C_{23}'C_{31}' - C_{13}'C_{21}'C_{32}' + C_{13}'C_{22}'C_{31}' + C_{11}'C_{23}'C_{32}' - C_{11}'C_{22}'C_{33}'}\;.
\end{aligned}  
\end{align}
We emphasize that none of the studied models features a physical fixed point candidate of the type $\text{FP}_{1}$, $\text{FP}_{12}$,  $\text{FP}_{13}$ or $\text{FP}_{123}$ with all couplings positive.\\
When interacting fixed point may exist, i.e., for $N_{q,{\rm BSM}}\le 4$, see Sec.~\ref{sec:construction}, we can state a general lower bound on $\alpha_3^*\big|_{\text{FP}_{3}}=B_3/C'_{33}$. From~\eqref{eq:modcoeff} it is clear that $C'_{33}<C_{33}$ and with the expressions of App.~\ref{app:model1} we can write
\begin{equation}
\alpha_3^*\big|_{\text{FP}_{3}}>\frac{6-N_{q,{\rm BSM}}}{28+\frac{34}{3}N_{q,{\rm BSM}}}>\frac{3}{110}\approx 0.027\;,
\label{eq:alpha3bound}
\end{equation}
where $N_{q,{\rm BSM}} \leq 4$ has been used.
\end{widetext}

\subsection{Beta Functions: New Quark Singlets}
\label{app:model1}
Here, we summarise formul\ae\ for the perturbative RG equations of type I of MSSM extensions introduced in Sec.~\ref{sec:model1}, with superpotential terms parametrised by (\ref{eq:model1yukawaparams}). In terms of these parameters, the lower bounds on BSM matter fields  \eq{eq:model1parameterscan} are given by
\begin{align}
\begin{aligned}
&n^{\text{min}}_{d}=\text{max}\{ I_{12}+I_{13}+I_d-2\;,\;0\}\;,\\
&n^{\text{min}}_{u}=\text{max}\{I_u-2\;,\;0\}\;,\\
&n_L^{\text{min}}=\text{max}\{2(I_{12}+I_{13})+I_d-3\;,\; I_u\}\,,
\end{aligned}
\end{align}
with $I_d=I_{1d}+I_{2d}+I_{3d}$ and $I_u=I_{1u}+I_{2u}+I_{3u}$.
Using (\ref{alphai})  for the  reduced BSM Yukawas $(i=4,\cdots,13)$, we conclude that the RG running of all models  with \eq{eq:model1yukawaparams}  is encoded by up to 15 different beta functions for the gauge couplings  $\{\alpha_1,\alpha_2,\alpha_3\}$,  and up to 12 Yukawa couplings given by the top and bottom Yukawas $\{\alpha_t,\alpha_b\}$, and nine beyond MSSM Yukawas $\{\alpha_4,\cdots,\alpha_{13} \}$, modulo additional copies due to flavor symmetries. The corresponding Yukawa beta functions are denoted as $\beta_{t},\beta_{b},\beta_{4},...,\beta_{13}$, with $\partial_t\alpha_i\equiv\beta_i$. The general beta functions for the Yukawa and gauge couplings are given in \eqref{generalYukawabetas} and \eqref{gauge2}, respectively, involving the loop coefficients $B, C, D, E$ and $F$. 
Using the gauge couplings \eq{alphaa} and 
Yukawa couplings \eq{alphai} with $\{y_i\}$ as in (\ref{eq:model1yukawaparams}) 
the one-loop gauge coefficients $B$ read
\begin{equation}
\begin{aligned}
B_3&=6-2(n_u+n_d)\;,\;\\
B_2&=-2-2n_L\;,\;\\
B_1&=-22-\s0{16}{3}n_u-\s0{4}{3}n_d-2n_L\;,
\label{eq:B123type1}
\end{aligned}
\end{equation}
The two-loop matrices $C$  and $D$, and the one-loop matrices $E$ and $F$ in the Yukawa sector are given by
\begin{widetext}
\begin{align}
C&=\left(\begin{array}{c|c|c}
\frac{398}{9}+\frac{256}{27}n_u+\frac{16}{27}n_d+2n_L\;\;&\;\;18+6n_L\;\;&\;\frac{176}{3}+\frac{256}{9}n_u+\frac{64}{9}n_d\\[1ex]
\hline
6+2n_L&50+14n_L&48\\[1ex]
\hline
\frac{22}{3}+\frac{32}{9}n_u+\frac{8}{9}n_d&18&28+\frac{68}{3}(n_u+n_d)\\
\end{array}\right)\;,\\[1ex]
D&=
\left(\begin{smallmatrix}
\frac{52}{3}x_t&\frac{28}{3}x_b&\frac{28}{3}I_{12}&\frac{28}{3}I_{1d}&\frac{28}{3}I_{12}
&\frac{28}{3}I_{2d}&\frac{28}{3}I_{13}&\frac{28}{3}I_{13}&\frac{28}{3}I_{3d}&\frac{28}{3}I_{1u}
&\frac{52}{3}I_{2u}&\frac{52}{3}I_{3u}\\[1ex]
12x_t&12x_b&12I_{12}&12I_{1d}&12I_{12}&12I_{2d}&12I_{13}&12I_{13}&12I_{3d}&12I_{1u}&12I_{2u}&12I_{3u}\\[1ex]
8x_t&8x_b&8I_{12}&8I_{1d}&8I_{12}&8I_{2d}&8I_{13}&8I_{13}&8I_{3d}&8I_{1u}&8I_{2u}&8I_{3u}
\end{smallmatrix}\right)\;.\\[1ex]
E&{=}
\left(
\begin{smallmatrix}
{\scriptstyle12}&2x_b&0&0&0&0&0&2I_{13}&2I_{3d}&0&0&2I_{3u}\\
2x_t&{\scriptstyle12}&0&0&0&0&0&2I_{13}&2I_{3d}&0&0&2I_{3u}\\
0&0&{\scriptstyle
10{+}2I_{12}}&2I_{1d}&4&0&2I_{13}&0&0&2I_{1u}&0&0\\
0&0&2I_{12}&{\scriptstyle10{+}2I_{1d}}&0&0&2I_{13}&0&0&2I_{1u}&0&0\\ 
0&0&4&0&{\scriptstyle10{+}2I_{12}}&2I_{2d}&0&0&0&0&2I_{2u}&0\\    
0&0&0&0&2I_{12}&{\scriptstyle10{+}2I_{2d}}&0&0&0&0&2I_{2u}&0\\
0&0&2I_{12}&2I_{1d}&0&0&{\scriptstyle10{+}2I_{13}}&4&0&2I_{1u}&0&0\\    
2x_t&2x_b&0&0&0&0&4&{\scriptstyle10{+}2I_{13}}&2I_{3d}&0&0&2I_{3u}\\ 
2x_t&2x_b&0&0&0&0&0&2I_{13}&{\scriptstyle10{+}2I_{3d}}&0&0&2I_{3u}\\   
0&0&2I_{12}&2I_{1d}&0&0&2I_{13}&0&0&{\scriptstyle10{+}2I_{1u}}&0&0\\
0&0&0&0&2I_{12}&2I_{2d}&0&0&0&0&{\scriptstyle10{+}2I_{2u}}&0\\
2x_t&2x_b&0&0&0&0&0&2I_{13}&2I_{3d}&0&0&{\scriptstyle10{+}2I_{3u}}\\
\end{smallmatrix}
\right)\;,\\[1ex]
F&=\left(\begin{smallmatrix}
\frac{26}{9}&\frac{14}{9}&\frac{14}{9}&\frac{14}{9}&\frac{14}{9}&\frac{14}{9}&\frac{14}{9}&\frac{14}{9}&\frac{14}{9}&\frac{26}{9}&\frac{26}{9}&\frac{26}{9}\\[1ex]
6&6&6&6&6&6&6&6&6&6&6&6\\[1ex]
\frac{32}{3}&\frac{32}{3}&\frac{32}{3}&\frac{32}{3}&\frac{32}{3}&\frac{32}{3}&\frac{32}{3}&\frac{32}{3}&\frac{32}{3}&\frac{32}{3}&\frac{32}{3}&\frac{32}{3}
\end{smallmatrix}\right)^\text{T}\;.
\end{align}
\end{widetext}

\subsection{Beta Functions: New Quark Doublets}
\label{app:model2}
Here we summarise formul\ae\ for the perturbative RG equations of all gauge and Yukawa couplings for type II of  MSSM extensions  Sec.~\ref{sec:model2}, up to two loop for the gauge and one loop for the Yukawa beta functions. We also give  specifics for the selection of superpotential couplings  (\ref{eq:model2yukawaparams}):

 \begin{itemize}
 \item[$(i)$] The parameters $0\le x_b, x_t \le 1$ determine whether respectively the MSSM bottom- and top-Yukawa couplings are switched on ($x=1$) or off ($x=0$).
 \item[$(ii)$] The first- and second generation quark doublets $Q_1$ and $Q_2$ can appear in terms involving up- and down quark singlets $\overline{d}_i$, $\overline{u}_i$. The parameter $1\le I_Q\le 2$ indicates whether all terms containing $Q_1$ and quark singlets are present while $Q_2$ is absent ($I_Q=1$), or whether both doublets appear simultaneously ($I_Q=2$). Which quark singlets are available is determined by parameters introduced below.
 \item[$(iii)$] With $0\le x_4,\overline{x}_4\le 1$, superpotential terms involving $Q_4$  and first- and second generation down quarks $\overline{d}_{1,2}$ ($x_4$) and up-quarks  $\overline{u}_{1,2}$ ($\overline{x}_4$) are switched on ($x=1$) or off ($x=0$).
\item[$(iv)$] The parameter $0\le I_d\le 2$ determines the first- and second generation down quark singlet content in the superpotential. For $I_d=0$, $\overline{d}_{1}$ and $\overline{d}_2$ are absent while for $I_d=1$ only $\overline{d}_1$ is present in terms involving the quark doublets $Q_1$, $Q_2$ and $Q_4$, depending on whether they are allowed according to the parameters $I_Q$, $x_4$ and $\overline{x}_4$. For $I_d=2$ the contruction is analogous to the case $I_d=1$ but this time all respective terms containing both $\overline{d}_1$ and $\overline{d}_2$ are present. 
\item[$(v)$] With $0\le I_u\le 2$, the presence of superpotential terms analogous to $I_d$ in $(iv)$ is determined, but here concerning the up quark singlets $\overline{u}_{1,2}$.
\item[$(vi)$] Terms containing the third generation quark singlets $\overline{d}_3$ and $\overline{u}_3$ as well as the first- and second generation quark doublets $Q_1$ and $Q_2$ are switched on and off with the parameters $x_3, \overline{x}_3$. Here, $x_3$ determines whether such terms with $\overline{d}_3$ are present ($x_3=1$) or not ($x_3=0$), while $\overline{x}_3$ analogously is responsible for the presence of terms containing $\overline{u}_3$.
\item[$(vii)$] Each admitted Yukawa term gets amended by a lepton $L$ or anti-lepton $\overline{L}$, with each of these appearing at most once in the superpotential. In consequence,   the number of BSM leptons needs to be larger than
$n_{L,\text{min}}=\text{max}\{(1+x_4)I_d+x_3-3,(1+\overline{x}_4)I_u+\overline{x}_3\}$.
\item[$(viii)$] The presence of the $n_S$ Yukawa terms involving the gauge singletts $S_i$ are determined by the parameter $0\le x_S\le 1$. For $x_S=0$ these terms do not appear in the superpotential while for $x_S=1$ they do.
 \end{itemize}
 The one-loop gauge coefficients are given by 
$B_3=2\;,\;
B_2=-14-2n_L\;,\;
B_1=-\s0{74}{3}-2n_L\;.$
Further, the two-loop matrices $C$ and $D$,
and  the Yukawa matrices $E$ and $F$ as in \eqref{generalYukawabetas} and \eqref{gauge2}  are given by
\begin{widetext}
\begin{align}
C&=\left(\begin{array}{c|c|c}
\frac{1358}{27}+2n_L&38+6n_L&\frac{368}{9}\\[1ex]
\hline
\frac{38}{3}+2n_L&134+14n_L&80\\[1ex]
\hline
\frac{70}{9}&30&\frac{220}{3}\\
\end{array}\right)\;,\\[1ex]
D&=
\begin{pmatrix}
\frac{52}{3}x_t&\frac{28}{3}x_b&\frac{28}{3}I_dI_Q&\frac{52}{3}I_uI_Q&\frac{28}{3}I_Qx_3&\frac{52}{3}I_Q\overline{x}_3&\frac{28}{3}I_dx_4&\frac{52}{3}I_u\overline{x}_4&\frac{4}{3}n_Sx_S\\
12x_t&12x_b&12I_dI_Q&12I_uI_Q&12x_3I_Q&12\overline{x}_3I_Q&12x_4I_d&12\overline{x}_4I_u&12x_Sn_S\\
8x_t&8x_b&8I_dI_Q&8I_uI_Q&8x_3I_Q&8\overline{x}_3I_Q&8x_4I_d&8\overline{x}_4I_u&8x_Sn_S\\
\end{pmatrix}\;,\\[1ex]
E&=
\left(
\begin{smallmatrix}
12 & 2x_b &0 & 0 &  0 &4I_Q\overline{x}_3 &  0 & 0 & 0\\
2x_t & 12&0 & 0 &  4I_Qx_3 &0&  0 & 0 & 0\\
 0 & 0  &(10{+}2I_d)I_Q &2I_uI_Q&  2I_Qx_3 &2I_Q\overline{x}_3 &  4x_4 & 0& 0\\
 0& 0 &2I_dI_Q&(10{+}2I_u)I_Q&  2I_Qx_3&2I_Q\overline{x}_3 &  0& 4\overline{x}_4& 0\\
  0 & 4x_b &2I_dI_Q &2I_uI_Q &  12I_Q &2I_Q\overline{x}_3 &  0 & 0 &0\\
 4x_t & 0 &2I_dI_Q &2I_uI_Q &  2I_Qx_3 &12I_Q &  0 & 0 & 0\\
0 & 0  & 4I_Q & 0 & 0 & 0 & 10{+}2I_d & 2I_u\overline{x}_4& 2n_Sx_S\\
 0& 0  & 0 &4I_Q & 0 & 0&  2I_dx_4 & 10{+}2I_u& 2n_Sx_S\\
 0 & 0 &  0 & 0 & 0 &0 & 2I_dx_4 & 2I_u\overline{x}_4& 16n_S \\
\end{smallmatrix}
\right)\;,\\[1ex]
F&=\begin{pmatrix}
\frac{26}{9}&\frac{14}{9}&\frac{14}{9}&\frac{26}{9}&\frac{14}{9}&\frac{26}{9}&\frac{14}{9}&\frac{26}{9}&\frac{2}{9}\\
6&6&6&6&6&6&6&6&6\\
\frac{32}{3}&\frac{32}{3}&\frac{32}{3}&\frac{32}{3}&\frac{32}{3}&\frac{32}{3}&\frac{32}{3}&\frac{32}{3}&\frac{32}{3}\\
\end{pmatrix}^\text{T}\;.
\end{align}
\end{widetext}
\subsection{Beta Functions: Fourth Generation}
\label{app:model3}
We summarise formul\ae\ for the perturbative RG equations of all gauge and Yukawa couplings for type III of  the MSSM extensions  introduced in Sec.~\ref{sec:model3}.
The labelling of Yukawa couplings and the $I$-parameters is  given by (\ref{model3}), 
with $i\neq 3$ and the top- and bottom Yukawas always present.
The free parameters in \eq{model3} parameterize the  superpotential as follows:
\begin{itemize}
\item[$(i)$] The number of times down-quark singlets appear exactly once in the superpotentials  in terms involving $Q_1$ (or $Q_3$) is denoted by $I_{1d}$ (or $I_{3d}$).
\item[$(ii)$] Appearances of down-quarks in exactly two superpotential terms involving $Q_1$ and $Q_2$, or $Q_1$ and $Q_3$, or $Q_1$ and $Q_4$ are  counted by the parameters $I_{12}$, $I_{13}$ and $I_{14}$, respectively.
\item[$(iii)$] We let each up-quark $\overline{u}_i\neq\overline{u}_3$ appear at most once in our investigated superpotentials. Then, $I_{1u}$ counts the number of such terms additionally involving $Q_1$, and $I_{4u}$ those additionally involving $Q_4$, with $I_{1u}+I_{4u}\le 3$.
\item[$(iv)$] Each lepton doublet $L_i$ and each anti-lepton doublet $\overline{L}$ (both MSSM and BSM) may appear at most once. To accomodate for all Yukawa terms, the lepton number counting parameter needs to fulfill
\beq
n_L\,\ge\, n_{L,\text{min}}
\eeq
with $n_{L,\text{min}}=\text{max}\{2I_1{+}I_{1d}{+}I_{3d}{-}4\;, I_{1u}{+}I_{4u}\}$ 
and $I_1=I_{12}{+}I_{13}{+}I_{14}$.
\end{itemize}
The one-loop gauge coefficients read
$B_3=2\;,\;B_2=-6-2n_L\;,\;B_1=-\s0{89}{3}-2n_L\; ,$
while for  the two-loop matrices $C,D$, and the Yukawa matrices $E,F$, \eqref{generalYukawabetas} and \eqref{gauge2}, we obtain
 \begin{widetext}
\begin{align}
C&=\left(\begin{array}{c|c|c}
\frac{1574}{27}+2n_L&\;\;16+6n_L\;\;&\frac{704}{9}\\[1ex]
\hline
\frac{22}{3}+2n_L&71+14n_L&64\\[1ex]
\hline
\frac{88}{9}&24&\frac{220}{3}
\end{array}\right)\;,\\[1ex]
D&=\left(\begin{smallmatrix}
\frac{52}{3}&\frac{28}{3}&\frac{28}{3}I_{12}&\frac{28}{3}I_{12}&\frac{28}{3}I_{13}&\frac{28}{3}I_{13}&\frac{28}{3}I_{14}&\frac{28}{3}I_{14}&\frac{28}{3}I_{1d}&\frac{28}{3}I_{32}&\frac{52}{3}I_{1u}&\frac{52}{3}I_{4u}\\[1ex]
12&12&12I_{12}&12I_{12}&12I_{13}&12I_{13}&12I_{14}&12I_{14}&12I_{1d}&12I_{32}&12I_{1u}&12I_{4u}\\[1ex]
8&8&8I_{12}&8I_{12}&8I_{13}&8I_{13}&8I_{14}&8I_{14}&8I_{1d}&8I_{32}&8I_{1u}&8I_{4u}
\end{smallmatrix}\right)\;.\\[1ex]
E&{=}
\left(
\begin{smallmatrix}
12&   2      & 0             & 0              & 0             & 2I_{13}     &   0            & 0               & 0        & 2I_{3d}   & 0        & 0      \\  
 2 &   12    &0              & 0              & 0            & 2I_{13}      &  0           & 0                 & 0       & 2I_{3d}    & 0        & 0        \\  
 0 &   0      &{\scriptstyle10{+}2I_{12} }& 4              & 2I_{13}       & 0           & 2I_{14}       &0                & 2I_{1d}I_{1d}   &, 0       & 2I_{1u}    & 0    \\    
 0 &   0      & 4             & {\scriptstyle10{+}2I_{12}}& 0              & 0           &  0             & 0                  & 0       & 0        & 0       & 0         \\
 0 &   0      &2I_{12}       & 0              & {\scriptstyle10{+}2I_{13} }& 4           &   2I_{14}    & 0                 & 2I_{1d}    & 0       & 2+I_{1u}   &, 0  \\           
 2 &   2      &0             &0               & 4              & {\scriptstyle10{+}2I_{13}}&   0          & 0                 & 0        & 2I_{3d}    & 0        & 0      \\  
0  &   0      & 2I_{12}     & 0              & 2I_{13}        & 0            &  {\scriptstyle10{+}2I_{14}}& 4               & 2I_{1d}   & 0      & 2I_{1u}    & 0     \\      
0  &   0      & 0            & 0             & 0              & 0            &   4              & {\scriptstyle10{+}2I_{14}} &0          & 0        & 0        & 2I_{4u}    \\ 
0  &   0      &2I_{12}      & 0              & 2I_{13}        & 0           &   2I_{14}       & 0               & {\scriptstyle10{+}2I_{1d}} & 0       & 2I_{1u}    &0   \\       
2  &   2      & 0            & 0              & 0            & 2I_{13}       &   0             & 0                & 0        &{\scriptstyle10{+}2I_{3d}} & 0        & 0    \\      
0  &   0      & 2I_{12}   & 0              & 2I_{13}       & 0            &   2I_{14}       & 0                & 2I_{1d}    & 0& {\scriptstyle10{+}2I_{1u} }& 0      \\ 
0  &   0      &0            &0               & 0              & 0            &   0             & 2I_{14}          & 0        & 0       & 0       &{\scriptstyle10{+}2I_{4u} } \\
\end{smallmatrix}
\right)\;,\\[1ex]
F&=\left(\begin{smallmatrix}
\frac{26}{9}&\frac{14}{9}&\frac{14}{9}&\frac{14}{9}&\frac{14}{9}&\frac{14}{9}&\frac{14}{9}&\frac{14}{9}&\frac{14}{9}&\frac{14}{9}&\frac{26}{9}&\frac{26}{9}\\[1ex]
6&6&6&6&6&6&6&6&6&6&6&6\\[1ex]
\frac{32}{3}&\frac{32}{3}&\frac{32}{3}&\frac{32}{3}&\frac{32}{3}&\frac{32}{3}&\frac{32}{3}&\frac{32}{3}&\frac{32}{3}&\frac{32}{3}&\frac{32}{3}&\frac{32}{3}
\end{smallmatrix}\right)^\text{T}\;.
\end{align}
\end{widetext}

\subsection{Two-Loop Relations}
\label{sec:appendixrelationFP3FP23}

In Sec.~\ref{sec:Inverted}, we have analysed a   condition for a reverted hierarchy 
\beq\label{invert}
\alpha_3^*|_\text{UV}>\alpha_3^*|_\text{IR}
\eeq
 to occur at two-loop order in perturbation theory, where $\alpha_3^*|_\text{UV}$ refers to the fixed point coupling at the partially interacting UV fixed point FP${}_3$, and $\alpha_3^*|_\text{IR}$ refers to the fixed point coupling at the  corresponding IR fixed point FP${}_{23}$.

In this appendix, we study this relation  in a more principled manner, starting with a semi-simple supersymmetric gauge theory with gauge group $G_2\times G_3$, coupled to matter and a superpotential involving superfields $A$, $B$, and  $C$ which we write schematically as
\begin{equation}
W= \sum_i y_i(ABC)_i\;,
\label{eq:generalsuperpotential}
\end{equation}
and $i$  counting the different Yukawa terms. We assume that $A$ and $B$ are charged under one of the gauge groups, and $B$ and $C$ are charged under the other.
 Then, from~\eqref{TwoLoopa}, we observe the ratio of off-diagonal two-loop gauge contributions $(a\neq b)$
\begin{align}\label{CabCba}
\begin{aligned}
\frac{C_{ab}}{C_{ba}}&
=\frac{\sum\limits_{c}\frac{d_c}{d_c(R_a)}S_2^{R_a}(c)C_2^{R_b}(c)}{\sum\limits_{c}\frac{d_c}{d_c(R_b)}S_2^{R_b}(c)C_2^{R_a}(c)}\,.
\end{aligned}
\end{align}
Here, $c$ counts all superfields simultaneously charged under the gauge groups $G_a$ and $G_b$ (without representation components), and $d_c$ is the product of all dimensionalities of groups under which the field counted by $c$ is charged. Using the  identity 
$S_2^{R_a}(c)\,d(G_a)=d_c(R_a)\, C_2^{R_a}(c)$,
 the ratio \eqref{CabCba} simplifies into 
\begin{align}\label{CabCbaF}
\begin{aligned}
\frac{C_{ab}}{C_{ba}}&
=\frac{d(G_b)}{d(G_a)}=\frac{N_b^2-1}{N_a^2-1}
\end{aligned}
\end{align}
for $a\neq b$. We further specified to  $G_i=\text{SU}(N_i)$ gauge groups in the last step. Evidently, the general result \eqref{CabCbaF} falls back onto  \eqref{C38}
for $N_2=2$ and $N_3=3$, as it must.

Next, we are interested in the ratio of the off-diagonal Yukawa-shifted two-loop coefficients, which by definition \eqref{eq:modcoeff} take the form
\begin{equation}\label{C'C'}
\frac{C'_{ab}}{C'_{ba}}
=\frac{(C-DE^{-1}F)_{ab}}{(C-DE^{-1}F)_{ba}}\,.
\end{equation}
Here, we assumed that the Yukawa and gauge beta functions can always be written as in \eqref{generalYukawabetas}, \eqref{gauge2}. The key observation  is that  for some superpotentials the off-diagonal Yukawa-induced shift terms $(DE^{-1}F)$
 simplify into expressions of the form $(E^{-1})(DF)$, meaning that any dependence on the specifics of the Yukawa nullclines, encoded in the matrix $E^{-1}$,  drop out in the ratio $(DE^{-1}F)_{ab}/(DE^{-1}F)_{ba}$. In our case, we are left with the loop coefficients
\begin{align}
D_{ai}&=\frac{4}{d(G_a)} \sum\limits_k C_2^{R_a}(k)\;,\\
F_{ia}&=4 \sum\limits_{\tilde{k}} C_2^{R_a}(\tilde{k})\;,
\end{align}
where $D_{ai}$ as defined in \eqref{Y4D}  comes from the term $2Y_{4,a}$ in \eqref{gaugeYukawa2}, and $F_{ia}$
 comes from the second term of the anomalous dimensions~\eqref{gammaa}. Both $D_{ai}$ and $F_{ia}$ are independent of the Yukawa index $i$, implying that any dependence on the structure of the Yukawa nullclines drops out,
 \begin{equation}\label{DFab}
\frac{(DE^{-1}F)_{ab}}{(DE^{-1}F)_{ba}}=
\frac{(DF)_{ab}}{(DF)_{ba}}
=\frac{d(G_b)}{d(G_a)}\;.
\end{equation}
Hence, since the ratio of the off-diagonal elements \eqref{CabCba} and the ratio of the shift terms \eqref{DFab} are identical, it follows that the shifted matrix elements \eqref{C'C'}, again, have the same ratio,
\begin{equation}\label{CC=C'C'}
\frac{C_{ab}}{C_{ba}}
=\frac{(DE^{-1}F)_{ab}}{(DE^{-1}F)_{ba}}=\frac{C'_{ab}}{C'_{ba}}\,.
\end{equation}
Since $C_{ab}$ for $a\neq b$ are positive numbers in any quantum field theory, it follows that $C'_{ab}<0<C'_{ba}$ is strictly out of reach for these types of theories. 

{For other explicit examples of perturbatively controlled  supersymmetric quantum field theories with interacting fixed points where the relations \eqref{CabCbaF}, \eqref{DFab}, and \eqref{CC=C'C'}  are realised in a large-$N$ Veneziano limit, 
we refer to the models in \cite{Bond:2017suy}.}

\bibliographystyle{JHEP}
\bibliography{MSSM}

\providecommand{\href}[2]{#2}\begingroup\raggedright\begin{thebibliography}{10}

\bibitem{ParticleDataGroup:2020ssz}
{\scshape Particle Data Group} collaboration, \emph{{Review of Particle
  Physics}}, \href{https://doi.org/10.1093/ptep/ptaa104}{\emph{PTEP} {\bfseries
  2020} (2020) 083C01}.

\bibitem{Barbieri:2000gf}
R.~Barbieri and A.~Strumia, \emph{{The 'LEP paradox'}},  in \emph{{4th
  Rencontres du Vietnam}: {Physics at Extreme Energies (Particle Physics and
  Astrophysics)}}, 7, 2000
  [\href{https://arxiv.org/abs/hep-ph/0007265}{{\ttfamily hep-ph/0007265}}].

\bibitem{Arkani-Hamed:2004ymt}
N.~Arkani-Hamed and S.~Dimopoulos, \emph{{Supersymmetric unification without
  low energy supersymmetry and signatures for fine-tuning at the LHC}},
  \href{https://doi.org/10.1088/1126-6708/2005/06/073}{\emph{JHEP} {\bfseries
  06} (2005) 073} [\href{https://arxiv.org/abs/hep-th/0405159}{{\ttfamily
  hep-th/0405159}}].

\bibitem{Baer:2020kwz}
H.~Baer, V.~Barger, S.~Salam, D.~Sengupta and K.~Sinha, \emph{{Status of weak
  scale supersymmetry after LHC Run 2 and ton-scale noble liquid WIMP
  searches}}, \href{https://doi.org/10.1140/epjst/e2020-000020-x}{\emph{Eur.
  Phys. J. ST} {\bfseries 229} (2020) 3085}
  [\href{https://arxiv.org/abs/2002.03013}{{\ttfamily 2002.03013}}].

\bibitem{Bond:2016dvk}
A.D.~Bond and D.F.~Litim, \emph{{Theorems for Asymptotic Safety of Gauge
  Theories}}, \href{https://doi.org/10.1140/epjc/s10052-017-4976-5}{\emph{Eur.
  Phys. J.} {\bfseries C77} (2017) 429}
  [\href{https://arxiv.org/abs/1608.00519}{{\ttfamily 1608.00519}}].

\bibitem{Bond:2018oco}
A.D.~Bond and D.F.~Litim, \emph{{Price of Asymptotic Safety}},
  \href{https://doi.org/10.1103/PhysRevLett.122.211601}{\emph{Phys. Rev. Lett.}
  {\bfseries 122} (2019) 211601}
  [\href{https://arxiv.org/abs/1801.08527}{{\ttfamily 1801.08527}}].

\bibitem{Litim:2014uca}
D.F.~Litim and F.~Sannino, \emph{{Asymptotic safety guaranteed}},
  \href{https://doi.org/10.1007/JHEP12(2014)178}{\emph{JHEP} {\bfseries 12}
  (2014) 178} [\href{https://arxiv.org/abs/1406.2337}{{\ttfamily 1406.2337}}].

\bibitem{Bond:2017tbw}
A.D.~Bond, D.F.~Litim, G.~Medina~Vazquez and T.~Steudtner, \emph{{UV conformal
  window for asymptotic safety}},
  \href{https://doi.org/10.1103/PhysRevD.97.036019}{\emph{Phys. Rev. D}
  {\bfseries 97} (2018) 036019}
  [\href{https://arxiv.org/abs/1710.07615}{{\ttfamily 1710.07615}}].

\bibitem{Bond:2019npq}
A.D.~Bond, D.F.~Litim and T.~Steudtner, \emph{{Asymptotic safety with Majorana
  fermions and new large $N$ equivalences}},
  \href{https://doi.org/10.1103/PhysRevD.101.045006}{\emph{Phys. Rev. D}
  {\bfseries 101} (2020) 045006}
  [\href{https://arxiv.org/abs/1911.11168}{{\ttfamily 1911.11168}}].

\bibitem{Bond:2021tgu}
A.D.~Bond, D.F.~Litim and G.M.~Vazquez, \emph{{Conformal windows beyond
  asymptotic freedom}},
  \href{https://doi.org/10.1103/PhysRevD.104.105002}{\emph{Phys. Rev. D}
  {\bfseries 104} (2021) 105002}
  [\href{https://arxiv.org/abs/2107.13020}{{\ttfamily 2107.13020}}].

\bibitem{Bond:2017lnq}
A.D.~Bond and D.F.~Litim, \emph{{More asymptotic safety guaranteed}},
  \href{https://doi.org/10.1103/PhysRevD.97.085008}{\emph{Phys. Rev. D}
  {\bfseries 97} (2018) 085008}
  [\href{https://arxiv.org/abs/1707.04217}{{\ttfamily 1707.04217}}].

\bibitem{Bond:2017suy}
A.D.~Bond and D.F.~Litim, \emph{{Asymptotic safety guaranteed in
  supersymmetry}},
  \href{https://doi.org/10.1103/PhysRevLett.119.211601}{\emph{Phys. Rev. Lett.}
  {\bfseries 119} (2017) 211601}
  [\href{https://arxiv.org/abs/1709.06953}{{\ttfamily 1709.06953}}].

\bibitem{Weinberg:1980gg}
S.~Weinberg, \emph{{Ultraviolet divergences in quantum theories of
  gravitation}}, {\emph{in: General Relativity: An Einstein centenary survey,
  Eds. Hawking, S.W., Israel, W; Cambridge University Press} (1979) 790}.

\bibitem{Bond:2017wut}
A.D.~Bond, G.~Hiller, K.~Kowalska and D.F.~Litim, \emph{{Directions for model
  building from asymptotic safety}},
  \href{https://doi.org/10.1007/JHEP08(2017)004}{\emph{JHEP} {\bfseries 08}
  (2017) 004} [\href{https://arxiv.org/abs/1702.01727}{{\ttfamily
  1702.01727}}].

\bibitem{Kowalska:2017fzw}
K.~Kowalska, A.~Bond, G.~Hiller and D.~Litim, \emph{{Towards an asymptotically
  safe completion of the Standard Model}},
  \href{https://doi.org/10.22323/1.314.0542}{\emph{PoS} {\bfseries EPS-HEP2017}
  (2017) 542}.

\bibitem{Bissmann:2020lge}
S.~Bi\ss{}mann, G.~Hiller, C.~Hormigos-Feliu and D.F.~Litim,
  \emph{{Multi-lepton signatures of vector-like leptons with flavor}},
  \href{https://doi.org/10.1140/epjc/s10052-021-08886-3}{\emph{Eur. Phys. J. C}
  {\bfseries 81} (2021) 101}
  [\href{https://arxiv.org/abs/2011.12964}{{\ttfamily 2011.12964}}].

\bibitem{Hiller:2019mou}
G.~Hiller, C.~Hormigos-Feliu, D.F.~Litim and T.~Steudtner, \emph{{Anomalous
  magnetic moments from asymptotic safety}},
  \href{https://doi.org/10.1103/PhysRevD.102.071901}{\emph{Phys. Rev. D}
  {\bfseries 102} (2020) 071901}
  [\href{https://arxiv.org/abs/1910.14062}{{\ttfamily 1910.14062}}].

\bibitem{Hiller:2020fbu}
G.~Hiller, C.~Hormigos-Feliu, D.F.~Litim and T.~Steudtner, \emph{{Model
  Building from Asymptotic Safety with Higgs and Flavor Portals}},
  \href{https://doi.org/10.1103/PhysRevD.102.095023}{\emph{Phys. Rev. D}
  {\bfseries 102} (2020) 095023}
  [\href{https://arxiv.org/abs/2008.08606}{{\ttfamily 2008.08606}}].

\bibitem{Bause:2021prv}
R.~Bause, G.~Hiller, T.~H\"ohne, D.F.~Litim and T.~Steudtner,
  \emph{{B-anomalies from flavorful U(1)$'$ extensions, safely}},
  \href{https://doi.org/10.1140/epjc/s10052-021-09957-1}{\emph{Eur. Phys. J. C}
  {\bfseries 82} (2022) 42} [\href{https://arxiv.org/abs/2109.06201}{{\ttfamily
  2109.06201}}].

\bibitem{Martin:2000cr}
S.P.~Martin and J.D.~Wells, \emph{{Constraints on Ultraviolet Stable Fixed
  Points in Supersymmetric Gauge Theories}},
  \href{https://doi.org/10.1103/PhysRevD.64.036010}{\emph{Phys. Rev.}
  {\bfseries D64} (2001) 036010}
  [\href{https://arxiv.org/abs/hep-ph/0011382}{{\ttfamily hep-ph/0011382}}].

\bibitem{Intriligator:2015xxa}
K.~Intriligator and F.~Sannino, \emph{{Supersymmetric Asymptotic Safety is Not
  Guaranteed}}, \href{https://doi.org/10.1007/JHEP11(2015)023}{\emph{JHEP}
  {\bfseries 11} (2015) 023}
  [\href{https://arxiv.org/abs/1508.07411}{{\ttfamily 1508.07411}}].

\bibitem{Luty:2012ww}
M.A.~Luty, J.~Polchinski and R.~Rattazzi, \emph{{The $a$-theorem and the
  Asymptotics of 4D Quantum Field Theory}},
  \href{https://doi.org/10.1007/JHEP01(2013)152}{\emph{JHEP} {\bfseries 01}
  (2013) 152} [\href{https://arxiv.org/abs/1204.5221}{{\ttfamily 1204.5221}}].

\bibitem{Allanach:1996nj}
B.C.~Allanach, G.~Amelino-Camelia and O.~Philipsen, \emph{{Infrared Fixed Point
  Structure Characterizing SUSY $SU(5)$ Symmetry Breaking}},
  \href{https://doi.org/10.1016/S0370-2693(96)01630-9}{\emph{Phys. Lett. B}
  {\bfseries 393} (1997) 349}
  [\href{https://arxiv.org/abs/hep-ph/9611286}{{\ttfamily hep-ph/9611286}}].

\bibitem{Lanzagorta:1995ai}
M.~Lanzagorta and G.G.~Ross, \emph{{Infrared Fixed Point Structure of Soft
  Supersymmetry Breaking Mass Terms}},
  \href{https://doi.org/10.1016/0370-2693(95)01053-X}{\emph{Phys. Lett. B}
  {\bfseries 364} (1995) 163}
  [\href{https://arxiv.org/abs/hep-ph/9507366}{{\ttfamily hep-ph/9507366}}].

\bibitem{Kobayashi:1996zu}
T.~Kobayashi and Y.~Yamagishi, \emph{{Quasiyukawa Fixed Point Due to Decoupling
  of SUSY Particles}},
  \href{https://doi.org/10.1016/0370-2693(96)00598-9}{\emph{Phys. Lett. B}
  {\bfseries 381} (1996) 169}
  [\href{https://arxiv.org/abs/hep-ph/9601374}{{\ttfamily hep-ph/9601374}}].

\bibitem{Codoban:1999fp}
S.~Codoban and D.I.~Kazakov, \emph{{Approximate Analytic Solutions of RG
  Equations for Yukawa and Soft Couplings in SUSY Models}},
  \href{https://doi.org/10.1007/s100520050726}{\emph{Eur. Phys. J. C}
  {\bfseries 13} (2000) 671}
  [\href{https://arxiv.org/abs/hep-ph/9906256}{{\ttfamily hep-ph/9906256}}].

\bibitem{Aulakh:2008sn}
C.S.~Aulakh and S.K.~Garg, \emph{{The New Minimal Supersymmetric GUT : Spectra,
  RG Analysis and Fermion Fits}},
  \href{https://doi.org/10.1016/j.nuclphysb.2011.12.003}{\emph{Nucl. Phys. B}
  {\bfseries 857} (2012) 101}
  [\href{https://arxiv.org/abs/0807.0917}{{\ttfamily 0807.0917}}].

\bibitem{Abel:1998yi}
S.A.~Abel and B.C.~Allanach, \emph{{Ruling Out the MSSM at the Low Tan Beta
  Fixed Point}},
  \href{https://doi.org/10.1016/S0370-2693(98)00584-X}{\emph{Phys. Lett. B}
  {\bfseries 431} (1998) 339}
  [\href{https://arxiv.org/abs/hep-ph/9803476}{{\ttfamily hep-ph/9803476}}].

\bibitem{Huang:2000rn}
C.-S.~Huang, W.~Liao, Q.-S.~Yan and S.-H.~Zhu, \emph{{Renormalization Group
  Equations and Infrared Quasifixed Point Behaviors of Nonuniversal Soft Terms
  in MSSM}}, \href{https://doi.org/10.1088/0954-3899/27/4/308}{\emph{J. Phys.
  G} {\bfseries 27} (2001) 833}
  [\href{https://arxiv.org/abs/hep-ph/0008166}{{\ttfamily hep-ph/0008166}}].

\bibitem{Nevzorov:2013ixa}
R.~Nevzorov, \emph{{Quasifixed Point Scenarios and the Higgs Mass in the E6
  Inspired Supersymmetric Models}},
  \href{https://doi.org/10.1103/PhysRevD.89.055010}{\emph{Phys. Rev. D}
  {\bfseries 89} (2014) 055010}
  [\href{https://arxiv.org/abs/1309.4738}{{\ttfamily 1309.4738}}].

\bibitem{Casas:1998vh}
J.A.~Casas, J.R.~Espinosa and H.E.~Haber, \emph{{The Higgs Mass in the MSSM
  Infrared Fixed Point Scenario}},
  \href{https://doi.org/10.1016/S0550-3213(98)00327-7}{\emph{Nucl. Phys. B}
  {\bfseries 526} (1998) 3}
  [\href{https://arxiv.org/abs/hep-ph/9801365}{{\ttfamily hep-ph/9801365}}].

\bibitem{Barger:1993vu}
V.D.~Barger, M.S.~Berger, P.~Ohmann and R.J.N.~Phillips,
  \emph{{Phenomenological Implications of the M(T) Rge Fixed Point for SUSY
  Higgs Boson Searches}},
  \href{https://doi.org/10.1016/0370-2693(93)91248-L}{\emph{Phys. Lett. B}
  {\bfseries 314} (1993) 351}
  [\href{https://arxiv.org/abs/hep-ph/9304295}{{\ttfamily hep-ph/9304295}}].

\bibitem{Bardeen:1993rv}
W.A.~Bardeen, M.~Carena, S.~Pokorski and C.E.M.~Wagner, \emph{{Infrared Fixed
  Point Solution for the Top Quark Mass and Unification of Couplings in the
  MSSM}}, \href{https://doi.org/10.1016/0370-2693(94)90832-X}{\emph{Phys. Lett.
  B} {\bfseries 320} (1994) 110}
  [\href{https://arxiv.org/abs/hep-ph/9309293}{{\ttfamily hep-ph/9309293}}].

\bibitem{Machacek:1983tz}
M.E.~Machacek and M.T.~Vaughn, \emph{{Two Loop Renormalization Group Equations
  in a General Quantum Field Theory. 1. Wave Function Renormalization}},
  \href{https://doi.org/10.1016/0550-3213(83)90610-7}{\emph{Nucl.Phys.}
  {\bfseries B222} (1983) 83}.

\bibitem{Martin:1993zk}
S.P.~Martin and M.T.~Vaughn, \emph{{Two Loop Renormalization Group Equations
  for Soft Supersymmetry Breaking Couplings}},
  \href{https://doi.org/10.1103/PhysRevD.50.2282}{\emph{Phys. Rev. D}
  {\bfseries 50} (1994) 2282}
  [\href{https://arxiv.org/abs/hep-ph/9311340}{{\ttfamily hep-ph/9311340}}].

\bibitem{Martin:1993yx}
S.P.~Martin and M.T.~Vaughn, \emph{{Regularization Dependence of Running
  Couplings in Softly Broken Supersymmetry}},
  \href{https://doi.org/10.1016/0370-2693(93)90136-6}{\emph{Phys. Lett. B}
  {\bfseries 318} (1993) 331}
  [\href{https://arxiv.org/abs/hep-ph/9308222}{{\ttfamily hep-ph/9308222}}].

\bibitem{Farrar:1978xj}
G.R.~Farrar and P.~Fayet, \emph{{Phenomenology of the Production, Decay, and
  Detection of New Hadronic States Associated with Supersymmetry}},
  \href{https://doi.org/10.1016/0370-2693(78)90858-4}{\emph{Phys. Lett. B}
  {\bfseries 76} (1978) 575}.

\bibitem{Dreiner:1997uz}
H.K.~Dreiner, \emph{{An Introduction to Explicit R-Parity Violation}},
  \href{https://doi.org/10.1142/9789814307505_0017}{\emph{Adv. Ser. Direct.
  High Energy Phys.} {\bfseries 21} (2010) 565}
  [\href{https://arxiv.org/abs/hep-ph/9707435}{{\ttfamily hep-ph/9707435}}].

\bibitem{Dawson:1985vr}
S.~Dawson, \emph{{R-Parity Breaking in Supersymmetric Theories}},
  \href{https://doi.org/10.1016/0550-3213(85)90577-2}{\emph{Nucl. Phys. B}
  {\bfseries 261} (1985) 297}.

\bibitem{Barbieri:1985ty}
R.~Barbieri and A.~Masiero, \emph{{Supersymmetric Models with Low-Energy Baryon
  Number Violation}},
  \href{https://doi.org/10.1016/0550-3213(86)90136-7}{\emph{Nucl. Phys. B}
  {\bfseries 267} (1986) 679}.

\bibitem{Barger:1989rk}
V.D.~Barger, G.~Giudice and T.~Han, \emph{{Some New Aspects of Supersymmetry
  R-Parity Violating Interactions}},
  \href{https://doi.org/10.1103/PhysRevD.40.2987}{\emph{Phys. Rev. D}
  {\bfseries 40} (1989) 2987}.

\bibitem{Godbole:1992fb}
R.M.~Godbole, P.~Roy and X.~Tata, \emph{{Tau Signals of R-Parity Breaking at
  Lep-200}}, \href{https://doi.org/10.1016/0550-3213(93)90298-4}{\emph{Nucl.
  Phys. B} {\bfseries 401} (1993) 67}
  [\href{https://arxiv.org/abs/hep-ph/9209251}{{\ttfamily hep-ph/9209251}}].

\bibitem{Bhattacharyya:1995pq}
G.~Bhattacharyya and D.~Choudhury, \emph{{D and Tau Decays: Placing New Bounds
  on R-Parity Violating Supersymmetric Coupling}},
  \href{https://doi.org/10.1142/S0217732395001812}{\emph{Mod. Phys. Lett. A}
  {\bfseries 10} (1995) 1699}
  [\href{https://arxiv.org/abs/hep-ph/9503263}{{\ttfamily hep-ph/9503263}}].

\bibitem{Domingo:2018qfg}
F.~Domingo, H.K.~Dreiner, J.S.~Kim, M.E.~Krauss, M.~Lozano and Z.S.~Wang,
  \emph{{Updating Bounds on $R$-Parity Violating Supersymmetry from Meson
  Oscillation Data}},
  \href{https://doi.org/10.1007/JHEP02(2019)066}{\emph{JHEP} {\bfseries 02}
  (2019) 066} [\href{https://arxiv.org/abs/1810.08228}{{\ttfamily
  1810.08228}}].

\bibitem{Novikov:1983uc}
V.~Novikov, M.A.~Shifman, A.~Vainshtein and V.I.~Zakharov, \emph{{Exact
  Gell-Mann-Low Function of Supersymmetric Yang-Mills Theories from Instanton
  Calculus}}, \href{https://doi.org/10.1016/0550-3213(83)90338-3}{\emph{Nucl.
  Phys. B} {\bfseries 229} (1983) 381}.

\bibitem{Novikov:1985rd}
V.~Novikov, M.A.~Shifman, A.~Vainshtein and V.I.~Zakharov, \emph{{The Beta
  Function in Supersymmetric Gauge Theories. Instantons Versus Traditional
  Approach}}, \href{https://doi.org/10.1016/0370-2693(86)90810-5}{\emph{Sov. J.
  Nucl. Phys.} {\bfseries 43} (1986) 294}.

\bibitem{Intriligator:2003jj}
K.A.~Intriligator and B.~Wecht, \emph{{The Exact superconformal R symmetry
  maximizes a}},
  \href{https://doi.org/10.1016/S0550-3213(03)00459-0}{\emph{Nucl. Phys.}
  {\bfseries B667} (2003) 183}
  [\href{https://arxiv.org/abs/hep-th/0304128}{{\ttfamily hep-th/0304128}}].

\bibitem{HLM22}
G.~Hiller, D.F.~Litim and K.~Moch, \emph{{in preparation}}, .

\bibitem{Anselmi:1997am}
D.~Anselmi, D.~Freedman, M.T.~Grisaru and A.~Johansen, \emph{{Nonperturbative
  Formulas for Central Functions of Supersymmetric Gauge Theories}},
  \href{https://doi.org/10.1016/S0550-3213(98)00278-8}{\emph{Nucl. Phys. B}
  {\bfseries 526} (1998) 543}
  [\href{https://arxiv.org/abs/hep-th/9708042}{{\ttfamily hep-th/9708042}}].

\bibitem{Cardy:1988cwa}
J.L.~Cardy, \emph{{Is There a c Theorem in Four-Dimensions?}},
  \href{https://doi.org/10.1016/0370-2693(88)90054-8}{\emph{Phys. Lett.}
  {\bfseries B215} (1988) 749}.

\bibitem{Osborn:1989td}
H.~Osborn, \emph{{Derivation of a Four-dimensional $c$ Theorem}},
  \href{https://doi.org/10.1016/0370-2693(89)90729-6}{\emph{Phys. Lett.}
  {\bfseries B222} (1989) 97}.

\bibitem{Hofman:2008ar}
D.M.~Hofman and J.~Maldacena, \emph{{Conformal collider physics: Energy and
  charge correlations}},
  \href{https://doi.org/10.1088/1126-6708/2008/05/012}{\emph{JHEP} {\bfseries
  05} (2008) 012} [\href{https://arxiv.org/abs/0803.1467}{{\ttfamily
  0803.1467}}].

\bibitem{Komargodski:2011vj}
Z.~Komargodski and A.~Schwimmer, \emph{{On Renormalization Group Flows in Four
  Dimensions}}, \href{https://doi.org/10.1007/JHEP12(2011)099}{\emph{JHEP}
  {\bfseries 12} (2011) 099} [\href{https://arxiv.org/abs/1107.3987}{{\ttfamily
  1107.3987}}].

\bibitem{tHooft:1979rat}
G.~'t~Hooft, \emph{{Naturalness, chiral symmetry, and spontaneous chiral
  symmetry breaking}},
  \href{https://doi.org/10.1007/978-1-4684-7571-5_9}{\emph{NATO Sci. Ser. B}
  {\bfseries 59} (1980) 135}.

\end{thebibliography}\endgroup

\end{document}